\documentclass[twocolumn,a4paper,pra,aps,nofootinbib,superscriptaddress]{revtex4-2}

\usepackage[utf8x]{inputenc}
\usepackage{ucs}
\usepackage{microtype}
\usepackage{graphicx}

\usepackage{amsmath}
\usepackage{amsfonts}
\usepackage{amssymb}
\usepackage{bm}
\usepackage{braket}
\usepackage{mathtools}

\usepackage[colorlinks=true, linkcolor=blue, citecolor=blue, urlcolor=blue]{hyperref}
\usepackage[capitalize]{cleveref}

\usepackage{tabularx}
\usepackage{cellspace}
\usepackage{booktabs}
\usepackage{newfloat}

\usepackage[per-mode=symbol,separate-uncertainty=true]{siunitx}
\usepackage[table]{xcolor}
\usepackage{chemformula}
\usepackage{xprintlen}
\usepackage{lipsum}
\usepackage{fancyhdr}
\usepackage{listings}

\usepackage[acronym,shortcuts]{glossaries}
\makeglossaries
\newacronym{SQUID}{SQUID}{superconducting quantum interference device}
\newacronym{DC}{DC}{direct current}
\newacronym{AWG}{AWG}{arbitrary waveform generator}
\newglossaryentry{RT}{name=room-temperature,description=room-temperature}
\newacronym{FIR}{FIR}{finite impulse response}
\newacronym{IIR}{IIR}{infinite impulse response}
\newacronym{LTI}{LTI}{linear time-invariant}
\makeglossaries
\glsenableentrycount
\glsdisablehyper

\DeclareMathOperator{\diag}{diag}

\renewcommand{\exp}[1]{\mathrm{e}^{#1}}
\DeclareMathOperator{\dif}{d\!}
\DeclareMathOperator*{\argmax}{argmax}
\newcommand{\Lapl}[1]{\mathcal{L}\left\{#1\right\}}

\DeclareSIUnit\permille{\text{\textperthousand}}

\crefformat{subequations}{Eqs.~(#2#1#3)}
\Crefformat{subequations}{Eqs.~(#2#1#3)}

\let\originalleft\left
\let\originalright\right
\renewcommand{\left}{\mathopen{}\mathclose\bgroup\originalleft}
\renewcommand{\right}{\aftergroup\egroup\originalright}

\newcommand{\fge}[1][]{f_\mathrm{01}^\mathrm{#1}}
\newcommand{\fef}[1][]{f_\mathrm{12}^\mathrm{#1}}
\newcommand{\fidle}[1][]{f_\mathrm{idle}^\mathrm{#1}}
\newcommand{\getrans}{$\ket{0}\leftrightarrow\ket{1}$}
\newcommand{\eftrans}{$\ket{1}\leftrightarrow\ket{2}$}
\newcommand{\textgetrans}{transition}
\newcommand{\textqbtrans}{qubit transition}
\newcommand{\dynphase}{\theta}
\newcommand{\neff}{\ensuremath{t_{\mathrm{int}}}}

\newcommand{\estimatesymbol}{\check}

\newcommand{\postsubscript}{p}
\newcommand{\VDC}{V_\mathrm{DC}}
\newcommand{\phiDC}{\Phi_\mathrm{DC}}
\newcommand{\nC}{N}
\newcommand{\phitotal}{\Phi}
\newcommand{\phichange}{\phi}
\newcommand{\hatphichange}{\estimatesymbol\phi}
\newcommand{\hatphipost}{{\estimatesymbol\phi}_\mathrm{\postsubscript}}
\newcommand{\ffunc}{\nu}
\newcommand{\hatffunc}{\estimatesymbol\ffunc}
\newcommand{\hatfge}[1][]{\estimatesymbol f_\mathrm{01}^\mathrm{#1}}
\newcommand{\hatfgespec}[1][]{\estimatesymbol f_\mathrm{01,spec}^\mathrm{#1}}
\newcommand{\hatfpost}{\estimatesymbol f_\mathrm{\postsubscript}}
\newcommand{\hats}{\estimatesymbol s}
\newcommand{\gpost}{g_\mathrm{\postsubscript}}
\newcommand{\Hi}{H_i}
\newcommand{\gainH}{\kappa}
\newcommand{\disct}{n}
\newcommand{\Hinvp}{1 / H(p)}
\newcommand{\Hinvnorm}{H_\mathrm{inv}}
\newcommand{\hinvnorm}{h_\mathrm{inv}}
\newcommand{\hinvFIR}[1][]{h_{\mathrm{inv}#1}}
\newcommand{\labelfpscopeSeq}{a}
\newcommand{\labelfpscopeTwoD}{b}
\newcommand{\labelfpscopeAppTwoD}{a}
\newcommand{\labelfpscopeAppRabi}{b}
\newcommand{\labelIIRmeas}{a}
\newcommand{\labelIIRresid}{b}
\newcommand{\labelIIRerr}{c}
\newcommand{\labelIIRfit}{d}
\newcommand{\tauHP}{\tau_\mathrm{HP}}
\newcommand{\Ts}{T_\mathrm{s}}
\newcommand{\tfp}{t_\mathrm{fl}}

\newcommand{\labelModelSigflow}{a}
\newcommand{\labelModelFreq}{b}
\newcommand{\statuncertainty}{\sigma_{\mathrm{f}}}
\newcommand{\statuncertaintyphase}{\sigma_{\dynphase}}
\newcommand{\nphases}{N_{\dynphase}}
\newcommand{\nruns}{N_{\mathrm{R}}}

\newcommand{\affiliationeth}{\affiliation{Department of Physics, ETH Zurich, CH-8093 Zurich, Switzerland}}
\newcommand{\affiliationqc}{\affiliation{Quantum Center, ETH Zurich, CH-8093 Zurich, Switzerland}}
\newcommand{\affiliationpsi}{\affiliation{ETH Zurich - PSI Quantum Computing Hub, Paul Scherrer Institute, CH-5232 Villigen, Switzerland}}
\newcommand{\nowataq}{\thanks{Present address: Atlantic Quantum, Cambridge, MA}}
\newcommand{\nowatzi}{\thanks{Present address: Zurich Instruments AG, Zurich, Switzerland}}
\newcommand{\nowatfau}{\thanks{Present address: Department of Physics, Friedrich-Alexander University Erlangen-Nürnberg (FAU), Erlangen, Germany}}
\newcommand{\nowattud}{\thanks{Present address: QuTech and Kavli Institute of Nanoscience, Delft University of Technology, Delft, The Netherlands}}

\begin{document}

\date{March 6, 2025}

\author{Christoph~Hellings}\email{christoph.hellings@phys.ethz.ch}\affiliationeth\affiliationqc
\author{Nathan~Lacroix}\affiliationeth\affiliationqc
\author{Ants~Remm}\nowataq\affiliationeth\affiliationqc
\author{Richard~Boell}\affiliationeth
\author{Johannes~Herrmann}\affiliationeth\affiliationqc
\author{Stefania~Laz\u{a}r}\nowatzi\affiliationeth\affiliationqc
\author{Sebastian~Krinner}\nowatzi\affiliationeth\affiliationqc
\author{Fran\c{c}ois~Swiadek}\affiliationeth\affiliationqc
\author{Christian~Kraglund~Andersen}\nowattud\affiliationeth
\author{Christopher~Eichler}\nowatfau\affiliationeth\affiliationqc
\author{Andreas~Wallraff}\affiliationeth\affiliationqc\affiliationpsi

\title{Calibrating Magnetic Flux Control in Superconducting Circuits by Compensating Distortions on Time Scales from Nanoseconds up to Tens of Microseconds}

\begin{abstract}\vspace*{-2mm}%
Fast tuning of the transition frequency of superconducting qubits using magnetic flux is essential, for example, for realizing high-fidelity two-qubit gates with low leakage or for reducing errors in dispersive qubit readout.
To apply accurately shaped flux pulses, signal distortions induced by the flux control lines need to be carefully compensated for.
This requires their \emph{in situ} characterization at the reference plane of the qubit.
However, many existing approaches are limited in time resolution or in pulse duration.
Here, we overcome these limitations and demonstrate accurate flux control with sub-permille residual frequency errors on time scales ranging from nanoseconds to tens of microseconds.
We achieve this by combining two complementary methods to characterize and compensate for pulse distortions.
We have deployed this approach successfully in a quantum error correction experiment  calibrating 24 flux-activated two-qubit gates.
Reliable calibration methods, as the ones presented here, are essential in experiments scaling up superconducting quantum processors.
\end{abstract}

\maketitle
\glsresetall

\section{Introduction}\vspace*{-2mm}%
\label{sec:intro}
The Josephson energy of a \cgls{SQUID} can be tuned \emph{in situ} by controlling the magnetic flux through the \cgls{SQUID} loop.
For a transmon qubit \cite{Koch2007,Schreier2008},
flux-tuning of the Josephson energy
gives control over its {\textgetrans} frequency.
This not only enables a variety of control and measurement protocols for individual transmon qubits, such as measuring the frequency dependence of the relaxation time $T_1$ \cite{Klimov2018},
tuning the parameters of dispersive qubit readout \cite{Krinner2022},
or resetting qubits to their ground state \cite{Zhou2021,McEwen2021a},
but also provides the basis for implementing fast entangling gates between pairs of qubits.
Flux-activated two-qubit gates include ones based on resonant interactions \cite{Strauch2003,DiCarlo2009,Rol2019,Negirneac2021},
gates parametrically activated by modulating the qubit frequency \cite{Royer2017,Caldwell2018},
and gates based on flux-tunable couplers \cite{Chen2014m,Yan2018b,Sung2021a,Xu2020c,Collodo2020,Foxen2020}.
As two-qubit gates frequently limit the performance of superconducting quantum processors,
for example in the context of implementing quantum error correction \cite{Krinner2022,Marques2021,Acharya2023,Sundaresan2023,Acharya2024a},
accurate flux control is of great importance.

Flux control waveforms created by an \cgls{AWG}
are subject to distortions in the
flux control line, stemming from
the \cgls{AWG} output stage,
the control wiring
leading up to the superconducting quantum device, its package, 
and the on-chip part of the flux line.
For implementing fast two-qubit gates with high fidelity and low leakage \cite{Rol2019,Negirneac2021,Krinner2022},
we aim at compensating distortions on time scales ranging from nanoseconds to tens of microseconds.
Compensation on short time scales avoids ringing and overshoots for rapidly changing flux profiles, while compensation on long time scales enables deep sequences of gates to be executed faithfully.

\emph{In situ} characterization methods \cite{Rol2020} account for all flux control line components \cite{Foxen2018a} as well as their operation at cryogenic temperatures \cite{Rol2020}.
While characterization of the \cgls{RT} part of the flux control line and its electronics could proceed using conventional methods \cite{Johnson2011PhD,Barends2014,Kelly2015a,Foxen2018a,Rol2020} using oscilloscopes, spectrum-, or network analyzers, \emph{in situ} methods can also compensate, for example, for small impedance mismatches between components and their wiring (Appendix~\ref{app:LTI}).  

In prior work, \emph{in situ} characterization was performed by
measuring the instantaneous qubit frequency as a function of the time after the rising or falling edge of a flux pulse using pulsed spectroscopy \cite{Hofheinz2009,Johnson2011PhD,Kelly2015a,Chen2022b}
or a combination of an $X_{\pi/2}$ and a $Y_{\pi/2}$ pulse applied to the qubit  \cite{Kelly2015a,Barends2014}.
For this class of methods, the time resolution is limited by the drive pulse duration
\cite{Johnson2011PhD,Hofheinz2009}.
An alternative method 
is based on characterizing the dynamic phase acquired by a qubit during flux pulses of various durations.
This approach was implemented using quantum state tomography
\cite{Chen2018j,Neill2018,Sung2021a,Braumuller2022}
or Ramsey-type experiments \cite{Rol2020}.
Measuring dynamic phases is inherently limited by the coherence time of the qubit and, thus, not well suited for long flux pulses. A similar limitation applies to the method from \cite{Jerger2019}, which measures the transfer function of the flux control line in the frequency domain by characterizing the qubit with a microwave drive applied to the line.
Other methods require a dedicated on-chip circuit for flux measurements \cite{Foxen2018a}, are designed for tunable couplers \cite{Li2025c}, or are only focused on repeatability of two-qubit gates \cite{Kelly2014}.

Here, we present an end-to-end framework for \emph{in situ} characterization and compensation of flux pulse distortions occuring
on time scales of tens of microseconds (Sec.~\ref{sec:IIR}) down to nanoseconds (Sec.~\ref{sec:FIR}). We extend known methods, which are either limited in their time resolution
or by the coherence times of the qubits employed.
By characterizing flux waveforms up to a duration of \SI{100}{\micro\second}, we go far beyond the flux pulse durations of up to a few microseconds in previous demonstrations of the characterization of flux pulse distortions with flux-tunable qubits
\cite{Johnson2011PhD,Barends2014,Hofheinz2009,Kelly2015a,Sung2021a,Chen2018j,Chen2022,Neill2018,Kelly2014,Rol2020}.
In our procedure, we calibrate a higher-order \cgls{IIR} filter based on a single series of measurements,
and \cgls{FIR} filters using robust regularized optimization at the sampling rate of the \acrlong{AWG}.
Using these methods, we demonstrate accurate flux control with sub-permille residual qubit-frequency errors (Sec.~\ref{sec:results}),
not requiring any prior room-temperature measurements of components.
The presented methods are applicable in settings in which the flux control line is in good approximation linear. 

\section{Flux-Tuning of Transmon Qubits}
\label{sec:model}
\begin{figure}[t!]%
	\includegraphics{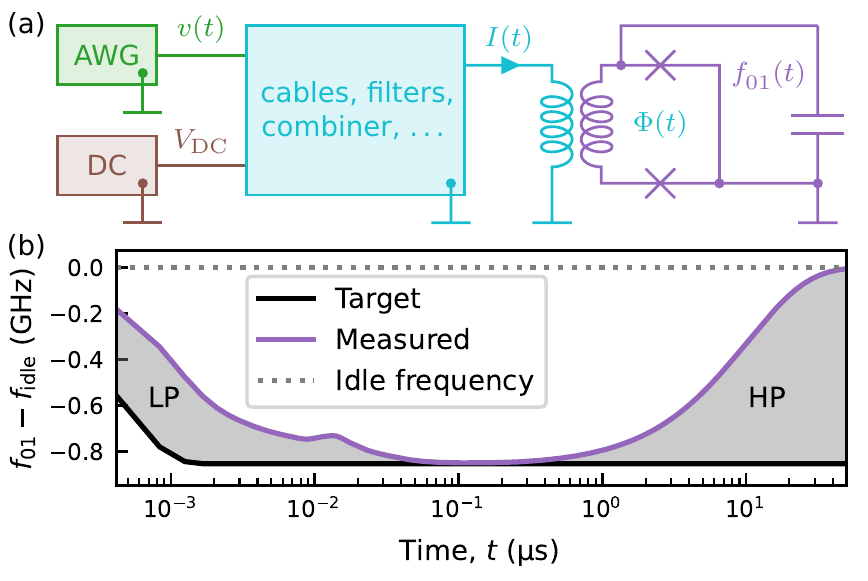}
	\caption{(a) Schematic of the experimental setup (details in Appendix~\ref{app:setup}) with a superconducting transmon qubit (purple) with {\textgetrans} frequency $\fge(t)$, which is tuned by a magnetic flux $\phitotal(t)=\phiDC+\phichange(t)$, consisting of a static bias $\phiDC$ and a time-varying component $\phichange(t)$.
	The magnetic flux is
		induced from an inductively coupled flux line (cyan), to which we apply a DC bias voltage $\VDC$ (brown) and pulses $v(t)$ generated by an arbitrary waveform generator (AWG, green).
		(b) Simulated transition frequency $\Delta\fge(t)$ relative to the idle frequency $\fidle$ in response to a Gaussian-filtered step function at the input of the flux control line
		(simulation based on the measured response of the flux control line, see Appendix~\ref{app:distort_simul}),
		showing distortions with low-pass (LP) and high-pass (HP) characteristic.
	}
	\label{fig:intro}
\end{figure}

We implement flux control of a transmon qubit with a dedicated flux line, which couples inductively to the \cgls{SQUID} loop of the transmon, see Fig.~\ref{fig:intro}\,(a).
A voltage applied at room temperature yields a current through the on-chip flux line, inducing magnetic flux through the \cgls{SQUID} loop, which tunes the {\textgetrans} frequency of the transmon \cite{Koch2007}, see Appendix~\ref{app:hamil} for a detailed model.
By combining a \cgls{DC} bias with flux pulses, we control both the idle frequency and time-dependent frequency excursions of the qubit.
Since flux noise has a detrimental effect on the coherence time of the qubit \cite{Ithier2005},
we reduce the noise from the \cgls{RT} electronics by employing filters in the flux control line. For a detailed description of the experimental setup, see Appendix~\ref{app:setup}.

In Fig.~\ref{fig:intro}\,(b), we show the excursion of the {\textgetrans} frequency $\fge(t)$ of the qubit (solid purple line) from its idle frequency $\fidle$ (dotted gray line) on a logarithmic time scale in response to a Gaussian-filtered step function programmed into the \cgls{AWG}.
The solid black line indicates the target qubit frequency
that would result from the same \cgls{AWG} waveform 
in the absence of distortions in the flux control line.
The dominant distortion at long time scales (beyond approximately $\SI{100}{\nano\second}$)
is an exponential decay of the flux waveform due to
the high-pass characteristic of a bias tee
(time constant $\tauHP=\SI{19.2}{\micro\second}$, see Appendix~\ref{app:biasT}),
which we employ to combine the constant bias voltage with the \cgls{AWG} signal. The bias tee
also suppresses $1/f$ noise from the \cgls{AWG}.
Further contributions at long time scales can originate from other filters,
either intentionally included in the flux control line (not the case in our setup) or formed by parasitic capacitances or inductances together with resistive elements (attenuators).
At short time scales, relevant distortions are caused by
low-pass filters in the output stage of the \cgls{AWG} and at the base plate of the cryogenic setup, see Appendix~\ref{app:setup}.
Due to the uncompensated low-pass characteristic, the qubit reaches the target frequency only after approximately $\SI{50}{\nano\second}$.
Further possible causes for distortions at short time scales are the skin effect and reflections due to impedance mismatches~\cite{Rol2020}.

To model the distortions, we describe them 
by the impulse response $h(t)$ of a \cgls{LTI} system,
which represents the entire flux control line.
The flux change $\phichange(t)$ with respect to the idle flux is given by the output of this \cgls{LTI} system,
see the signal flow diagram in Fig.~\ref{fig:model}\,(\labelModelSigflow)
and the analytical description in Appendix~\ref{app:LTI}.
We introduce a nonlinear function $\ffunc$ to describe the functional dependence of
the {\textgetrans} frequency $\fge(t)$ on the flux change as $\fge(t)=\ffunc(\phichange(t))$.

To characterize the flux-dependence of the {\textqbtrans} frequency $\fge(t)=\ffunc(\phichange(t))$,
we first measure the {\textgetrans} frequency
from the ground state to the first excited state of the transmon
in a series of Ramsey experiments
as a function of a static flux bias,
see Fig.~\ref{fig:model}\,(\labelModelFreq).
The asymmetric \cgls{SQUID} loop of the transmon leads to two first-order flux-insensitive points, see the first and the last data point.
At the upper first-order flux-insensitive point, we also measure the transition frequency
from the first to the second excited state.
We then fit the model to the data as described in Appendix~\ref{app:hamil}, see Fig.~\ref{fig:model}\,(\labelModelFreq).

The distortions of flux control pulses can be characterized by
measuring the {\textqbtrans} frequency in response to a known control waveform $v(t)$
and converting the measurement result $\hatfge(t)$ to a flux change $\hatphichange(t)$ by means of the inverse $\hatffunc^{-1}$ of the qubit frequency model $\ffunc$,
see Fig.~\ref{fig:model}\,(\labelModelSigflow).
Note that we use the symbol~$\estimatesymbol~$ to indicate quantities inferred from measured data,
in contrast to their respective unknown true values,
which appear in the mathematical model without this symbol.

To compensate for the characterized distortions, we employ digital prefiltering (also called predistortion, e.g., \cite{Rol2020,Rice2009}).
That is,
we program an intentionally distorted version $v(t)$
of the desired waveform $a(t)$ into the \cgls{AWG},
in a way that
distortions in the flux line wiring and electronics are compensated for.
The calibration of this predistortion consists in designing an \cgls{LTI} system with impulse response $g(t)$ such that
the cascade of the predistortion filter $g(t)$ and the physical system $h(t)$ form a scaled identity operation.
This means that we aim at 
making $\phichange(t)$ equal to the input waveform $a(t)$ up to a scaling factor.

\begin{figure}[tb]
	\centering
	\includegraphics{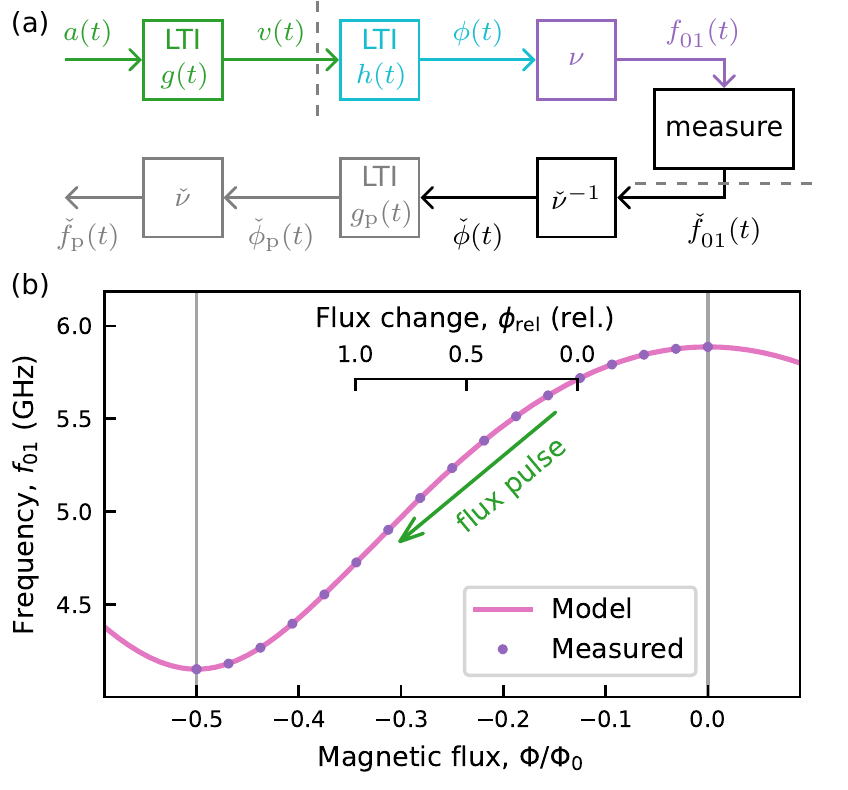}
	\caption{%
		(\labelModelSigflow) Signal flow diagram (details in the main text), showing the flux control (green, cyan) of the {\textgetrans} frequency (purple) of a transmon, the characterization measurement (black), and optional analysis steps (gray, see Sec.~\ref{sec:IIR}).
		The signal flow between (outside) the dashed lines is in hardware (software).
		(\labelModelFreq) Measured {\textqbtrans} frequency (purple points) and fitted model (pink solid line)
		plotted vs.\ a static magnetic flux bias $\phitotal=\phiDC$ normalized by the flux quantum $\Phi_0$.
		The inset axis illustrates the flux tuning (relative to the maximal flux change) with a flux pulse starting from a given DC bias (origin of the inset axis)
		during a measurement for characterizing flux pulse distortions (see Sec.~\ref{sec:IIR}).
	}
	\label{fig:model}
\end{figure}

\section{Distortions at Long Time Scales}
\label{sec:IIR}
\begin{figure}[t!]
	\includegraphics[width=\linewidth]{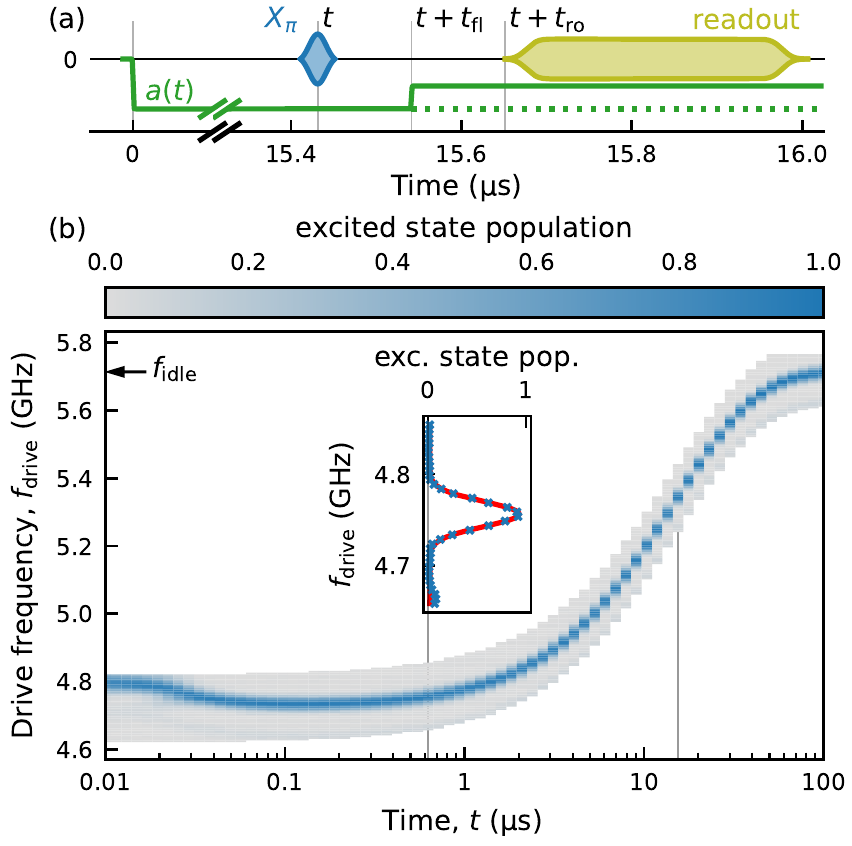}
	\caption{
		\emph{In situ} characterization of the flux line step response.
		(\labelfpscopeSeq) Pulse sequence showing 
		envelopes of the qubit drive pulse (blue) and the readout pulse (olive), as well as the flux waveform (solid green),
		which corresponds to a Gaussian-filtered step function (dotted green)
		with an adjusted flux value for achieving consistent readout (details in main text).
		The times correspond to
		the measurement indicated by the right gray vertical line in (\labelfpscopeTwoD).
		(\labelfpscopeTwoD) Measured excited state population in a two-dimensional sweep of drive pulse time $t$ and drive frequency $f_\mathrm{drive}$.
		The inset shows the data (markers) for a fixed drive pulse time indicated by the left gray vertical line, along with a Gaussian fit (solid line) of the qubit spectral line.
	}
	\label{fig:fpscope}
\end{figure}

For characterizing the distortions in the flux control line,
we avoid measuring close to the first-order flux-insensitive points, where the 
dependence of the qubit transition frequency $\hatffunc$ on the magnetic flux cannot be inverted.
Thus, we first adjust the \cgls{DC} bias such that the qubit is tuned to a flux-sensitive point, see the origin of the relative flux change axis in Fig.~\ref{fig:model}\,(\labelModelFreq).
At the selected flux bias $-0.127\,\Phi_0$,
the slope of the flux-frequency-curve of the considered qubit is
approximately at half
its maximal value.
We then use an \cgls{AWG} to apply a step function, smoothed by a Gaussian filter of width $\sigma_\mathrm{FP} = \SI{0.5}{\nano\second}$, see Fig.~\ref{fig:fpscope}\,(\labelfpscopeSeq).
The step height of $0.217\,\Phi_0$
is chosen to be similar to typical pulse amplitudes for implementing two-qubit gates on this device.
We do not apply a predistortion filter at this point, i.e., $v(t)=a(t)$ in Fig.~\ref{fig:model}\,(\labelModelSigflow).

We measure the resulting
time dependence of the {\textgetrans} frequency $\hatfge(t)$
by probing in a two-dimensional sweep at which combinations of drive time $t$ and drive frequency $f_\mathrm{drive}$ a qubit transition is resonantly driven.
The drive pulse has a Gaussian envelope of width $\sigma_\mathrm{drive} = \SI{10}{\nano\second}$,
and we adapt the pulse amplitude as a function of $f_\mathrm{drive}$
to implement a $\pi$ pulse at any $f_\mathrm{drive}$.
Details of the drive pulse generation are presented in Appendix~\ref{app:comp_pulse}.

To achieve consistent qubit readout despite large frequency excursions of the qubit, 
we tune the qubit back to its initial bias point $\phichange=0$ at the time $t+\tfp=t+\SI{110}{\nano\second}$
before starting the readout at $t+t_\mathrm{ro}=t+\SI{220}{\nano\second}$.
Due to the long-time distortions we need to apply a compensation pulse with negative amplitude to reset the flux to its idling value, see Appendix~\ref{app:comp_pulse}.
We restrict the sweep to drive frequencies $f_\mathrm{drive}$ in a range of $\pm\SI{100}{\mega\hertz}$ around the {\textgetrans} frequency calculated based on a theoretical model of the dominant distortion, see Fig.~\ref{fig:fpscope}\,(\labelfpscopeTwoD).
To characterize distortions across several orders of magnitude in time,
we measure the excited state population for $70$ logarithmically spaced values of the drive time $t$ from $t=\SI{10}{\nano\second}$ up to $t=\SI{100}{\micro\second}\approx5\,\tauHP$
and average the results over $2048$ experimental runs.

\begin{figure}[t!]
	\includegraphics[width=\linewidth]{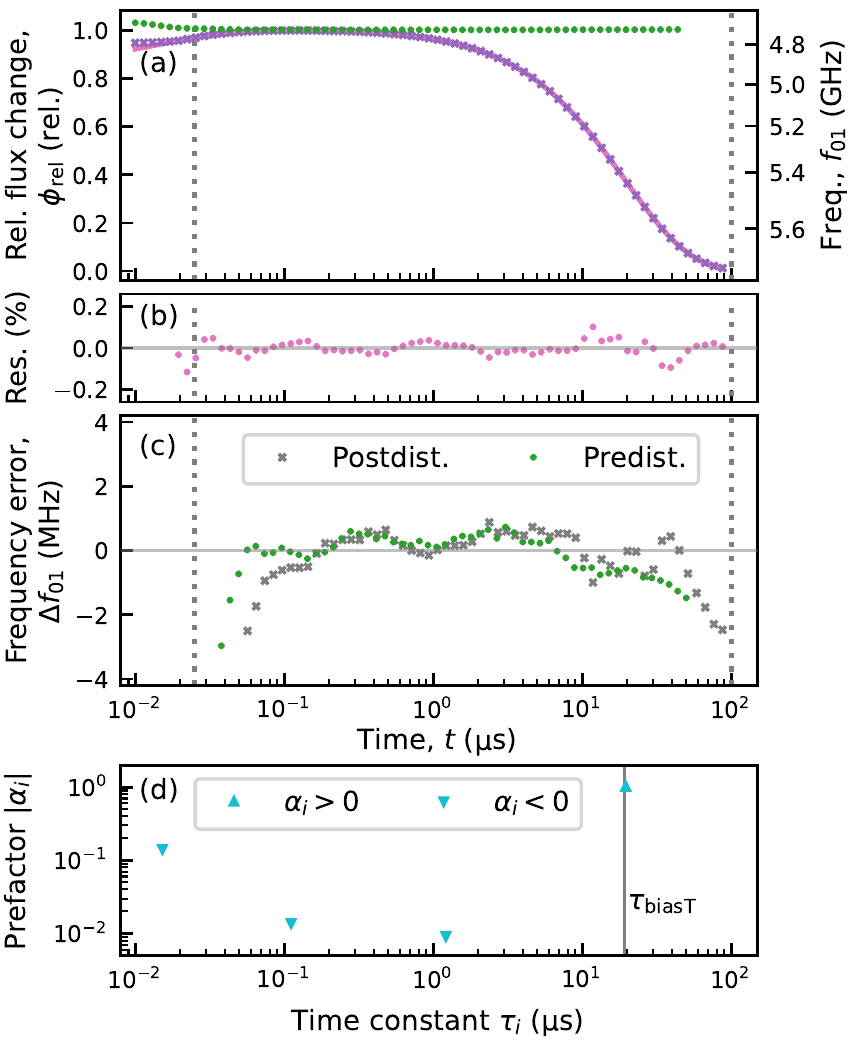}
	\caption{%
(\labelIIRmeas)~Flux change at the qubit (relative to the maximum change) and corresponding {\textgetrans} frequency when programming a step function into the AWG (blue crosses),
and fit of a sum of exponentials (pink solid line) to the data points between the dotted gray lines.
The green dots result from repeating the measurement after calibrating an \acrlong{IIR} filter
as predistortion filter.
(\labelIIRresid)~Residuals of the fit in (\labelIIRmeas).
(\labelIIRerr)~Difference between the measured {\textgetrans} frequency and its targeted value ($\Delta\fge=\fge-f_\mathrm{target}$) when compensating the distortions in the flux control line with an inverse \cgls{IIR} filter based on the fit in (\labelIIRmeas):
expected performance (gray crosses) calculated via postdistortion (details in the main text),
and verification measurement (green points) when applying the \cgls{IIR} filter for predistortion.
(\labelIIRfit)~Parameters of the sum-of-exponentials model fitted in (\labelIIRmeas).
	}
	\label{fig:IIRfitting}
\end{figure}

Unlike a continuous drive tone, a drive pulse combined with a subsequent measurement only leads to a weak power dependence of the measured transition linewidth of the qubit \cite{Vitanov2001}. In this case the effective linewidth observed in a spectroscopic measurement is determined by the spectral width of the drive pulse.
This is observed in the data for a single time $t$ shown in the inset of Fig.~\ref{fig:fpscope}\,(\labelfpscopeTwoD),
which is well approximated with a Gaussian,
whose width is in good agreement with the
spectral width $\sigma_{\mathrm{drive},f}
= (2\pi\sigma_\mathrm{drive})^{-1}
= \SI{16}{\mega\hertz}$
of the Gaussian drive pulse.
The center of the Gaussian fit yields the measured {\textqbtrans} frequency $\hatfge(t)$, see right axis of Fig.~\ref{fig:IIRfitting}\,(\labelIIRmeas).
Standard error estimates calculated from the variance of the fit residuals
are on average
$\SI{0.2}{\mega\hertz}$
in the presented data.
With the inverse qubit frequency model $\hatffunc^{-1}$ from Sec.~\ref{sec:model}, we calculate the flux change $\hatphichange(t)$
and plot it relative to the largest flux excursion observed in the measurement, see left axis of Fig.~\ref{fig:IIRfitting}\,(\labelIIRmeas).
This curve corresponds to a normalized estimate of the step response of the flux control line, i.e., of the integral over the impulse response $h(t)$, see~Appendix~\ref{app:LTI}.

We model this step response as
\begin{equation}
\label{eq:sumexp}
s(t) = \left(\alpha_0 + \sum_{i=1}^\nC \alpha_i \, \exp{-\frac{t}{\tau_i}}\right)u(t)
\end{equation}
with the unit step function $u(t)$.
This model accounts for distortions from an overdamped $\nC$th order linear dynamic circuit \cite{Chua1987}
with $\nC$ linearly independent reactive elements, i.e., capacitors and inductors, and an arbitrary number of linear resistive elements (see Appendix~\ref{app:LTI}).
Similar models have previously been used in \cite{Foxen2018a,Sung2021a,Braumuller2022}.
We choose $\alpha_0 = 0$ (accounting for the high-pass in the bias tee),
and fit the remaining parameters
$\alpha_1,\dots,\alpha_\nC$ and $\tau_1,\dots,\tau_\nC$, see Appendix~\ref{app:IIR}.
Fig.~\ref{fig:IIRfitting}\,(\labelIIRfit) shows the parameters obtained with $\nC=4$, yielding the fit in panel~(\labelIIRmeas) with the fit residuals in panel~(\labelIIRresid).
The fitted time constant with the largest prefactor is in good agreement with a \cgls{RT} measurement
of the time constant of the high-pass in the bias tee (gray line, see Appendix~\ref{app:biasT}).

Note that we have restricted the fit to the time range between the gray dotted lines in Fig.~\ref{fig:IIRfitting}
because the spectroscopic approach 
is not able to track changes of the {\textqbtrans} frequency on time scales significantly below the duration of the drive pulse.
In the measurement presented here, the drive pulse duration was $\SI{40}{\nano\second}$,
resulting
from the width $\sigma_\mathrm{drive} = \SI{10}{\nano\second}$ of the Gaussian pulse shape and a truncation of the pulse to $4\sigma_\mathrm{drive}$.

As discussed in Appendix~\ref{app:IIR}, we use the Laplace transform to derive a filter with impulse response $\hinvnorm(t)$, which inverts the fitted step response up to a scaling factor.
Using a root finding method and the $z$-transform, we obtain a numerically stable digital implementation
of the filter as a cascade of second-order \acrlong{IIR} filters, so-called second-order sections (SOS) \cite{Oppenheim1999},
each of which is implemented digitally as a recursive equation.

Applying $\hinvnorm$ as a predistortion filter, we re-measure the {\textqbtrans}
frequency with the spectroscopic method for waveform durations up to $\SI{50}{\micro\second}$,
see the green dots in Fig.~\ref{fig:IIRfitting}\,(\labelIIRmeas).
For most of the time interval that was taken into account for the fit (between dotted gray lines),
we observe residual frequency errors of less than $\SI{2}{\mega\hertz}$,
see Fig.~\ref{fig:IIRfitting}\,(\labelIIRerr),
with larger deviations at times smaller than $\SI{40}{\nano\second}$.

As a qualitative indicator of whether this data set suffers from significant systematic errors of the measurement method, we 
compare the results to post-processed data from the initial characterization measurement,
where we apply $\hinvnorm$ as a postdistortion filter
$\gpost(t) = \hinvnorm(t)$
as indicated in Fig.~\ref{fig:model}\,(\labelModelSigflow).
We convert the resulting flux $\hatphipost(t)$ to the post-processed {\textqbtrans} frequency $\hatfpost(t)$, see the crosses in Fig.~\ref{fig:IIRfitting}\,(\labelIIRerr).
Under ideal conditions,
the commutativity of linear filtering  implies that the effect of the postdistortion is equivalent to applying $g(t)=\hinvnorm(t)$ as a predistortion filter
and remeasuring the {\textgetrans} frequency as done above.
However, when errors propagate through the post-processing,
which includes two nonlinear transformations $\hatffunc^{-1}$ and $\hatffunc$
as shown in Fig.~\ref{fig:model}\,(\labelModelSigflow),
this equivalence breaks down.
We thus expect measurement errors or model mismatches to manifest themselves
in the post-processed $\hatfpost(t)$
in a different way than in the directly measured $\hatfge(t)$ with predistortion.
The good agreement of most of the data points in Fig.~\ref{fig:IIRfitting}\,(\labelIIRerr)
hints at
small measurement errors, indicating that $\hatphichange(t) \approx \phichange(t)$,
and at accurate model assumptions,
including the assumption that distortions are linear
and the assumption that the 
qubit frequency $\ffunc(\phichange)$ is well described by the model $\hatffunc(\phichange)$.
Residual errors in the qubit frequency model $\hatffunc$ can explain discrepancies at long time scales,
where the measurements with postdistortion and with predistortion are performed at significantly different {\textgetrans} frequencies, see Fig.~\ref{fig:IIRfitting}\,(\labelIIRmeas).
Furthermore, we observe a slight discrepancy at short times, which we attribute to errors due to the limited time resolution of the measurement.
These errors, which affect $\hatfge(t)$ at times that are smaller than the drive pulse duration,
propagate to later points in time in $\hatfpost(t)$
due to error propagation in the postdistortion filter $\gpost$.

To further reduce the residual frequency errors of approximately $\SI{2}{\mega\hertz}$ observed in Fig.~\ref{fig:IIRfitting}\,(\labelIIRerr),
we treat the cascade of digital predistortion and physical flux control line
as an effective \acrlong{LTI} system.
By computing the flux change from the measured frequencies with predistortion,
we obtain the step response of this effective \cgls{LTI} system.
From this step response,
which describes the residual distortions that remain despite the predistortion,
we calculate another inverse filter and apply it as an additional predistortion filter, i.e., cascaded with the initially calibrated predistortion filter.
This additional predistortion step reduces the residual errors to approximately $\SI{0.5}{\mega\hertz}$, as we discuss
in Sec.~\ref{sec:results}.

Unlike methods based on state tomography or Ramsey experiments,
the spectroscopic method presented in this section characterizes the step response of the flux control line at time scales longer than the coherence time of the qubit.
However, the time resolution is inherently limited by the drive pulse duration as discussed above.
We thus characterize distortions on the nanosecond time scale in a separate measurement described in Sec.~\ref{sec:FIR}.

\section{Distortions at Short Time Scales}
\label{sec:FIR}

After calibrating \acrfull{IIR} filters for time scales
from $\SI{100}{\micro\second}$ down to $\SI{25}{\nano\second}$, see dashed vertical lines in Fig.~\ref{fig:IIRfitting},
we employ the cryoscope method \cite{Rol2020} to characterize distortions at time scales below $\SI{50}{\nano\second}$.
To this end, we model the cascade of the \cgls{IIR} predistortion filters and the physical flux control line as an effective \acrlong{LTI} system
with impulse response $h(t)$,
and aim at reducing short-time distortions with an additional \acrfull{FIR} filter $g(t)$.
We first characterize the response of the {\textqbtrans} frequency to a rectangular flux pulse $a(t)$ of duration $\SI{100}{\nano\second}$
without any \cgls{FIR} filter applied to the input signal, i.e., $v(t)=a(t)$.

\begin{figure}[t!]
	\includegraphics[width=\linewidth]{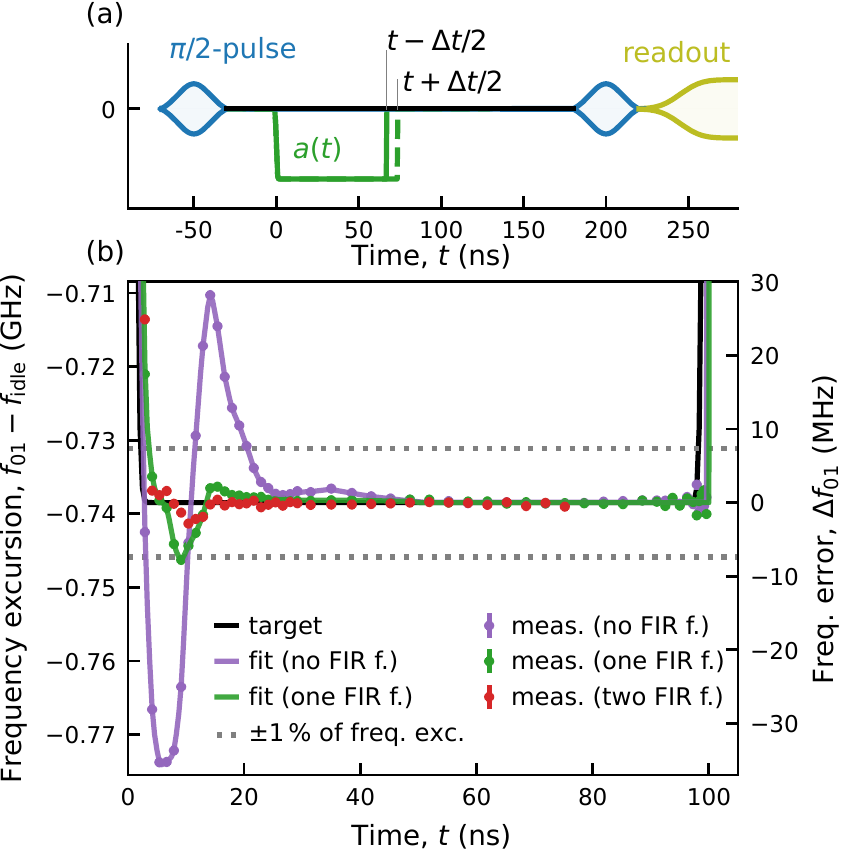}
	\caption{Characterization and correction of  distortions at time scales below $\SI{50}{\nano\second}$. (a) Pulse sequence showing envelopes of the $\pi/2$ drive pulses (blue), the readout pulse (olive), and two flux pulses differing in duration by $\Delta t$ (solid and dashed green),	see main text for details. (b) Measured (dots) and fitted qubit frequency excursion (solid lines) with respect to the idle frequency $\fidle$
	when applying a rectangular flux pulse
	without an \cgls{FIR} filter (purple),
	with a single \cgls{FIR} filter (green),
	or with two \cgls{FIR} filters (red).}
	\label{fig:cryoscope}
\end{figure}

Following \cite{Rol2020},
we measure the dynamic phase accumulated by the qubit during a flux pulse
by enclosing the flux pulse between two $\pi/2$ pulses and sweeping the phase of the second $\pi/2$ pulse.
The pulse sequence of this Ramsey-type measurement is illustrated in Fig.~\ref{fig:cryoscope}\,(a).
By truncating the flux pulse first at $t-\Delta t/2$ and then, in a second experiment, at $t+\Delta t/2$,
yielding the dynamic phase $\dynphase_{t-\Delta t/2}$ and $\dynphase_{t+\Delta t/2}$, respectively,
we approximate the excursion of the qubit from its idle transition frequency as
\begin{equation} \label{eq:cryoscope}
	\hatfge(t) - \fidle = \frac{\dynphase_{t+\Delta t/2} - \dynphase_{t-\Delta t/2}}{2\pi\Delta t},
\end{equation}
see the purple markers in Fig.~\ref{fig:cryoscope}\,(b).

As long as there are uncompensated distortions,
the method leads to systematic errors if the difference between
$\fge(t-\Delta t/2)$ and $\fge(t+\Delta t/2)$ is large,
i.e., in case of steep rising or falling edges \cite{Rol2020}.
This is because any difference in the turn-off transients when truncating the flux pulse
at times $t-\Delta t/2$ and $t+\Delta t/2$
affects the difference of measured dynamic phases in \cref{eq:cryoscope}.
We partially mitigate these errors
by choosing the idle frequency $\fidle$ to be at
a first-order flux-insensitive bias point,
which weakens the effect of flux variations due to turn-off transients
on the measured dynamic phase \cite{Rol2020}.

The time resolution of this method to determine the {\textqbtrans} frequency is given by the step size $\Delta t$, which can be as small as the AWG sampling interval $\Ts$, i.e., 
$\Delta t\geq \Ts=1/(\SI{2.4}{\giga\hertz}) = \SI{0.42}{\nano\second}$ in our setup, see Appendix~\ref{app:setup}.
On the one hand, 
the difference between $\fge(t-\Delta t/2)$ and $\fge(t+\Delta t/2)$ during edges of the pulse,
which causes the systematic errors discussed above,
becomes smaller when reducing the step size $\Delta t$.
On the other hand, increasing $\Delta t$ reduces the number of data points to be measured and reduces statistical errors as it acts as an effective low pass filter, see Appendix~\ref{app:cryoscope}.

We obtained the purple data points in Fig.~\ref{fig:cryoscope}\,(b) by applying the described procedure with $16$ different phases of the second $\pi/2$ pulse and averaged over $65536$ repetitions,
while varying $\Delta t$ as function of $t$ in the range between one and eight times the AWG sampling period $\Ts$, see details in Appendix~\ref{app:cryoscope}.
This leads to standard errors between
$\SI{0.25}{\mega\hertz}$ (for $\Delta t=8\Ts=\SI{3.3}{\nano\second}$) and
$\SI{0.73}{\mega\hertz}$ (for $\Delta t=\Ts=\SI{0.42}{\nano\second}$).

We observe distortions not compensated for by the long time-scale methods in the initial part of the flux pulse,
which are most pronounced in the first $\SI{25}{\nano\second}$, see purple data points in Fig.~\ref{fig:cryoscope}\,(b).
This is the interval that was excluded from the fit in the \cgls{IIR} calibration (see dashed line in Fig.~\ref{fig:IIRfitting})
due to the limited time resolution of the spectroscopic measurement.
In fact, the \cgls{IIR} filters for compensating distortions at longer time scales can contribute
to additional distortions at time scales that cannot be resolved with the method used in Sec.~\ref{sec:IIR}, such as the large overshoot after the initial falling edge, see the purple markers in Fig.~\ref{fig:cryoscope}\,(b).
The \cgls{FIR} filter calibration described below compensates for short-time distortions independently of whether they have a physical origin
or whether they are artifacts caused by the previously applied
filters.

Since the distortions at short time scales have multiple origins,
see Sec.~\ref{sec:model},
we describe the transmission through the cascade of \cgls{IIR} predistortion filters and the physical flux control line by a generic \cgls{FIR} impulse response
\begin{equation}
\label{eq:FIRmodel}
h(t) = \sum_{\ell = 0}^{L-1}{h_\ell \,\delta(t-\ell\Ts)}
\end{equation}
with $L=120$ free parameters $h_\ell$ spaced by the \cgls{AWG} sampling period $\Ts$,
where $\delta$ is the Dirac impulse.
To fit this model to the acquired data set of measured frequencies $\hatfge(t)$,
we use a regularized optimization \cite{Milton1991, Borchers13} (see Appendix~\ref{app:FIR})
to overcome the following challenges.
First, unlike for the sum-of-exponentials model used in Sec.~\ref{sec:IIR}, the high number of coefficients in \cref{eq:FIRmodel} makes the model susceptible to overfitting to noise.
Second, as the qubit idles at a first-order flux-insensitive bias point to mitigate errors due to turn-off transients as discussed above,
the function $\ffunc$ mapping flux change $\phichange$ to {\textqbtrans} frequency $\fge$ is not invertible at $\fidle$.
This makes the inverse problem of inferring $\hatphichange$ from the measured frequencies $\hatfge$ ill-conditioned for frequencies close to the idle point.
For these frequencies, even small errors in $\hatfge$ would lead to
severe errors in the inferred flux change $\hatphichange$
when explicitly using the inverse qubit frequency model $\hatffunc^{-1}$.
The regularized optimization overcomes both problems because it avoids an explicit inversion and provides robustness to overfitting, see Appendix~\ref{app:FIR}.

The purple line in Fig.~\ref{fig:cryoscope}\,(b) shows the result of
filtering the rectangular input pulse with the fitted impulse response of the residual distortions, $h(t)$, and
applying the qubit frequency model $\hatffunc$.
Further, we compute an inverse \cgls{FIR} filter $\hinvFIR$ via a second regularized optimization, see Appendix~\ref{app:FIR}.
An alternative approach based on a Savitzky-Golay filter and the CMA-ES algorithm is presented in \cite{Rol2020} with extensions in \cite{Ferreira2024}.

With the inverse filter applied as a predistortion filter $g(t)=\hinvFIR(t)$,
we remeasure $\hatfge(t)$ (green markers in Fig.~\ref{fig:cryoscope}\,(b)) and
perform the same fitting procedure as before (solid green line),
revealing residual frequency deviations of up to approximately $\SI{8}{\mega\hertz}$ from the target,
corresponding to $1\,\%$ of the total frequency excursion.
We attribute the deviations mostly to the fact that
we calibrated the \cgls{FIR} predistortion filter 
based on a characterization data set that suffers from systematic errors of the method from \cite{Rol2020}.

As the systematic errors in the method from \cite{Rol2020} are caused by turn-off transients,
they are expected to be reduced by the calibrated \cgls{FIR} filter, which reduces these transients.
We thus used the data set obtained with this filter to calibrate a second \cgls{FIR} filter in order to reduce residual distortions,
see the red markers in Fig.~\ref{fig:cryoscope}\,(b), which show the response to a Gaussian-filtered step function with all predistortion filters (two IIR and two FIR filters) applied.

\section{Performance Verification and Application for Two-Qubit Gates}
\label{sec:results}

\begin{figure}[t!]
	\includegraphics[width=\linewidth]{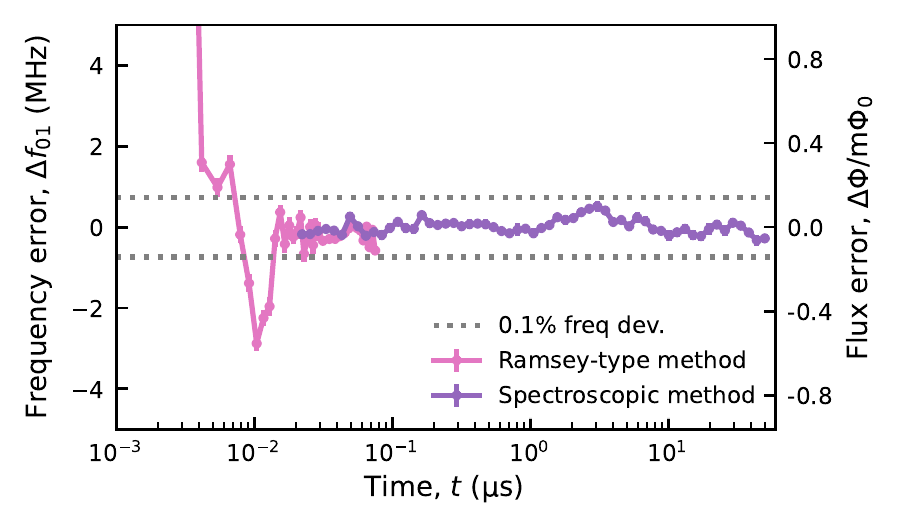}
	\caption{Performance verification with predistortion filters:
		{\textgetrans} frequency over time in response to a Gaussian-filtered step function at the input of the predistortion filters.
		The dashed lines indicate a deviation of $0.1\,\%$ of the intended frequency change of $-\SI{738}{\mega\hertz}$ (flux change $\approx\Phi_0/4$).
		}
	\label{fig:eval}
\end{figure}

After two iterations each of calibrating \acrlong{IIR} filters and \acrlong{FIR} filters,
we evaluate the compensation of distortions at long and short time scales using the methods from Sec.~\ref{sec:IIR} and Sec.~\ref{sec:FIR}, respectively.
For both measurements, when the qubit idles at the upper first-order flux-insensitive bias point
we apply a flux pulse with the same amplitude of approximately $\Phi_0/4$ as for the calibration of the FIR filters, yielding
a frequency change of $-\SI{738}{\mega\hertz}$.
This is a representative pulse amplitude
for flux pulses implementing two-qubit gates on the device used for the present study \cite{Krinner2022}.
From approximately $\SI{15}{\nano\second}$ after the rising edge, the deviation from the target frequency is below $0.1\,\%$ of the total frequency excursion,
see Fig.~\ref{fig:eval}.
This corresponds to a relative flux error of approximately $5\cdot 10^{-4}$, which is similar to the value of $10^{-3}$ reported in \cite{Rol2020}.
We attribute the slightly larger deviations at times before $\SI{15}{\nano\second}$ to miscalibrations resulting from systematic errors of the characterization method at steep edges, see Sec.~\ref{sec:FIR}.
For times at which data from both characterization methods is available,
the good agreement between the two methods
hints at small systematic errors
since we would expect the two conceptually different methods to display different systematic errors.

To study a particularly relevant use case of flux control,
we have implemented controlled phase (CZ) gates via resonant $\ket{11}\leftrightarrow\ket{20}$ interactions \cite{DiCarlo2009,Negirneac2021} between
the $24$ pairs of
neighboring qubits on a 17-qubit device \cite{Krinner2022},
see Appendix~\ref{app:CZ}.
We perform interleaved randomized benchmarking \cite{Magesan2012,Corcoles2013,Barends2014} with extensions to quantify leakage \cite{Rol2019,Wood2017},
yielding gate errors on individual qubit pairs between 0.006 and 0.054
with a mean of 0.015(10) across all 24 pairs \cite{Krinner2022}
and a mean leakage per gate of 0.0011(10).
Dominant contributions to the gate errors originate from crossing defect modes
and from increased dephasing rates at the interaction frequencies due to flux noise,
see \cite{Krinner2022}.

The compensation of distortions on short time scales enables
steep rising and falling edges,
which is crucial to mitigate interactions when crossing defect modes \cite{Krinner2022}
or other transitions accessible in a pair of transmon qubits
(e.g., in a case where the
$\ket{11}\leftrightarrow\ket{02}$ and
$\ket{01}\leftrightarrow\ket{10}$ resonances are crossed to reach the 
$\ket{11}\leftrightarrow\ket{20}$ resonance \cite{Negirneac2021}).
The overshoot of $\SI{3}{\mega\hertz}$, which we observe around $t=\SI{10}{\nano\second}$ in Fig.~\ref{fig:eval},
is nearly two orders of magnitude below the
anharmonicity of $\SI{174}{\mega\hertz}$, which is a typical anharmonicity for transmon qubits.
Thus, this overshoot does not cause undesired $\ket{10}\leftrightarrow\ket{01}$ interactions.
As a reference value to judge residual errors at intermediate time scales,
we determine the coupling strength of the $\ket{11}\leftrightarrow\ket{20}$ interaction for the considered transmon and one of its neighbors as $J_2/2\pi = \SI{7.4}{\mega\hertz}$,
see Appendix~\ref{app:CZ}.
This is a representative value for a qubit-qubit coupling that allows two-qubit gates with a duration below $\SI{100}{\nano\second}$.
The residual errors after the overshoot in Fig.~\ref{fig:eval}
are clearly small enough to avoid significant off-resonant interactions.

Finally, the above interleaved randomized benchmarking results obtained with gate sequences lasting up to $\SI{50}{\micro\second}$ also demonstrate the successful mitigation of long-time distortions which would otherwise affect two-qubit gate performance.
This mitigation is achieved by the long-time \cgls{IIR} filters and by making use of net-zero pulses \cite{Rol2019,Negirneac2021,Karamlou2022},
consisting of two shorter pulses with opposite polarity
such that the time average of the control signal vanishes, see Fig.~\ref{fig:gate}\,(a) in Appendix~\ref{app:CZ}.
The vanishing time average leads to robustness against distortions on time scales much longer than the duration of an individual two-qubit gate.
This can be intuitively understood in the Fourier domain by noting that the exact values of the transfer function of the flux control line
at frequencies close to zero
do not matter for control signals that do not have any spectral content in that frequency range.

Accurate compensation of long-time distortions by means of filtering as demonstrated in this paper is indispensable when unipolar pulses are applied to quantum devices, e.g., in scenarios in which the idle point is not a symmetry point. Such scenarios occur, for example, when defect modes (two-level systems) \cite{Mueller2019}) restrict the choice of the idle frequency \cite{Sung2021a,Acharya2023}. They also result from design choices, for example, for fluxonium qubits that implement a $\Lambda$~system \cite{Earnest2018b}, for tunable couplers idling at the point of minimal coupling \cite{Xu2020c,Collodo2020,Foxen2020}, or for tunable resonators used as flux-sensitive devices \cite{Grover2020}.

We have demonstrated the compensation of flux distortions on long and short time scales by combining two
methods for characterizing the qubit frequency as a function of time while applying flux control pulses.
At long time scales, we have extended a spectroscopic  method 
\cite{Hofheinz2009,Johnson2011PhD} which enables the characterization of long flux waveforms of up to $\SI{100}{\micro\second}$ with large excursions of the qubit frequency up to $\SI{1}{\giga\hertz}$.
Our approach allows us to calibrate compensations for multiple time scales 
at once by directly characterizing the step response
and by applying a numerically robust implementation of an inverse, higher-order infinite impulse response (IIR) filter.
To compensate for short-time distortions,
we have combined the characterization method from \cite{Rol2020}
with a
method for fitting finite impulse response (FIR) filter coefficients based on regularized optimization.
Our combination of methods achieves accurate flux control with sub-permille residual frequency errors on time scales from nanoseconds to tens of microseconds.
We have demonstrated that the methods presented in this paper allow us to implement fast flux-activated two-qubit gates with high fidelity and low leakage, which are essential for example in  quantum error correction experiments such as the one we presented in Ref.~\onlinecite{Krinner2022}.
Furthermore, compensation of flux distortions 
is a key ingredient for achieving high fidelity in 
experiments which require long or repeated unipolar flux pulses.

\section*{Author Contributions}
C.H., N.L., A.R., R.B., and C.K.A.\ developed the methodology,
C.H.\ and N.L.\ planned and performed the measurements,
and C.H., N.L., and A.R. analyzed the data.
C.H., N.L., A.R., and S.L.\ developed control and calibration software routines,
and C.H.\ and R.B.\ performed numerical simulations.
J.H., A.R., R.B., C.H., S.K., and F.S.\ designed and built elements of the experimental setup.
F.S., A.R., and C.K.A. designed the device,
and S.K. and A.R. fabricated the device.
C.H.\ and N.L.\ prepared the figures and wrote the manuscript with input from all co-authors.
C.E.\ and A.W.\ supervised the work.

\section*{Acknowledgments}
We acknowledge the contributions of G.J.~Norris to device fabrication,
and of M.~Kerschbaum to early work on the two-qubit gate implementation.
Research was sponsored by IARPA and the Army Research Office,
under the Entangled Logical Qubits program,
and was accomplished under Cooperative Agreement Number W911NF-16-1-0071 and W911NF-23-2-0212,
by the Swiss National Science Foundation under R'equip grant 206021-170731,
by the Baumgarten foundation
and by ETH Zurich.
The views and conclusions contained in this document are those of the authors
and should not be interpreted as representing the official policies,
either expressed or implied, of IARPA,
the Army Research Office, or the U.S. Government.
The U.S. Government is authorized to reproduce and distribute reprints
for Government purposes notwithstanding any copyright notation herein.

\appendix

\section{Linear Time-Invariant Systems and Linear Dynamic Circuits}
\label{app:LTI}
The output $y(t)$ of a \acrfull{LTI} system with impulse response $h(t)$ can be calculated from the convolution
\begin{equation}
\label{eq:filters:model}
y(t) = (h \ast v)(t)  = \int_{-\infty}^\infty v(t') h(t-t') \mathrm{d}t',
\end{equation}
where $v(t)$ is the input signal.
When using this model to describe dynamic distortions in a flux control line,
the input $v(t)$ is the voltage programmed to the AWG, and we interpret the change of the flux through the \cgls{SQUID} loop $\phichange(t)$ as the output,
i.e., $y(t)=\phichange(t)$.
An \cgls{LTI} system can be equivalently described by its step response $s(t)$
via
\begin{align}
\label{eq:stepres_impres_transfer}
h(t)&=\frac{\dif}{\dif t}s(t).
\end{align}

When cascading LTI systems, the impulse response of the combined system is the convolution of the individual impulse responses,
e.g,. $(h\ast g)(t)$ for the cascade of predistortion filter and flux control line in Fig.~\ref{fig:model}\,(\labelModelSigflow).
An identity operation $y(t)=v(t)$ is obtained with the Dirac impulse $\delta(t)$ as impulse response.
The aim of the calibration of the predistortion in this paper is to achieve
$(h\ast g)(t) \approx \alpha \delta(t)$, corresponding to inversion up to a scaling factor $\alpha$.

An important origin of linear distortions are reactive (capacitive or inductive) circuit elements forming linear dynamic circuits \cite{Chua1987}.
Consider a circuit that consists of only linear resistors, capacitors, and inductors as well as of voltage sources to model input signals and constant biases.
The dynamics of an $\nC$th order circuit with $\nC$ linearly independent reactive elements and an arbitrary number of resistive elements
are described by \cite{Chua1987}
\begin{equation}
\label{eq:circuit_diff_eq}
\dot{x}(t) = A x(t) + b(t)
\end{equation}
with a system matrix $A\in\mathbb{R}^{\nC\times \nC}$ and an input vector $b(t)\in\mathbb{R}^\nC$, which both depend on the circuit elements,
and with the state vector $x(t)\in\mathbb{R}^\nC$,
which contains the voltages of all (linearly independent) capacitors and the currents trough all (linearly independent) inductors.

Since constant bias voltages can be accounted for by a shift of the coordinate system,
we consider a unit step function with step height $V_\mathrm{step}$ as the only input to the system.
Solving \cref{eq:circuit_diff_eq} for $t\geq0$ with a constant excitation $b(t) = b, ~ t\geq 0$ yields
\begin{align}
x(t) &= \exp{A t} (x(0) - x_\infty) + x_\infty 
\\&= Q \diag_i(\exp{\lambda_i t}) \, Q^{-1} (x(0) - x_\infty) + x_\infty.
\end{align}
with $x_\infty = -A^{-1} b$.
The diagonalization $A= Q \diag_i(\lambda_i t) \,Q^{-1} $ with the modal matrix $Q$ and negative real eigenvalues $\lambda_i$ (so-called  natural frequencies \cite{Chua1987})
is possible in any stable, overdamped system.

Due to linearity, every voltage or current in the system can be expressed by a linear combination of the components of $x(t)$ and $b$, and thus in the form
\begin{equation}
\label{eq:vout}
y(t) = \alpha_0 V_\mathrm{step} + \sum_{i=1}^\nC \alpha_i V_\mathrm{step}\, \exp{-\frac{t}{\tau_i}}, ~~t\geq 0
\end{equation}
with the time constants $\tau_i = - 1/\lambda_i$ and the scalar factors $\alpha_i$ for $i=0,\dots,\nC$.
For the initial state $x(0) = 0$,
we obtain the step response
\begin{equation}
\label{eq:app:stepres}
s(t)=\frac{y(t)}{V_\mathrm{step}} = \left(\alpha_0 + \sum_{i=1}^\nC \alpha_i \, \exp{-\frac{t}{\tau_i}}\right)u(t)
\end{equation}
with the unit step function $u(t)$.
This model is used in Sec.~\ref{sec:IIR}
to describe the long-time distortions, see \cref{eq:sumexp}.
In case of an underdamped or critically damped system, the model would need to be extended
by accounting for complex eigenvalues or a non-diagonalizable system matrix, respectively \cite{Chua1987}.

While each matrix element in $A$ usually depends only on few circuit elements
(with details depending on the structure of the circuit),
the time constants in \cref{eq:app:stepres} depend on the eigenvalues of $A$ and can, thus, be influenced by the device parameters of all circuit elements.
Thus, when characterizing
the \cgls{RT} flux line with
a measurement instrument that nominally has the same load resistance as the cryogenic flux line
(in our setup approximately $50\,\mathrm{\Omega}$ due to the attenuator at the $\SI{4}{\kelvin}$ stage),
differences of the resistance due to device tolerances can propagate to all time constants $\tau_i$.
Even though some $\tau_i$ might be influenced only weakly, the dependence can be linear in extreme cases
(such as in a first-order $RC$ circuit with time constant $\tau=RC$),
meaning that resistance variations on the percent level can directly translate to errors on the percent level in the characterization of a time constant.
Such inaccuracies can be avoided by performing in situ characterization as an end-to-end approach for the whole flux control line, including the AWG output stage and the room-temperature wiring.

The description of a linear dynamic circuit in \cref{eq:app:stepres} corresponds to an \acrfull{IIR},
as can be verified by taking the derivative of the step response $s(t)$, treating $\delta(t)$ as the derivative of $u(t)$.
When modeling other kinds of distortions, $h(t)$ in the \cgls{LTI} model in \cref{eq:filters:model} can be a \acrfull{FIR}.
For example, a single discrete echo of a signal after a delay $T_\mathrm{e}$ can be described by
$h(t)=\delta(t) + \alpha_\mathrm{e} \delta(t-T_\mathrm{e})$
with a scaling factor $\alpha_\mathrm{e}$.
On the other hand, the model from \cref{eq:filters:model} and the presented compensation methods do not account for any nonlinear distortions.
We have, thus, carefully designed the experimental setup in a way that nonlinearities in the flux control line are negligible.

\section{Qubit Frequency Model}
\label{app:hamil}

To calculate the frequency of a transmon qubit with charge operator $\hat{n}$ and phase operator $\hat\varphi$, we consider the Hamiltonian \cite{Koch2007,Blais2021}
\begin{align}
\hat{H}_0 \,&= 4 E_\mathrm{C} \hat{n}^2 - E_\mathrm{J}(\phitotal)\cos(\hat\varphi)
\end{align}
with charging energy  $E_\mathrm{C}$ and total Josephson energy $E_\mathrm{J}(\phitotal)$.
The latter depends on the external flux $\phitotal$ through the SQUID loop via
\begin{align}
E_\mathrm{J}(\phitotal) = E_{\mathrm{J}\Sigma} \left| \cos\left(\pi \frac{\phitotal}{\Phi_0}\right)\right| \sqrt{1 + d^2 \,\tan\left(\pi \frac{\phitotal}{\Phi_0}\right)^2},
\end{align}
where $E_{\mathrm{J}\Sigma}=E_{\mathrm{J},1}+E_{\mathrm{J},2}$ is the sum of the Josephson energies of the two junctions
and $d=(E_{\mathrm{J},2}-E_{\mathrm{J},1}) / E_{\mathrm{J}\Sigma}$ the junction asymmetry.
We can write the Josephson potential in the charge basis as
\begin{align}
\cos\left(\hat\varphi\right) \stackrel{N\to\infty}{=} \frac{1}{2}\sum_{n=-N}^N (\ket{n}\bra{n+1} + \ket{n+1}\bra{n}),
\end{align}
where $\ket{n}$ are the eigenstates of $\hat{n}$.

We model the coupled system of the transmon and the readout resonator with the Hamiltonian \cite{Koch2007}
\begin{align}
\label{eq:app:totalHamil}
\hat{H} &= \hat{H}_0 + \hat{H}_\mathrm{R}, &
\hat{H}_\mathrm{R} &= \hbar \omega_\mathrm{R} \, \hat{a}^\dag \hat{a} + \hbar g\,  \hat{n}(\hat{a}^\dag + \hat{a})
\end{align}
where $\omega_\mathrm{R}$ is the bare resonance frequency of the readout resonator, $\hat{a}$ ($\hat{a}^\dag$) is the annihilation (creation) operator for the resonator mode,
and $g$ characterizes the coupling strength between the transmon and the resonator.
We solve the total Hamiltonian $\hat{H}$ numerically in the charge basis, truncated to $N=15$ and to three photon number states in the resonator,
and we identify the dressed qubit frequencies by finding the eigenstates with the largest overlap with the bare eigenstates of $\hat{H}_0$. 

Finally, we convert between magnetic flux $\phitotal$ and \cgls{DC} bias voltage $\VDC$ applied to the flux control line using
\begin{align}
\label{eq:app:phitotal}
\Big.\phitotal(t)\Big|_{\phichange(t)=0} = \alpha_\Phi (\VDC - V_0),
\end{align}
where the factor $\alpha_\Phi$ depends on the total resistance of the flux control line as well as on its mutual inductance with the SQUID loop,
the term $\alpha_\Phi V_0$ models a flux offset,
and the time-dependent flux change $\phichange(t)$ due to flux pulses $v(t)$ from an \cgls{AWG} is zero during the calibration of the qubit frequency model.
Note that using \cref{eq:app:phitotal} requires a high degree of linearity between the output voltage requested from the \cgls{DC} source and the true output voltage.

The transmon model as presented here has 7 free parameters: $E_\mathrm{C}$, $E_{\mathrm{J}\Sigma}$, $d$, $\omega_\mathrm{R}$, $g$, $\alpha_\Phi$ and $V_0$.
We fix $\omega_\mathrm{R}$ to the resonator frequency known from readout tuneup.
To find the remaining parameters, we set the time-dependent part $\phichange(t)$ to zero by choosing $v(t)=0$,
and we perform a series of Ramsey experiments to characterize the {\textgetrans} frequency $\fge(\VDC)$ as a function of the \cgls{DC} bias $\VDC$.

First, we need to determine the bias voltages corresponding to the first-order flux-insensitive points $\phitotal=0$ and $\phitotal=-\Phi_0/2$,
so that $\alpha_\Phi$ and $V_0$ can be obtained by solving \cref{eq:app:phitotal}.
To this end, we sweep $\VDC$ in close vicinity of $\phitotal\in\{0, -\Phi_0/2\}$,
and we fit a second-order polynomial model to the measured value of $\fge(\VDC)$ in each of the two data sets.
The polynomial fits correspond to second-order Taylor approximations of $\fge(\VDC)$ around the bias voltages that yield $\phitotal\in\{0, -\Phi_0/2\}$.

In order to find the remaining four parameters $E_\mathrm{C}$, $E_{\mathrm{J}\Sigma}$, $d$, and $g$,
we obtain four data points by measuring $\fge$ at $\phitotal\in\{0, -\Phi_0/4, -\Phi_0/2\}$ and $\fef$ at $\phitotal=0$.
The fit is then performed 
by iteratively solving the Hamiltonian from \cref{eq:app:totalHamil}
inside the loop of a nonlinear least-squares minimization.

\begin{figure}[t!]
	\includegraphics[width=0.9\linewidth]{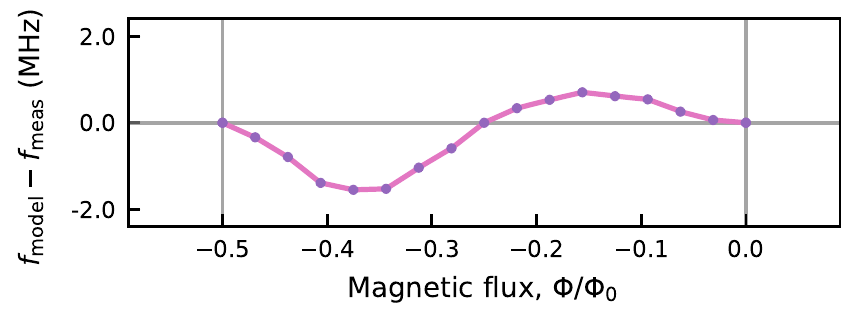}
	\caption{Difference between the Hamiltonian model and the measured values (markers), and interpolation $\Delta f_\mathrm{model}(\phitotal)$.} \label{fig:model_residuals}
\end{figure}

We performs further measurements of $\fge$ with $\phitotal\in[-\Phi_0/2, 0]$ to verify the model,
and we plot the residual errors $\Delta f_\mathrm{model}(\phitotal)= f_\mathrm{model}(\phitotal) - f_\mathrm{meas}(\phitotal)$ in Fig.~\ref{fig:model_residuals}.
While the error level in principle allows to employ the Hamiltonian model in the presented calibration of flux distortions,
it is insufficient for reaching the high accuracy of flux control that we report in Sec.~\ref{sec:results}.
We thus improve the accuracy of the model by using the interpolation
$\Delta f_\mathrm{model}(\phitotal)$
shown in Fig.~\ref{fig:model_residuals}
as an empirical correction that we apply after calculating frequencies with the Hamiltonian model $f_\mathrm{model}(\phitotal)$
or before applying the inverse model to measured frequencies.

The qubit frequency model $\hatffunc$ used in the main text can now be summarized as
\begin{align}
\hatffunc(\phichange) &= f_\mathrm{model}(\phitotal(\phichange)) - \Delta f_\mathrm{model}(\phitotal(\phichange))
\\
\text{with}~~\phitotal(\phichange) &= \alpha_\Phi \VDC+\phichange - V_0
\end{align}
for a fixed \cgls{DC} bias $\VDC$.
For the case of applying the inverse model $\hatffunc^{-1}$ to spectroscopic data as in Sec.~\ref{sec:IIR},
we calibrate an additional minor correction as described below in order to reduce systematic errors in the spectroscopy data,
which are caused by the influence of the second excited state of the transmon.

With the qubit tuned to the same idle point as in Fig.~\ref{fig:fpscope}\,(\labelfpscopeTwoD),
we measure the idle frequency (no flux pulse applied) with both the spectroscopic method and a Ramsey experiment.
We observe a frequency shift of $\Delta f_\mathrm{spec}= f_\mathrm{spec}-f_\mathrm{Ramsey}=\SI{1.1}{\mega\hertz}$,
which is in good agreement with the following simulation study.
We perform a unitary evolution simulation to obtain the excited state population of a transmon with {\textgetrans} frequency $\fge$
when applying a drive pulse, whose frequency $f_\mathrm{drive}$ is swept as in the spectroscopic measurement.
We then fit a Gaussian curve to the simulated excited state population in the same manner as in the inset of Fig.~\ref{fig:fpscope}\,(\labelfpscopeTwoD),
and we observe a similar shift of the center of the Gaussian compared to the nominal {\textqbtrans} frequency $\fge$ used in the simulation.
The simulated shift decreases with increasing anharmonicity $\fef-\fge$ and vanishes when restricting the simulation to the computational subspace,
indicating that the effect can be attributed to a partial excitation of the second excited state.

Since $\Delta f_\mathrm{spec}$ depends only weakly on the flux bias via the flux-dependence of the anharmonicity,
we interpret $\Delta f_\mathrm{spec}$ as a flux-independent offset and subtract it from all frequencies $\hatfgespec(t)$
extracted with the spectroscopic method, i.e.,
we convert to a flux change via
\begin{equation}
\hatphichange=\hatffunc^{-1}(\hatfgespec(t) - \Delta f_\mathrm{spec}).
\end{equation}

\section{Experimental Setup}
\label{app:setup}
\begin{figure}[t!]
	\includegraphics[width=0.9\linewidth]{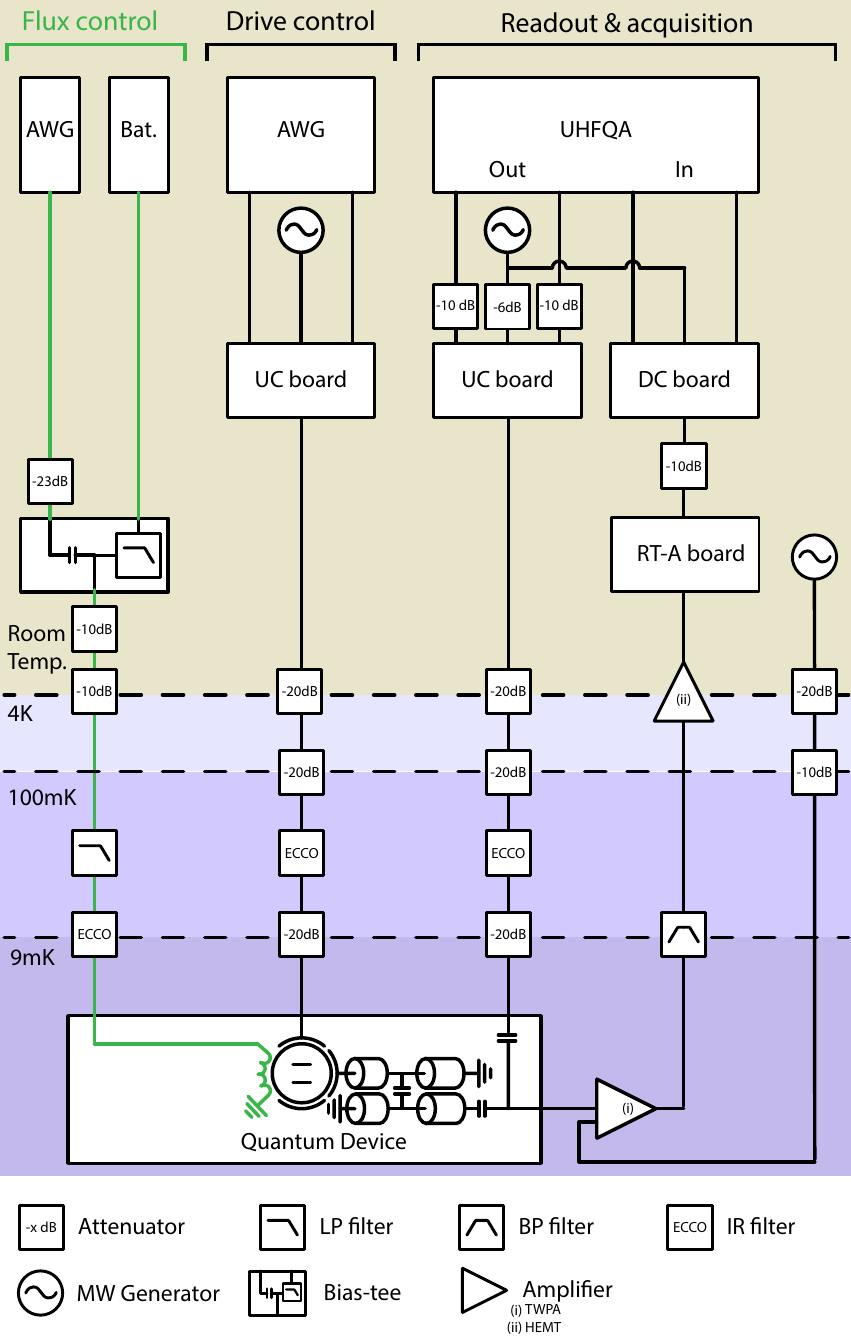}
	\caption{Simplified schematics of the experimental setup. The flux control (green), drive control, and readout lines connect signal generators to the quantum device (see text for details). Signals are routed through a series of bandpass filters (BP), lowpass filters (LP), Eccosorb filters and amplifiers.} \label{fig:setup}
\end{figure}
All experiments described in this manuscript are performed on the 17-qubit quantum device introduced in Ref.~\cite{Krinner2022}. The device is installed in a sample holder mounted at the base plate of a dilution refrigerator~\cite{Krinner2019} and connected to the control electronics setup located at room temperature as summarized in Fig.~\ref{fig:setup}. 

To control the transmon qubit frequency, a voltage source (Bat.) generates a DC current, which is combined with voltage pulses generated by an \cgls{AWG} using a custom bias tee, see Appendix~\ref{app:biasT}.
The dominant time constant $\tauHP=\SI{19.2}{\micro\second}$ of the high-pass part of the bias tee
corresponds to a $\SI{3}{\decibel}$ cutoff frequency of approximately $\SI{8.3}{\kilo\hertz}$. The output amplifier of the \cgls{AWG} has a frequency cutoff of $\SI{300}{\mega\hertz}$, and the low-pass filter in the flux line located at the base plate has a frequency cutoff of $\SI{780}{\mega\hertz}$. The predistortion of the voltage pulses sent through the flux line is performed in software before uploading the waveforms to the \cgls{AWG} (as, e.g., in \cite{Ferreira2024}).

To realize single-qubit gates, drive pulses are generated at an intermediate frequency in the range of 0-\SI{500}{\mega\hertz} by an \cgls{AWG} and subsequently up-converted (UC) to microwave frequencies with an analog IQ mixer located in the UC board using the continuous-wave signal provided by a microwave (MW) generator. 
We operate the drive \cgls{AWG} in a direct output mode bypassing the output amplifier,
yielding a relatively weak dependence of the $\pi$~pulse amplitude on the drive frequency, see Fig.~\ref{fig:fpscopeApp}(\labelfpscopeAppRabi).

We use an ultra-high frequency quantum analyzer (\mbox{UHFQA}) to generate readout pulses. The readout signal at each output port of the device is  amplified using a wideband near-quantum-limited traveling-wave parametric amplifier (TWPA) \cite{Macklin2015}, a high-electron-mobility transistor (HEMT) amplifier, and low-noise amplifiers operating at room temperature (RT-A board)~\cite{Krinner2022}. Thereafter, the amplified signal is down-converted with an IQ mixer in a down-conversion (DC) board and then both demodulated and integrated by the acquisition unit of the UHFQA.

Using standard spectroscopy and time-domain methods, we measure the coherence properties of the qubit used in the main text for the flux line characterization, see \cref{tab:qb_params}.

\begin{table}[]
    \centering
    \caption{Parameters and coherence properties of the qubit used for the flux line characterization measurements.}
\begin{tabular}{lr}
\toprule
Parameter &          Value     \\
\midrule
Qubit idle frequency, $\omega_Q/2\pi$ (GHz)					        	        & 5.887 \\ 
Minimum qubit frequency, $\omega_{Q, \mathrm{min}}/2\pi$ (GHz)            & 4.151 \\
Qubit anharmonicity, $\alpha/2\pi$ (MHz)							                    & -174\\
Lifetime, $T_1$ (\si{\micro s})							                    & 16.2 \\
Ramsey decay time, $T_2^*$  (\si{\micro s})				        	        & 31.4\\
Echo decay time, $T_2^\mathrm{e}$   (\si{\micro s})		        		    & 31.8 \\
Readout frequency, $\omega_\mathrm{RO/2\pi}$ (GHz)				        	  & 7.556 \\
Three-state readout error, $\epsilon_{\mathrm{RO}}^{(3)}$ (\%)			         & 2.5 \\
\bottomrule

    \end{tabular}
    \label{tab:qb_params}
\end{table}

\section{Custom Bias Tee}
\label{app:biasT}
\begin{figure}[t!]
	\includegraphics{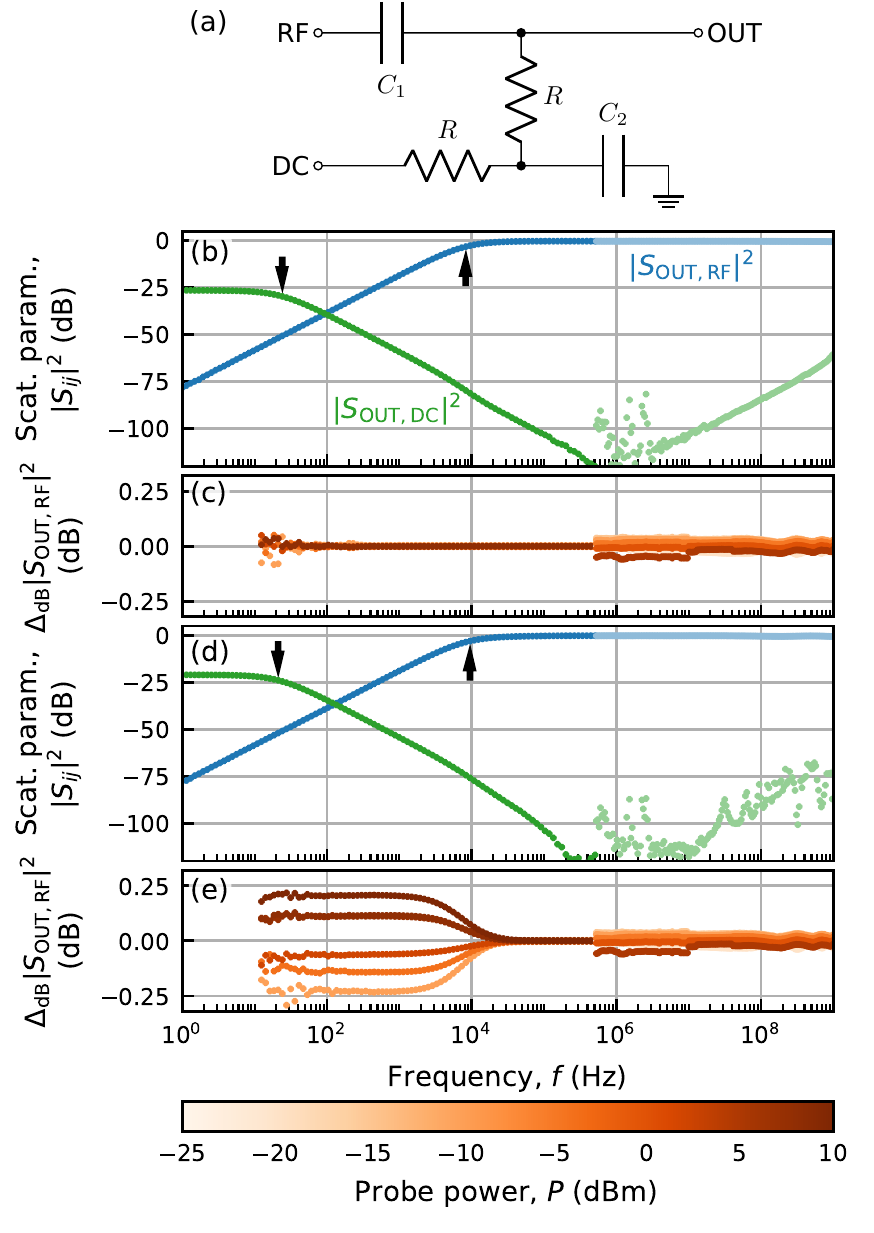}
	\caption{Custom bias tee. (a) Circuit schematic, where $C_1=\SI{200}{\nano\farad}$, $C_2=\SI{13.6}{\micro\farad}$, and $R=\SI{1}{\kilo\ohm}$.
		(b) Transmission $|S_{\mathrm{OUT},\mathrm{RF}}|$ (blue) and $|S_{\mathrm{OUT},\mathrm{DC}}|$ (green) measured with a lock-in amplifier (dark markers)
		and a vector network analzer (shaded markers). Black arrows indicate $\SI{3}{\decibel}$ cutoff frequencies.
		(c) Transmission $|S_{\mathrm{OUT},\mathrm{RF}}|$ for different input powers, plotted relative to the mean (in the dB domain)
		of the measured transmission curves.
		(d,e) Same as (b,c), but for a commercial bias tee (MC ZFBT-4R2GW).
	} \label{fig:biasT}
\end{figure}

We use a custom bias tee, which we designed
based on a low number of reactive components to keep the number of terms in the step response in \cref{eq:app:stepres} low,
see circuit schematic in Fig.~\ref{fig:biasT}\,(a).
We have built the bias tee from components that were chosen with particular attention to ensuring a linear transfer characteristic.

In Fig.~\ref{fig:biasT}\,(b), we show the transmission $|S_{\mathrm{OUT},\mathrm{RF}}|$ (blue) and $|S_{\mathrm{OUT},\mathrm{DC}}|$ (green)
of a typical instance of the custom bias tee,
measured with a lock-in amplifier up to $\SI{0.5}{\mega\hertz}$ (dark markers)
and with a vector network analzer at the higher frequencies (shaded markers).
For the measurement of $|S_{\mathrm{OUT},\mathrm{RF}}|$ with an impedance of $\SI{50}{\ohm}$ at the RF and OUT ports, the DC input was shorted.
To verify the linearity of the transfer characteristic, we perform the measurement with different input powers,
and subtract the mean of the curves computed in the dB domain, see Fig.~\ref{fig:biasT}\,(c).
The small nonlinearity in the data from the vector network analzer is attributed to imperfections in the calibration of this measurement instrument.
The reduced signal-to-noise ratio at low frequencies comes from the reduction of signal power in the stop band.

In Fig.~\ref{fig:biasT}\,(d) and (e), we repeat the same series of measurements for a commercial bias tee,
revealing a power-dependent residual transmission in the stop band.
Such a nonlinearity in the stop band might be irrelevant for many applications that only rely on the bias tee to suppress frequency components in the stop band.
However, we operate the bias tee in the transient regime, and a linear transfer characteristic of frequency components in the stop band is crucial to
allow equalizing the transient response.

For the bias tee used in the experiments in this paper, we have determined a $\SI{3}{\decibel}$ cutoff frequency of $\SI{8.3}{\kilo\hertz}$
by intersecting low-frequency and high-frequency asymptotes to the $|S_{\mathrm{OUT},\mathrm{RF}}|$ spectrum (not shown here).
This corresponds to a time constant of $\tauHP=\SI{19.2}{\micro\second}$ for the high-pass part of the bias tee.

\section{Illustration of the Time Dependence of the Qubit Transition Frequency}
\label{app:distort_simul}
To obtain the time dependence of the {\textgetrans} frequency in Fig.~\ref{fig:intro},
we first measure the long-time impulse response $h_\mathrm{long}(t)=\mathrm{d}s(t)/\mathrm{d}t$ by fitting $s(t)$ from \cref{eq:sumexp} as described in Sec.~\ref{sec:IIR},
and we subsequently characterize the residual short-time distortions by determining the impulse response $h_\mathrm{short}(t)$ from \cref{eq:FIRmodel} as described in Sec.~\ref{sec:FIR}
while already compensating for distortions from $h_\mathrm{long}$.
The overall impulse response of the flux control line is then given by $h_\mathrm{total}(t) = (h_\mathrm{long} \ast h_\mathrm{short})(t)$,
where $\ast$ denotes convolution.
We can then compute the flux at the qubit by means of a convolution of $h_\mathrm{total}$ and the waveform programmed to the \cgls{AWG},
and we convert this to the time dependence of the {\textgetrans} frequency using the model $\hatffunc$ from Fig.~\ref{fig:model}.

\section{Drive Pulse Generation, Readout, and High-Pass Compensation while Measuring Long-Time Distortions}
\label{app:comp_pulse}
\begin{figure}[t!]
	\includegraphics[width=\linewidth]{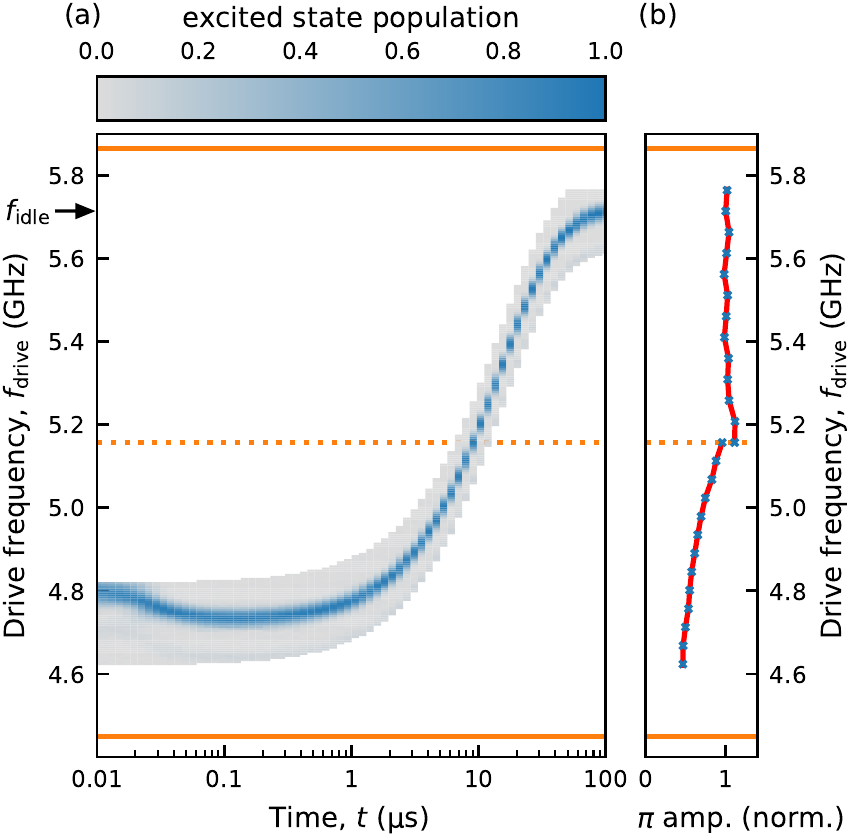}
	\caption{
		Drive pulse generation for the \emph{in situ} characterization of the flux line step response in Fig.~\ref{fig:fpscope}.
		(\labelfpscopeAppTwoD) Local oscillator frequencies $f_\mathrm{LO}$ (solid orange lines) shown along with the measured excited state population reproduced from Fig.~\ref{fig:fpscope}\,(\labelfpscopeTwoD).
		The dotted orange line indicates the boundary between using the lower or upper side band with $f_\mathrm{LO}$ at the higher or lower solid orange line, respectively.
		(\labelfpscopeAppRabi) $\pi$ pulse amplitudes normalized to the value at the idle frequency, plotted for the drive frequency range and local oscillator configuration used in (\labelfpscopeAppTwoD):
		calibration data (markers)
		and 
		interpolation (solid lines).
	}
	\label{fig:fpscopeApp}
\end{figure}

To perform the spectroscopic measurement described in Sec.~\ref{sec:IIR},
we generate a drive pulse
at an intermediate frequency $f_\mathrm{IF}$
and then perform a frequency upconversion to the desired drive frequency $f_\mathrm{drive} = f_\mathrm{LO} \pm f_\mathrm{IF}$ by an analog IQ mixer with a
microwave generator at frequency $f_\mathrm{LO}$ used as local oscillator.
To avoid impairments from driving the {\getrans} or {\eftrans} transition of the transmon with local oscillator leakage \cite{Sandia2002},
we restrict $f_\mathrm{LO}$ to be at one of the solid orange lines in Fig.~\ref{fig:fpscopeApp}\,(\labelfpscopeAppTwoD)
and sweep $f_\mathrm{IF}$.
In addition, we calibrate a suppression of local oscillator leakage individually for both choices of $f_\mathrm{LO}$, using a similar method as described in \cite{Herrmann2022a}.
For the sake of a simple measurement protocol, we do not suppress sideband leakage \cite{Jolin2020},
which would need to be calibrated individually for each drive frequency $f_\mathrm{drive}$ in the sweep shown in Fig.~\ref{fig:fpscopeApp}.
An alternative to calibrating sideband leakage suppression would be to use a double frequency conversion scheme as described in \cite{Herrmann2022a}
as such a scheme provides a higher spurious-free dynamic range without the need for calibrating out mixer imperfections.
According to the discussion in Sec.~\ref{sec:IIR}, the presented compensation of flux pulse distortions does not appear to be limited by 
systematic errors in the spectroscopic characterization measurement, including potential systematic errors from mixer imperfections.

To implement a $\pi$ pulse for each $f_\mathrm{drive}$, the drive amplitude is adjusted
by interpolating between calibration data, which was obtained in a series of Rabi experiments with the qubit tuned to the desired frequencies using a DC bias,
see Fig.~\ref{fig:fpscopeApp}\,(\labelfpscopeAppRabi).

To infer the excited state population
in the spectroscopic measurement in Sec.~\ref{sec:IIR},
we project the measured integrated transmission through the feed line
to a line in the complex plane, which is defined by calibration points for the $\ket{0}$ and $\ket{1}$ states of the qubit.
The calibration points are measured at the idle frequency (no flux pulse applied)
and using an accurately calibrated $\pi$ pulse amplitude given by the respective data point in Fig.~\ref{fig:fpscope}\,(\labelfpscopeAppRabi).

When applying this method during the first characterization of the step response of the flux control line,
the uncompensated long-time distortions need to be taken into account
since accurate readout is only achieved when the qubit is tuned to the frequency for which the calibration points have been measured.
To bring the qubit back to the idle frequency in the presence of flux distortions,
we output an appropriate amplitude $a_\mathrm{during\_RO}$ from the \cgls{AWG} after truncating the flux waveform, see Fig.~\ref{fig:fpscope}\,(\labelfpscopeSeq).
Since the required amplitude $a_\mathrm{during\_RO}$ depends on the impulse response of the flux control line, which is yet to be characterized,
we resort to the model-based choice
\begin{equation}
\label{eq:filters:fp_during_ro}
a_\mathrm{during\_RO} = a_\mathrm{step}  (1 - \exp{-(t+\tfp) / \tau} )
\end{equation}
where
$a_\mathrm{step}$ is the height of the step function applied to the flux control line,
$t+\tfp$ is the moment at which the step function is truncated,
and $\tau$ is a time constant.
By choosing $\tau=\SI{20}{\micro\second}$,
we approximately compensate for the dominant distortion in the setup,
which is the high-pass with time constant $\tauHP\approx \SI{20}{\micro\second}$ in the bias tee.
This choice brings the qubit to the idle point only approximately,
but since the readout parameters depend on the flux bias only indirectly via the coupling between qubit and resonator,
this is sufficient to reach consistent readout throughout the sweep shown in Fig.~\ref{fig:fpscope}\,(\labelfpscopeTwoD).
Note that this modeled-based adaptation of the flux waveform only aims at ensuring high readout quality and does not affect the qubit frequency during the drive pulse,
i.e., it does not influence the quantity that is to be determined by the measurement.
When repeating the spectroscopic measurement from Sec.~\ref{sec:IIR}
after having calibrated a first \acrlong{IIR} filter that compensates for the dominant distortion, we can instead choose $a_\mathrm{during\_RO}=0$.

We switch the additional flux amplitude $a_\mathrm{during\_RO}$ off after the readout,
to allow the capacitor of the high-pass filter in the bias tee to discharge
before the next measurement.
Due to the linearity of the system, the idle time relevant for this discharging is the time
between the end of the flux waveform and the start of the drive pulse of the following measurement,
even though the flux waveform of the following measurement might already start during this idle time.
By triggering measurements with a period of $\SI{300}{\micro\second}$, we achieve an idle time of at least $
15\,\tauHP$.
When recording data in a cyclic averaging mode in a way that measurements for different values of the time $t$ are performed consecutively,
sufficient idle time can be ensured by
arranging the measurements in the order of increasing $t$ and interleaving the calibration points (without flux pulses)
between the measurements with highest and lowest $t$.

\begin{figure}
\includegraphics{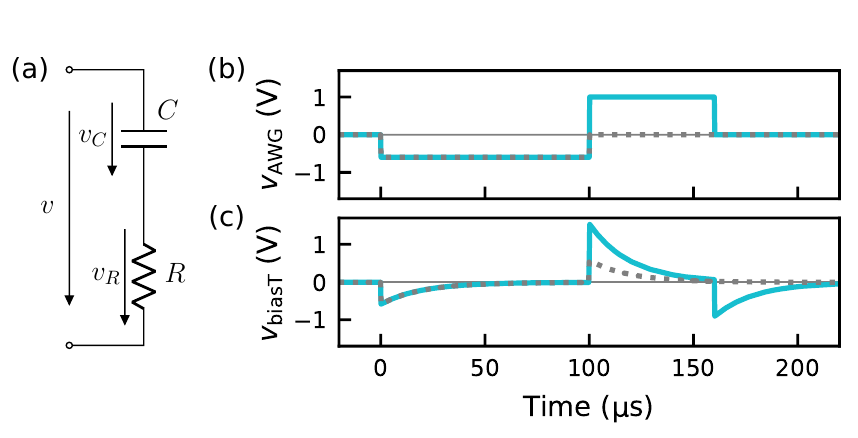}
\caption{Study of discharging pulses in the presence of distortions.
(a) Simplified high-pass circuit formed of a capacitor $C$ and a load resistor $R$ to illustrate the discussion in Appendix~\ref{app:comp_pulse}.
(b) Waveform consisting of a long pulse with or without a discharging pulse, programmed to an AWG that is connected to the radio frequency (RF) input of a bias tee.
(c) Measured output voltage of the bias tee for the two cases in (b),
showing that a net-zero waveform that is long compared to the high-pass time constant 
does not achieve the desired discharging
(see Appendix~\ref{app:comp_pulse}).}
\label{fig:discharge}
\end{figure}

{\newcommand{\tend}{t_1}
\newcommand{\tstart}{0}
A method to discharge the capacitor of the high-pass filter rapidly consists in appending a flux pulse after the final readout of each experiment,
with amplitude and duration chosen in a way that the integral over the whole flux waveform vanishes.
Such a discharging pulse is known to be effective if the waveform is short compared to the high-pass time constant \cite{Johnson2011PhD}.
Since this condition is not fulfilled for the measurements in Sec.~\ref{sec:IIR},
we study the effectiveness of this approach for long waveforms in the presence of distortions.

To this end, we consider the simplified high-pass circuit with time constant $\tau=RC$ in Fig.~\ref{fig:discharge}\,(a).
The inverse filter that compensates for the distortions between the input $v(t)$ and the output $v_R(t)$ 
has the impulse response
$h_\mathrm{inv}(t)= \delta(t) + u(t)/\tau$ with the Dirac impulse $\delta(t)$ and the unit step function $u(t)$.
This can be verified by taking the inverse Laplace transform of \cref{eq:transfer_sumexp_inv} in the special case of $N=1$ and $\alpha_0=0$.

For a waveform $a(t)$ with vanishing integral in $t\in[\tstart,\tend]$ and final amplitude $a(\tend)=0$, the predistortion with $g=h_\mathrm{inv}$ yields
\begin{equation}
v(\tend) =
   (g\ast a)(\tend) =
a(\tend) + \frac{1}{\tau}\int_{\tstart}^{\tend} a(t') \dif t' = 0.
\end{equation}
In case of ideally compensated distortions, where $v_R(t)$ is proportional to $a(t)$,
this implies $v_R(\tend)=0$ and, by Kirchhoff's voltage law, $v_C(\tend)=v(\tend)-v_R(\tend)=0$.
This demonstrates that a net-zero target waveform $a(t)$ is effective in discharging the capacitor in the high-pass filter at the end of the waveform if distortions are properly compensated for,
and we thus employ this approach in all measurements with predistortion.

However, the above argument does not hold when distortions are not compensated yet, i.e., when $v(t)=a(t)$ and $v_R(t)$ is not proportional to $a(t)$.
For the initial characterization measurement, we indeed observe that a discharging pulse leads to a higher charge in the capacitor than letting the circuit idle as described earlier in this section,
see the comparison in Fig.~\ref{fig:discharge}\,(b) and (c).
}

\section{Details of the \cgls{IIR} Filter Calibration}
\label{app:IIR}
After measuring the {\textgetrans} frequency $\hatfge(f)$ with the spectroscopic method described in Sec.~\ref{sec:IIR}
and converting the result to the flux change $\hatphichange(t)$,
we obtain samples of the step response
\begin{equation}
\hats(t) = \frac{\hatphichange(t)}{\hatphichange(t_\mathrm{max})} ~~\text{with}~~ t_\mathrm{max} = \argmax_t |\hatphichange(t)|.
\end{equation}
Expressing the flux change relative to the maximum flux excursion is a pragmatic choice
and is valid because the aim is to compensate for distortions up to a scaling factor (see Sec.~\ref{sec:model}).

The low number of coefficients in the sum of $N$ exponentials from \cref{eq:sumexp} makes this model inherently robust to noise in the following sense.
While a model with many free parameters might get adjusted too closely to a noisy data set
to which it is fitted and then fail to capture the true structure of the underlying data (so-called \emph{overfitting}), the sum-of-exponentials model used here has much less parameters than the $70$ data points we measure, effectively preventing overfitting.
On the other hand,
we have observed that the nonlinear least-squares problem of fitting the sum of $N$ exponentials from \cref{eq:sumexp} to $\hats(t)$ is very sensitive to the choice of initial parameter values.
We initialize the offset $\alpha_0$ with $\alpha_0=0$ when fitting to the first characterization data to account for the high-pass in the bias tee,
and with $\alpha=1$ when fitting to data from a measurement with predistortion, where the dominant high-pass characteristic has already been corrected.
For the other parameters, we increase the model order $N$ successively
by considering only data points above a threshold $t>t_{\mathrm{min},N}$ in the $N$th iteration
and using the parameters $\alpha_1,\dots,\alpha_{N-1}$ and $\tau_1,\dots,\tau_{N-1}$ fitted in the previous iteration as initial values.
Using a fixed initial value $\alpha_N=1$ in each iteration, the
procedure is configured with the threshold $t_{\mathrm{min},N}$ and the initial time constants $\tau_N$ for each iteration, which we choose as decreasing sequences.
Best results have been obtained by choosing $t_{\mathrm{min},N}$ and the initial $\tau_N$ manually based on visual inspection of the data.
Further automation of the procedure is a topic for future research.
However, it is noteworthy that the iteration based on human input is only during data analysis without requiring new measurements inside the loop.
The procedure is terminated when significant residual errors above the noise floor remain only on time scales below the time resolution of the spectroscopic method.

Having obtained a fit of the step response model from \cref{eq:sumexp}, the inverse filter can be calculated via the Laplace transform $\mathcal{L}$.
Using \cref{eq:stepres_impres_transfer}, we transform the step response $s(t)$ to the impulse response $h(t)$ and further to the transfer function
\begin{equation}
H(p) = \Lapl{h}(p) = \frac{\Lapl{\phichange}(p)}{\Lapl{v}(p)}
\end{equation}
with the complex variable $p$ of the Laplace transform.
We obtain
\begin{equation}
\label{eq:transfer_sumexp}
H(p) = \sum_{i=0}^\nC \Hi(p)
\end{equation}
with
\begin{equation}
\Hi(p)=  
\begin{cases}
\displaystyle
\Lapl{\frac{\dif}{\dif t} \alpha_0 u(t)}(p)
=
\alpha_0
, & i=0,
\\[3mm]
\displaystyle
\Lapl{\frac{\dif}{\dif t}  \alpha_i \exp{-\frac{t}{\tau_i}} u(t) }(p)=
\frac{ \alpha_i \tau_i p }{\tau_i p + 1}  
,
& i>0.
\end{cases}
\end{equation}
We have used Laplace transform pairs from \cite{Abramowitz1972} and the property that a derivative in the time domain corresponds to a multiplication by $p$ in the frequency domain.
Such a description of the transfer function $H(p)$ of the flux control line was also used in 
\cite{Neill2018}
for the special case $N=1$ and in \cite{Chen2018j}.

The transfer function of the inverse filter of \cref{eq:transfer_sumexp},
\begin{multline}
\label{eq:transfer_sumexp_inv}
\frac{1}{H(p)} =\\ \frac{\prod_{i=1}^{\nC}(\tau_i p + 1)}{\alpha_0\prod_{i=1}^{\nC} (\tau_i p + 1) +  \sum_{j=1}^{\nC} \alpha_j \tau_j p \prod_{i\neq j}(\tau_i p + 1)},
\end{multline}
is fully characterized by its zeros $p_{0,i}$, poles $p_{\infty,i}$, and gain factor $\gainH$.
For the zeros and gain of $\Hinvp$, we have
\begin{align}
p_{0,i} &= 1/\tau_i, &
\gainH &= \left(\sum_{i=0}^\nC \alpha_i\right)^{-1},
\end{align}
and we find the poles $p_{\infty,i}$ of $\Hinvp$ by applying the root finding method from \cite{Collins1992} to the denominator polynomial of \cref{eq:transfer_sumexp_inv}.
Since we aim at inverting distortions only up to a scaling factor, we opt for a normalized inverse filter $\Hinvnorm(p)=(\Hinvp)/\gainH$.

For a digital implementation of the filter, we transform the poles and zeros into the domain of the $z$-transform \cite{Oppenheim1997}
using 
\begin{align}
z_{0,i}&=\exp{p_{0,i}\, \Ts},  &  z_{\infty,i}&=\exp{p_{\infty,i}\, \Ts},
\end{align}
where $\Ts=1/(\SI{2.4}{\giga\hertz}) = \SI{0.4167}{\nano\second}$ is the sampling interval of the used AWG.
To avoid numerical instabilities that are known to occur in digital implementations of higher-order \acrfull{IIR} filters \cite{Oppenheim1999},
we follow the established practice of splitting the \cgls{IIR} filter into a cascade of second-order \cgls{IIR} filters, so-called second-order sections (SOS) \cite{Oppenheim1999}.
To this end, the poles and zeros are divided into groups of two poles and two zeros
by pairing poles with the nearest zeros, starting with the poles closest to the unit circle.
{\newcommand{\vin}{v_\mathrm{in}}%
	\newcommand{\vout}{v_\mathrm{out}}%
	\newcommand{\Vin}{V_\mathrm{in}}%
	\newcommand{\Vout}{V_\mathrm{out}}%
	Letting $z_{0,1}$, $z_{0,2}$, $z_{\infty,1}$, and $z_{\infty,2}$ denote the zeros and poles in such a group,
	the transfer function $\Vout(z)/\Vin(z)$ ($z$-transform of the discrete-time impulse response) of the corresponding second-order section is given by
	\begin{align}
	\frac{\Vout(z)}{\Vin(z)}  =
	\frac{(z-z_{0,1})(z-z_{0,2})}{(z-z_{\infty,1})(z-z_{\infty,2})} = \frac{1 + B_1 z^{-1} + B_2 z^{-2}}{1 + A_1 z^{-1} + A_2 z^{-2}}
	\end{align}
	with
	\begin{align}
	A_1 &= -z_{\infty,1}-z_{\infty,2}, & 
	A_2 &= z_{\infty,1}\,z_{\infty,2}, \\
	B_1 &= -z_{0,1}-z_{0,2}, & 
	B_2 &= z_{0,1}\,z_{0,2}.
	\end{align}
	We realize each second-order section in a so-called transposed direct form II structure,
	which is a canonical and efficient way of implementing discrete-time filters \cite{Oppenheim1999}.
	This structure is described by the recursive equation
	\begin{multline}
	\vout(\disct) = \vin(\disct) + B_1 \,\vin(\disct-1) + B_2 \,\vin(\disct-2) \\- A_1 \,\vout(\disct-1) - A_2 \,\vout(\disct-2)
	\end{multline}
	with the discrete time variable $\disct$.
	The digital implementation of 
	$\hinvnorm = \mathcal{L}^{-1}\{\Hinvnorm(p)\}(t)$ is then obtained by cascading these sections, i.e., by using the output $\vout(\disct)$ of one section as the input $\vin(\disct)$ of the next.
}

\section{Resolution, Accuracy, and Precision of the Cryoscope Method}
\label{app:cryoscope}
To correct flux pulse distortions at short time scales as described in Sec.~\ref{sec:FIR}, it is crucial to obtain a precise estimate of the qubit frequency during the pulse. 
The uncertainty of the phase $\statuncertaintyphase$ in the Ramsey measurement from which the qubit frequency is inferred scales approximately as 
\begin{equation}
\label{eq:sigma_theta}
\statuncertaintyphase \propto \frac{1}{\sqrt{\nphases \nruns}} 
\end{equation}
where $\nphases$ is the number of phases for the second $\pi/2$ pulse of the Ramsey sequence (see Sec.~\ref{sec:FIR} for details) and $\nruns$ is the number of experimental runs over which we average each data point.
To motivate \cref{eq:sigma_theta}, note that \cite[Eq.~(22)]{Gavin2024} yields that this proportionality becomes exact in the simplified scenario
of fitting $\theta$ as the only unknown parameter (assuming known frequency, amplitude, and mean value of the fitted cosine)
to a data set with $\nphases$ phase values that come in pairs offset by $\pi/2$,
assuming independent and identically distributed noise.
By propagating the uncertainty through \cref{eq:cryoscope}, the uncertainty of the qubit frequency excursion $\statuncertainty$ yields
\begin{equation} \label{eq:statuncertainty}
\statuncertainty = \frac{\sqrt{2}\statuncertaintyphase}{2\pi\Delta t}  \propto \frac{1}{\Delta t \sqrt{\nphases \nruns}} \equiv \frac{1}{\neff}
\end{equation}
where we have defined $\neff$ as an effective integration time.  

\begin{figure}[t!]
	\includegraphics[width=\linewidth]{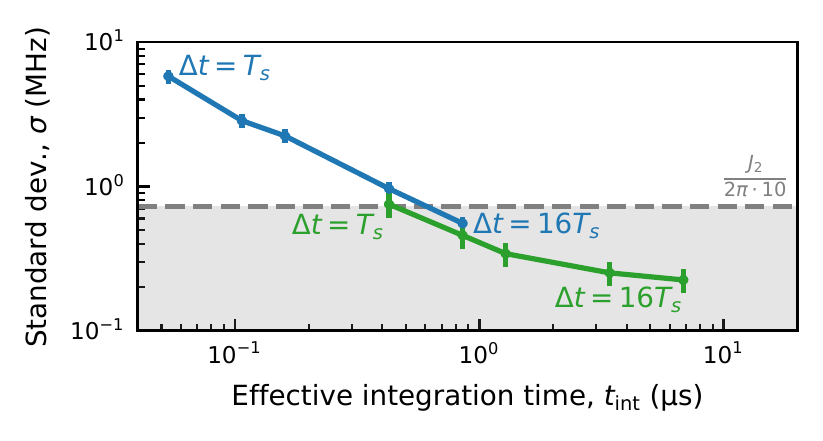}
	\caption{Qubit frequency standard deviation, $\statuncertainty$, estimated from 25 cryoscope measurements as a function of the effective integration time, $\neff$. The measurements are performed for $\nruns = 1024$ (blue line) and $\nruns=65536$ (green line) while $\Delta t$ is varied from $\Ts$ to $16\Ts$. $\nphases = 16$ is kept constant for all measurements.
	The gray region indicates deviations smaller than $1/10$ of the coupling strength of the $\ket{11}\leftrightarrow\ket{20}$ interaction $J_2/2\pi$.
	The error bars correspond to the standard error for 25 measurements.
	}
	\label{fig:cryoscope_averages}
\end{figure}

We estimate $\statuncertainty$ in practice by calculating the standard deviation $\sigma$ of 25 repetitions of the cryoscope measurement for a single time point $t = \SI{50}{\nano\second}$ and for different values of $\Delta t$ and $\nruns$.
We observe an inverse linear scaling of $\statuncertainty$ with respect to \neff{} (linear scaling with a slope of $-1$ in a double-logarithmic plot),  see Fig.~\ref{fig:cryoscope_averages}. For large \neff, the scaling approaches a plateau which indicates further disturbances not captured by the model of \cref{eq:statuncertainty}, which should be investigated in future work.

Note that we have to quadratically increase the measurement time to linearly decrease $\statuncertainty $ when adjusting $\nruns$ or $\nphases$. However, the same effect can also be achieved by linearly increasing $\Delta t$, which is independent of the measurement duration for a single time value $t$. On the other hand, since the estimate of the qubit frequency excursion is averaged over $\Delta t$, see \cref{eq:cryoscope}, a larger $\Delta  t$ leads to a lower time resolution of the inferred qubit frequency. In addition, larger step sizes during rapidly falling or rising edges lead to larger systematic errors caused by the larger phase difference accumulated during the low-pass-like turn-off transients, see Ref.~\cite{Rol2020} for details. 

To accommodate for this tradeoff, we adapt $\Delta t$ in different parts of the measurement. Near the flux pulse edges ($\pm \SI{2.4}{\nano\second}$), which is where we expect the largest systematic errors, we use $\Delta t = \Ts $ with the \cgls{AWG} sampling interval $\Ts\approx \SI{0.42}{\nano\second}$. From $\SI{2.4}{\nano\second}$ to $\SI{30}{\nano\second}$ after the rising edge, i.e. the time interval in which residual distortions are typically visible after the tuneup of the \acrlong{IIR} filter, we use $\Delta t = 3 \Ts$, which is sufficient to resolve features on the order of $\sim1.2\,$ns while decreasing the measurement noise in this interval. Finally, in the time interval in which we neither expect strong distortions nor large systematic errors (from 30ns until the falling edge of the pulse), we use $\Delta t = 8 \Ts$ to reduce the total measurement time. 

To ensure low errors for all $t$ in the cryoscope measurements presented in the main text, we use $\nphases = 16$ and $ \nruns  = 65536$.
This guarantees that $\sigma \lesssim \SI{0.75}{\mega\hertz}$ is an order of magnitude smaller than
the coupling strength of the $\ket{11}\leftrightarrow\ket{20}$ interaction
between neighboring qubits on this device
(see the typical value of $J_2/2\pi = \SI{7.4}{\mega\hertz}$ determined in Appendix~\ref{app:CZ}), such that residual distortions below the noise floor have a limited effect on the two-qubit gate fidelity.  

Finally, to further mitigate the systematic errors, we operate the asymmetric transmon at the upper first-order flux-insensitive bias point ($\phiDC = 0$) during the cryoscope measurement, which has a lower second-order flux sensitivity than the lower first-order flux-insensitive bias point ($\phiDC = 0.5$), thereby reducing the phase accumulated during the low-pass-like turn-off transients.

When using the parameters given above and
triggering measurements with a period of hundreds of microseconds ($\approx10T_1$) to wait for qubit relaxation,
the total measurement time of the cryoscope method is several hours.
This time can be reduced by an order of magnitude
by using active qubit reset, e.g., \cite{Salathe2018} to allow for a shorter triggering period.

\section{Details of the \cgls{FIR} Filter Calibration}
\label{app:FIR}

Compensating the distortions of the flux control line at short time scales (see Sec.~\ref{sec:FIR}) involves deconvolution operations, which are known to be ill-posed problems in practice~\cite{Riad1986}, combined with the inversion of the nonlinear qubit frequency model. 
In this appendix, we detail how we apply regularized optimization, a method which has previously been used for deconvolution~\cite{Milton1991, Borchers13}. Choosing regularization terms that yield a good balance between accuracy and robustness can be seen as an a-priory modeling decision justified by insights about the specific inverse problem~\cite{Milton1991, Neumaier1998}.
Accordingly, the regularization terms described below are a possible choice that worked well on our experimental setup, but future research might find alternative choices that potentially further improve the compensation of flux pulse distortions.

We decompose the computation of the filter that inverts the impulse response of the flux control line into two steps. The first optimization step is formulated based on the forward qubit frequency model $\hatffunc$ to avoid an explicit inversion $\hatffunc^{-1}$.
In particular, we find the coefficients $h_0,\dots,h_L$ in \cref{eq:FIRmodel} from
\begin{equation}
\min_{h_0,\dots,h_L}\, \left(w(t) \left\|\hatfge(t) - \hatffunc\big[(h \ast a)(t)\big] \right\|_2^2 + R_1 + R_2\right)
\end{equation}
with regularization terms $R_1$ and $R_2$, where $w(t)$ is a weight function proportional to \neff{} (see Appendix~\ref{app:cryoscope}) of each sample such that samples with low statistical uncertainty are trusted more than samples with higher statistical uncertainty. 
As a first regularization term, we choose a Tikhonov regularizer~\cite{Tikhonov1963} $R_1 = \lambda_1 ||h(t)||_2^2 $  with regularization parameter $\lambda_1$, which limits the total energy of the impulse response. In addition, we choose $R_2 = \lambda_2 x(t) ||h(t)||_2^2$ with regularization parameter $\lambda_2$ and prefactor $x(t)$  which scales exponentially with $t$. This additional regularizer smoothly forces the trailing filter coefficients to vanish, motivated by the physical intuition that the impulse response of a physical system has finite support.
We implement the optimization in discrete time by resampling $\hatffunc\big[(h \ast a)(t)\big]$ at the time values $t$ at which $\hatfge(t)$ is measured and we compute the Euclidean norm $\|\bullet\|_2$ at these sampling points only.

Note that the linear time-invariant model used in this optimization cannot capture the residual systematic errors of the cryoscope measurement caused by the turn-off transients (see Appendix~\ref{app:cryoscope} and Ref.~\cite{Rol2020} for more details) because they are different for a step from $\fidle$ to a flux sensitive $f_\mathrm{target}$ and for a step from $f_\mathrm{target}$ to the flux-insensitive $\fidle$. 
To mitigate this effect, we choose a rectangular flux pulse $a(t)$ (see Fig.~\ref{fig:cryoscope} in the main text), which encourages the optimizer to find a tradeoff between the systematic errors of rising and falling edges.

Having obtained the coefficients $h_\ell$ of the \acrfull{FIR} model in \cref{eq:FIRmodel},
we compute an inverse \cgls{FIR} filter by performing a second regularized optimization	
\begin{equation}
\min_{\hinvFIR[,0],\dots,\hinvFIR[,L']}~ \left(\left\|d(t) - (h \ast \hinvFIR)(t)\right\|_2^2 + R_1'\right),
\end{equation}
where $\hinvFIR(t)$ is defined in analogy to \cref{eq:FIRmodel}
and the target shape $d(t)$
is a realistic approximation of a Dirac impulse, which we choose to be a Gaussian pulse of width $\sigma=\SI{0.75}{\nano\second}$.
The limit case of a Dirac impulse would correspond to ideal inversion.  Here, we use a Sobolev regularizer~\cite{Adams2003}  $R_1' =\lambda_1' ||\nabla \hinvFIR(t)||^2$ with hyperparameter $\lambda_1'$ and numerical derivative operator $\nabla$, to limit rapid changes in $\hinvFIR(t)$.

\section{Two-Qubit Gate Calibration}
\label{app:CZ}
\begin{figure}
\includegraphics[scale=1]{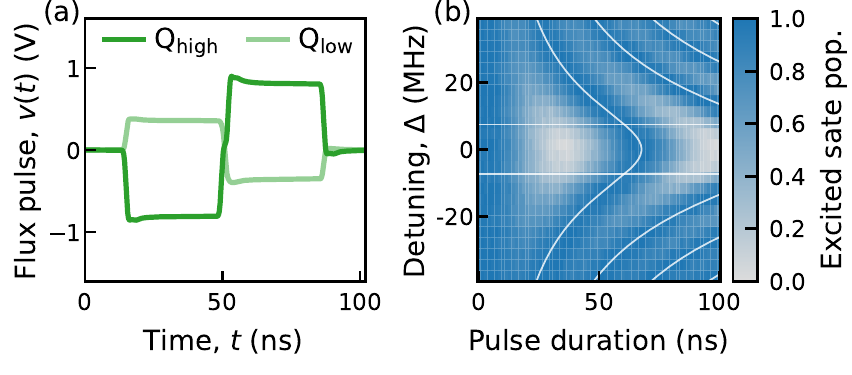}
\caption{(a) Waveforms (with predistortion applied) to implement a CZ gate between a high-frequency transmon $\mathrm{Q}_\mathrm{high}$ and a low-frequency transmon $\mathrm{Q}_\mathrm{low}$.
(b) Excited state population of $\mathrm{Q}_\mathrm{high}$ after a $\ket{11}\leftrightarrow\ket{20}$ interaction.
The curves indicate maximum recovery of the population to the $\ket{11}$ state, and the horizontal lines indicate the fitted coupling strength $J_2$.}
\label{fig:gate}
\end{figure}

With flux pulse predistortion applied,
we calibrate a controlled phase (CZ) gate between a pair of coupled transmons idling at a high and a low frequency (here $\SI{5.89}{\giga\hertz}$ and $\SI{3.88}{\giga\hertz}$), respectively,
as described in detail in \cite{Krinner2022} and summarized in the following.
We apply rectangular flux pulses (smoothed by a Gaussian filter of width $\sigma_\mathrm{FP} = \SI{0.5}{\nano\second}$) to both transmons in order to tune the {\eftrans} transition of the high-frequency transmon $\mathrm{Q}_\mathrm{high}$
and the {\getrans} transition of the low-frequency transmon $\mathrm{Q}_\mathrm{low}$ to a common, intermediate interaction frequency (here $\SI{4.4}{\giga\hertz}$).
The duration of the resonant interaction is chosen to be half a period of the vacuum Rabi oscillation between the $\ket{11}$ state and the noncomputational $\ket{20}$ state,
and a second pulse with opposite polarity is appended for each qubit to implement the second half of the oscillation.
To calibrate the conditional phase acquired by the $\ket{11}$ state to the target value of~$\pi$,
we tune the flux amplitude of $\mathrm{Q}_\mathrm{low}$ during a short transition phase interleaved between the two halves of the net-zero waveform \cite{Negirneac2021,Krinner2022}.
Fig.~\ref{fig:gate}\,(a) shows the predistorted waveforms for a gate between the qubit studied in the main text ($\mathrm{Q}_\mathrm{high}$) and one of its neighbors ($\mathrm{Q}_\mathrm{low}$).

The net-zero pulses employed for implementing the CZ gate
make the execution of long gate sequences robust
against residual distortions on time scales that are long compared to the duration of a gate.
Unlike for residual distortions on short and intermediate time scales,
it is not feasible to account for long-time distortions during gate
calibration as they depend on the history, i.e., on previously applied flux pulses.
Thus, when not using net-zero pulses,
even the small frequency errors observed for long time scales in Fig.~\ref{fig:eval} could lead to undesirably large gate errors in long gate sequences.
For instance, the dynamic phase error of a qubit due to a frequency error of $\SI{0.5}{\mega\hertz}$ during a $\SI{100}{\nano\second}$ flux pulse would be $18^\circ$.
This reveals that the presented methods should be further refined in future research
in order to further reduce residual long-term distortions for use cases where net-zero pulses are not applicable, as discussed in Sec.~\ref{sec:results}.

As a basis for the discussion of residual short-time distortions in Sec.~\ref{sec:results},
we characterize the coupling strength between the $\ket{11}$ and $\ket{20}$ states as follows.
After preparing a $\ket{11}$ state, we
apply unipolar pulses, similar to the first half of the waveforms in Fig.~\ref{fig:gate}\,(a), to both transmons
and measure the $\ket{1}$ state population of $\mathrm{Q}_\mathrm{high}$.
We sweep the amplitude of the pulse for $\mathrm{Q}_\mathrm{low}$ as well as the duration $t$ of both pulses.
To the resulting data set, see Fig.~\ref{fig:gate}\,(b), we fit the model
\begin{equation}
P_{\ket{11}}(\Delta, t) = \frac{\Delta^2 + 2 J_2^2\left(1 + \cos\left(t \sqrt{4J_2^2 +\Delta^2}\right)\right)}{4J_2^2 + \Delta^2}
\end{equation}
where we approximate the detuning $\Delta$ between the $\ket{11}$ and $\ket{20}$ states
as an affine function of the applied pulse amplitude (first-order Taylor approximation around the point of zero detuning)
and fit the parameters of this affine model to the data.
We obtain $J_2/2\pi = \SI{7.4}{\mega\hertz}$ for the coupling strength of the $\ket{11}\leftrightarrow\ket{20}$ interaction, as illustrated by the horizontal lines in Fig.~\ref{fig:gate}\,(b).

\bibliography{refs}

\begin{thebibliography}{66}%
\makeatletter
\providecommand \@ifxundefined [1]{%
 \@ifx{#1\undefined}
}%
\providecommand \@ifnum [1]{%
 \ifnum #1\expandafter \@firstoftwo
 \else \expandafter \@secondoftwo
 \fi
}%
\providecommand \@ifx [1]{%
 \ifx #1\expandafter \@firstoftwo
 \else \expandafter \@secondoftwo
 \fi
}%
\providecommand \natexlab [1]{#1}%
\providecommand \enquote  [1]{``#1''}%
\providecommand \bibnamefont  [1]{#1}%
\providecommand \bibfnamefont [1]{#1}%
\providecommand \citenamefont [1]{#1}%
\providecommand \href@noop [0]{\@secondoftwo}%
\providecommand \href [0]{\begingroup \@sanitize@url \@href}%
\providecommand \@href[1]{\@@startlink{#1}\@@href}%
\providecommand \@@href[1]{\endgroup#1\@@endlink}%
\providecommand \@sanitize@url [0]{\catcode `\\12\catcode `\$12\catcode
  `\&12\catcode `\#12\catcode `\^12\catcode `\_12\catcode `\%12\relax}%
\providecommand \@@startlink[1]{}%
\providecommand \@@endlink[0]{}%
\providecommand \url  [0]{\begingroup\@sanitize@url \@url }%
\providecommand \@url [1]{\endgroup\@href {#1}{\urlprefix }}%
\providecommand \urlprefix  [0]{URL }%
\providecommand \Eprint [0]{\href }%
\providecommand \doibase [0]{https://doi.org/}%
\providecommand \selectlanguage [0]{\@gobble}%
\providecommand \bibinfo  [0]{\@secondoftwo}%
\providecommand \bibfield  [0]{\@secondoftwo}%
\providecommand \translation [1]{[#1]}%
\providecommand \BibitemOpen [0]{}%
\providecommand \bibitemStop [0]{}%
\providecommand \bibitemNoStop [0]{.\EOS\space}%
\providecommand \EOS [0]{\spacefactor3000\relax}%
\providecommand \BibitemShut  [1]{\csname bibitem#1\endcsname}%
\let\auto@bib@innerbib\@empty
\bibitem [{\citenamefont {Koch}\ \emph {et~al.}(2007)\citenamefont {Koch},
  \citenamefont {Yu}, \citenamefont {Gambetta}, \citenamefont {Houck},
  \citenamefont {Schuster}, \citenamefont {Majer}, \citenamefont {Blais},
  \citenamefont {Devoret}, \citenamefont {Girvin},\ and\ \citenamefont
  {Schoelkopf}}]{Koch2007}%
  \BibitemOpen
  \bibfield  {author} {\bibinfo {author} {\bibfnamefont {J.}~\bibnamefont
  {Koch}}, \bibinfo {author} {\bibfnamefont {T.~M.}\ \bibnamefont {Yu}},
  \bibinfo {author} {\bibfnamefont {J.}~\bibnamefont {Gambetta}}, \bibinfo
  {author} {\bibfnamefont {A.~A.}\ \bibnamefont {Houck}}, \bibinfo {author}
  {\bibfnamefont {D.~I.}\ \bibnamefont {Schuster}}, \bibinfo {author}
  {\bibfnamefont {J.}~\bibnamefont {Majer}}, \bibinfo {author} {\bibfnamefont
  {A.}~\bibnamefont {Blais}}, \bibinfo {author} {\bibfnamefont {M.~H.}\
  \bibnamefont {Devoret}}, \bibinfo {author} {\bibfnamefont {S.~M.}\
  \bibnamefont {Girvin}},\ and\ \bibinfo {author} {\bibfnamefont {R.~J.}\
  \bibnamefont {Schoelkopf}},\ }\bibfield  {title} {\bibinfo {title}
  {Charge-insensitive qubit design derived from the {Cooper} pair box},\ }\href
  {https://doi.org/10.1103/PhysRevA.76.042319} {\bibfield  {journal} {\bibinfo
  {journal} {Phys. Rev. A}\ }\textbf {\bibinfo {volume} {76}},\ \bibinfo {eid}
  {042319} (\bibinfo {year} {2007})}\BibitemShut {NoStop}%
\bibitem [{\citenamefont {Schreier}\ \emph {et~al.}(2008)\citenamefont
  {Schreier}, \citenamefont {Houck}, \citenamefont {Koch}, \citenamefont
  {Schuster}, \citenamefont {Johnson}, \citenamefont {Chow}, \citenamefont
  {Gambetta}, \citenamefont {Majer}, \citenamefont {Frunzio}, \citenamefont
  {Devoret}, \citenamefont {Girvin},\ and\ \citenamefont
  {Schoelkopf}}]{Schreier2008}%
  \BibitemOpen
  \bibfield  {author} {\bibinfo {author} {\bibfnamefont {J.~A.}\ \bibnamefont
  {Schreier}}, \bibinfo {author} {\bibfnamefont {A.~A.}\ \bibnamefont {Houck}},
  \bibinfo {author} {\bibfnamefont {J.}~\bibnamefont {Koch}}, \bibinfo {author}
  {\bibfnamefont {D.~I.}\ \bibnamefont {Schuster}}, \bibinfo {author}
  {\bibfnamefont {B.~R.}\ \bibnamefont {Johnson}}, \bibinfo {author}
  {\bibfnamefont {J.~M.}\ \bibnamefont {Chow}}, \bibinfo {author}
  {\bibfnamefont {J.~M.}\ \bibnamefont {Gambetta}}, \bibinfo {author}
  {\bibfnamefont {J.}~\bibnamefont {Majer}}, \bibinfo {author} {\bibfnamefont
  {L.}~\bibnamefont {Frunzio}}, \bibinfo {author} {\bibfnamefont {M.~H.}\
  \bibnamefont {Devoret}}, \bibinfo {author} {\bibfnamefont {S.~M.}\
  \bibnamefont {Girvin}},\ and\ \bibinfo {author} {\bibfnamefont {R.~J.}\
  \bibnamefont {Schoelkopf}},\ }\bibfield  {title} {\bibinfo {title}
  {Suppressing charge noise decoherence in superconducting charge qubits},\
  }\href {https://doi.org/10.1103/PhysRevB.77.180502} {\bibfield  {journal}
  {\bibinfo  {journal} {Phys. Rev. B}\ }\textbf {\bibinfo {volume} {77}},\
  \bibinfo {pages} {180502} (\bibinfo {year} {2008})}\BibitemShut {NoStop}%
\bibitem [{\citenamefont {Klimov}\ \emph {et~al.}(2018)\citenamefont {Klimov},
  \citenamefont {Kelly}, \citenamefont {Chen}, \citenamefont {Neeley},
  \citenamefont {Megrant}, \citenamefont {Burkett}, \citenamefont {Barends},
  \citenamefont {Arya}, \citenamefont {Chiaro}, \citenamefont {Chen},
  \citenamefont {Dunsworth}, \citenamefont {Fowler}, \citenamefont {Foxen},
  \citenamefont {Gidney}, \citenamefont {Giustina}, \citenamefont {Graff},
  \citenamefont {Huang}, \citenamefont {Jeffrey}, \citenamefont {Lucero},
  \citenamefont {Mutus}, \citenamefont {Naaman}, \citenamefont {Neill},
  \citenamefont {Quintana}, \citenamefont {Roushan}, \citenamefont {Sank},
  \citenamefont {Vainsencher}, \citenamefont {Wenner}, \citenamefont {White},
  \citenamefont {Boixo}, \citenamefont {Babbush}, \citenamefont {Smelyanskiy},
  \citenamefont {Neven},\ and\ \citenamefont {Martinis}}]{Klimov2018}%
  \BibitemOpen
  \bibfield  {author} {\bibinfo {author} {\bibfnamefont {P.~V.}\ \bibnamefont
  {Klimov}}, \bibinfo {author} {\bibfnamefont {J.}~\bibnamefont {Kelly}},
  \bibinfo {author} {\bibfnamefont {Z.}~\bibnamefont {Chen}}, \bibinfo {author}
  {\bibfnamefont {M.}~\bibnamefont {Neeley}}, \bibinfo {author} {\bibfnamefont
  {A.}~\bibnamefont {Megrant}}, \bibinfo {author} {\bibfnamefont
  {B.}~\bibnamefont {Burkett}}, \bibinfo {author} {\bibfnamefont
  {R.}~\bibnamefont {Barends}}, \bibinfo {author} {\bibfnamefont
  {K.}~\bibnamefont {Arya}}, \bibinfo {author} {\bibfnamefont {B.}~\bibnamefont
  {Chiaro}}, \bibinfo {author} {\bibfnamefont {Y.}~\bibnamefont {Chen}},
  \bibinfo {author} {\bibfnamefont {A.}~\bibnamefont {Dunsworth}}, \bibinfo
  {author} {\bibfnamefont {A.}~\bibnamefont {Fowler}}, \bibinfo {author}
  {\bibfnamefont {B.}~\bibnamefont {Foxen}}, \bibinfo {author} {\bibfnamefont
  {C.}~\bibnamefont {Gidney}}, \bibinfo {author} {\bibfnamefont
  {M.}~\bibnamefont {Giustina}}, \bibinfo {author} {\bibfnamefont
  {R.}~\bibnamefont {Graff}}, \bibinfo {author} {\bibfnamefont
  {T.}~\bibnamefont {Huang}}, \bibinfo {author} {\bibfnamefont
  {E.}~\bibnamefont {Jeffrey}}, \bibinfo {author} {\bibfnamefont
  {E.}~\bibnamefont {Lucero}}, \bibinfo {author} {\bibfnamefont {J.~Y.}\
  \bibnamefont {Mutus}}, \bibinfo {author} {\bibfnamefont {O.}~\bibnamefont
  {Naaman}}, \bibinfo {author} {\bibfnamefont {C.}~\bibnamefont {Neill}},
  \bibinfo {author} {\bibfnamefont {C.}~\bibnamefont {Quintana}}, \bibinfo
  {author} {\bibfnamefont {P.}~\bibnamefont {Roushan}}, \bibinfo {author}
  {\bibfnamefont {D.}~\bibnamefont {Sank}}, \bibinfo {author} {\bibfnamefont
  {A.}~\bibnamefont {Vainsencher}}, \bibinfo {author} {\bibfnamefont
  {J.}~\bibnamefont {Wenner}}, \bibinfo {author} {\bibfnamefont {T.~C.}\
  \bibnamefont {White}}, \bibinfo {author} {\bibfnamefont {S.}~\bibnamefont
  {Boixo}}, \bibinfo {author} {\bibfnamefont {R.}~\bibnamefont {Babbush}},
  \bibinfo {author} {\bibfnamefont {V.~N.}\ \bibnamefont {Smelyanskiy}},
  \bibinfo {author} {\bibfnamefont {H.}~\bibnamefont {Neven}},\ and\ \bibinfo
  {author} {\bibfnamefont {J.~M.}\ \bibnamefont {Martinis}},\ }\bibfield
  {title} {\bibinfo {title} {Fluctuations of energy-relaxation times in
  superconducting qubits},\ }\href
  {https://doi.org/10.1103/PhysRevLett.121.090502} {\bibfield  {journal}
  {\bibinfo  {journal} {Phys. Rev. Lett.}\ }\textbf {\bibinfo {volume} {121}},\
  \bibinfo {pages} {090502} (\bibinfo {year} {2018})}\BibitemShut {NoStop}%
\bibitem [{\citenamefont {Krinner}\ \emph {et~al.}(2022)\citenamefont
  {Krinner}, \citenamefont {Lacroix}, \citenamefont {Remm}, \citenamefont
  {Paolo}, \citenamefont {Genois}, \citenamefont {Leroux}, \citenamefont
  {Hellings}, \citenamefont {Lazar}, \citenamefont {Swiadek}, \citenamefont
  {Herrmann}, \citenamefont {Norris}, \citenamefont {Andersen}, \citenamefont
  {M\"{u}ller}, \citenamefont {Blais}, \citenamefont {Eichler},\ and\
  \citenamefont {Wallraff}}]{Krinner2022}%
  \BibitemOpen
  \bibfield  {author} {\bibinfo {author} {\bibfnamefont {S.}~\bibnamefont
  {Krinner}}, \bibinfo {author} {\bibfnamefont {N.}~\bibnamefont {Lacroix}},
  \bibinfo {author} {\bibfnamefont {A.}~\bibnamefont {Remm}}, \bibinfo {author}
  {\bibfnamefont {A.~D.}\ \bibnamefont {Paolo}}, \bibinfo {author}
  {\bibfnamefont {E.}~\bibnamefont {Genois}}, \bibinfo {author} {\bibfnamefont
  {C.}~\bibnamefont {Leroux}}, \bibinfo {author} {\bibfnamefont
  {C.}~\bibnamefont {Hellings}}, \bibinfo {author} {\bibfnamefont
  {S.}~\bibnamefont {Lazar}}, \bibinfo {author} {\bibfnamefont
  {F.}~\bibnamefont {Swiadek}}, \bibinfo {author} {\bibfnamefont
  {J.}~\bibnamefont {Herrmann}}, \bibinfo {author} {\bibfnamefont {G.~J.}\
  \bibnamefont {Norris}}, \bibinfo {author} {\bibfnamefont {C.~K.}\
  \bibnamefont {Andersen}}, \bibinfo {author} {\bibfnamefont {M.}~\bibnamefont
  {M\"{u}ller}}, \bibinfo {author} {\bibfnamefont {A.}~\bibnamefont {Blais}},
  \bibinfo {author} {\bibfnamefont {C.}~\bibnamefont {Eichler}},\ and\ \bibinfo
  {author} {\bibfnamefont {A.}~\bibnamefont {Wallraff}},\ }\bibfield  {title}
  {\bibinfo {title} {Realizing repeated quantum error correction in a
  distance-three surface code},\ }\href
  {https://doi.org/10.1038/s41586-022-04566-8} {\bibfield  {journal} {\bibinfo
  {journal} {Nature}\ }\textbf {\bibinfo {volume} {605}},\ \bibinfo {pages}
  {669} (\bibinfo {year} {2022})}\BibitemShut {NoStop}%
\bibitem [{\citenamefont {Zhou}\ \emph {et~al.}(2021)\citenamefont {Zhou},
  \citenamefont {Zhang}, \citenamefont {Yin}, \citenamefont {Huai},
  \citenamefont {Gu}, \citenamefont {Xu}, \citenamefont {Allcock},
  \citenamefont {Liu}, \citenamefont {Xi}, \citenamefont {Yu}, \citenamefont
  {Zhang}, \citenamefont {Zhang}, \citenamefont {Li}, \citenamefont {Song},
  \citenamefont {Wang}, \citenamefont {Zheng}, \citenamefont {An},
  \citenamefont {Zheng},\ and\ \citenamefont {Zhang}}]{Zhou2021}%
  \BibitemOpen
  \bibfield  {author} {\bibinfo {author} {\bibfnamefont {Y.}~\bibnamefont
  {Zhou}}, \bibinfo {author} {\bibfnamefont {Z.}~\bibnamefont {Zhang}},
  \bibinfo {author} {\bibfnamefont {Z.}~\bibnamefont {Yin}}, \bibinfo {author}
  {\bibfnamefont {S.}~\bibnamefont {Huai}}, \bibinfo {author} {\bibfnamefont
  {X.}~\bibnamefont {Gu}}, \bibinfo {author} {\bibfnamefont {X.}~\bibnamefont
  {Xu}}, \bibinfo {author} {\bibfnamefont {J.}~\bibnamefont {Allcock}},
  \bibinfo {author} {\bibfnamefont {F.}~\bibnamefont {Liu}}, \bibinfo {author}
  {\bibfnamefont {G.}~\bibnamefont {Xi}}, \bibinfo {author} {\bibfnamefont
  {Q.}~\bibnamefont {Yu}}, \bibinfo {author} {\bibfnamefont {H.}~\bibnamefont
  {Zhang}}, \bibinfo {author} {\bibfnamefont {M.}~\bibnamefont {Zhang}},
  \bibinfo {author} {\bibfnamefont {H.}~\bibnamefont {Li}}, \bibinfo {author}
  {\bibfnamefont {X.}~\bibnamefont {Song}}, \bibinfo {author} {\bibfnamefont
  {Z.}~\bibnamefont {Wang}}, \bibinfo {author} {\bibfnamefont {D.}~\bibnamefont
  {Zheng}}, \bibinfo {author} {\bibfnamefont {S.}~\bibnamefont {An}}, \bibinfo
  {author} {\bibfnamefont {Y.}~\bibnamefont {Zheng}},\ and\ \bibinfo {author}
  {\bibfnamefont {S.}~\bibnamefont {Zhang}},\ }\bibfield  {title} {\bibinfo
  {title} {Rapid and unconditional parametric reset protocol for tunable
  superconducting qubits},\ }\href {https://doi.org/10.1038/s41467-021-26205-y}
  {\bibfield  {journal} {\bibinfo  {journal} {Nature Communications}\ }\textbf
  {\bibinfo {volume} {12}},\ \bibinfo {pages} {5924} (\bibinfo {year}
  {2021})}\BibitemShut {NoStop}%
\bibitem [{\citenamefont {McEwen}\ \emph {et~al.}(2021)\citenamefont {McEwen},
  \citenamefont {Kafri}, \citenamefont {Chen}, \citenamefont {Atalaya},
  \citenamefont {Satzinger}, \citenamefont {Quintana}, \citenamefont {Klimov},
  \citenamefont {Sank}, \citenamefont {Gidney}, \citenamefont {Fowler},
  \citenamefont {Arute}, \citenamefont {Arya}, \citenamefont {Buckley},
  \citenamefont {Burkett}, \citenamefont {Bushnell}, \citenamefont {Chiaro},
  \citenamefont {Collins}, \citenamefont {Demura}, \citenamefont {Dunsworth},
  \citenamefont {Erickson}, \citenamefont {Foxen}, \citenamefont {Giustina},
  \citenamefont {Huang}, \citenamefont {Hong}, \citenamefont {Jeffrey},
  \citenamefont {Kim}, \citenamefont {Kechedzhi}, \citenamefont {Kostritsa},
  \citenamefont {Laptev}, \citenamefont {Megrant}, \citenamefont {Mi},
  \citenamefont {Mutus}, \citenamefont {Naaman}, \citenamefont {Neeley},
  \citenamefont {Neill}, \citenamefont {Niu}, \citenamefont {Paler},
  \citenamefont {Redd}, \citenamefont {Roushan}, \citenamefont {White},
  \citenamefont {Yao}, \citenamefont {Yeh}, \citenamefont {Zalcman},
  \citenamefont {Chen}, \citenamefont {Smelyanskiy}, \citenamefont {Martinis},
  \citenamefont {Neven}, \citenamefont {Kelly}, \citenamefont {Korotkov},
  \citenamefont {Petukhov},\ and\ \citenamefont {Barends}}]{McEwen2021a}%
  \BibitemOpen
  \bibfield  {author} {\bibinfo {author} {\bibfnamefont {M.}~\bibnamefont
  {McEwen}}, \bibinfo {author} {\bibfnamefont {D.}~\bibnamefont {Kafri}},
  \bibinfo {author} {\bibfnamefont {Z.}~\bibnamefont {Chen}}, \bibinfo {author}
  {\bibfnamefont {J.}~\bibnamefont {Atalaya}}, \bibinfo {author} {\bibfnamefont
  {K.~J.}\ \bibnamefont {Satzinger}}, \bibinfo {author} {\bibfnamefont
  {C.}~\bibnamefont {Quintana}}, \bibinfo {author} {\bibfnamefont {P.~V.}\
  \bibnamefont {Klimov}}, \bibinfo {author} {\bibfnamefont {D.}~\bibnamefont
  {Sank}}, \bibinfo {author} {\bibfnamefont {C.}~\bibnamefont {Gidney}},
  \bibinfo {author} {\bibfnamefont {A.~G.}\ \bibnamefont {Fowler}}, \bibinfo
  {author} {\bibfnamefont {F.}~\bibnamefont {Arute}}, \bibinfo {author}
  {\bibfnamefont {K.}~\bibnamefont {Arya}}, \bibinfo {author} {\bibfnamefont
  {B.}~\bibnamefont {Buckley}}, \bibinfo {author} {\bibfnamefont
  {B.}~\bibnamefont {Burkett}}, \bibinfo {author} {\bibfnamefont
  {N.}~\bibnamefont {Bushnell}}, \bibinfo {author} {\bibfnamefont
  {B.}~\bibnamefont {Chiaro}}, \bibinfo {author} {\bibfnamefont
  {R.}~\bibnamefont {Collins}}, \bibinfo {author} {\bibfnamefont
  {S.}~\bibnamefont {Demura}}, \bibinfo {author} {\bibfnamefont
  {A.}~\bibnamefont {Dunsworth}}, \bibinfo {author} {\bibfnamefont
  {C.}~\bibnamefont {Erickson}}, \bibinfo {author} {\bibfnamefont
  {B.}~\bibnamefont {Foxen}}, \bibinfo {author} {\bibfnamefont
  {M.}~\bibnamefont {Giustina}}, \bibinfo {author} {\bibfnamefont
  {T.}~\bibnamefont {Huang}}, \bibinfo {author} {\bibfnamefont
  {S.}~\bibnamefont {Hong}}, \bibinfo {author} {\bibfnamefont {E.}~\bibnamefont
  {Jeffrey}}, \bibinfo {author} {\bibfnamefont {S.}~\bibnamefont {Kim}},
  \bibinfo {author} {\bibfnamefont {K.}~\bibnamefont {Kechedzhi}}, \bibinfo
  {author} {\bibfnamefont {F.}~\bibnamefont {Kostritsa}}, \bibinfo {author}
  {\bibfnamefont {P.}~\bibnamefont {Laptev}}, \bibinfo {author} {\bibfnamefont
  {A.}~\bibnamefont {Megrant}}, \bibinfo {author} {\bibfnamefont
  {X.}~\bibnamefont {Mi}}, \bibinfo {author} {\bibfnamefont {J.}~\bibnamefont
  {Mutus}}, \bibinfo {author} {\bibfnamefont {O.}~\bibnamefont {Naaman}},
  \bibinfo {author} {\bibfnamefont {M.}~\bibnamefont {Neeley}}, \bibinfo
  {author} {\bibfnamefont {C.}~\bibnamefont {Neill}}, \bibinfo {author}
  {\bibfnamefont {M.}~\bibnamefont {Niu}}, \bibinfo {author} {\bibfnamefont
  {A.}~\bibnamefont {Paler}}, \bibinfo {author} {\bibfnamefont
  {N.}~\bibnamefont {Redd}}, \bibinfo {author} {\bibfnamefont {P.}~\bibnamefont
  {Roushan}}, \bibinfo {author} {\bibfnamefont {T.~C.}\ \bibnamefont {White}},
  \bibinfo {author} {\bibfnamefont {J.}~\bibnamefont {Yao}}, \bibinfo {author}
  {\bibfnamefont {P.}~\bibnamefont {Yeh}}, \bibinfo {author} {\bibfnamefont
  {A.}~\bibnamefont {Zalcman}}, \bibinfo {author} {\bibfnamefont
  {Y.}~\bibnamefont {Chen}}, \bibinfo {author} {\bibfnamefont {V.~N.}\
  \bibnamefont {Smelyanskiy}}, \bibinfo {author} {\bibfnamefont {J.~M.}\
  \bibnamefont {Martinis}}, \bibinfo {author} {\bibfnamefont {H.}~\bibnamefont
  {Neven}}, \bibinfo {author} {\bibfnamefont {J.}~\bibnamefont {Kelly}},
  \bibinfo {author} {\bibfnamefont {A.~N.}\ \bibnamefont {Korotkov}}, \bibinfo
  {author} {\bibfnamefont {A.~G.}\ \bibnamefont {Petukhov}},\ and\ \bibinfo
  {author} {\bibfnamefont {R.}~\bibnamefont {Barends}},\ }\bibfield  {title}
  {\bibinfo {title} {Removing leakage-induced correlated errors in
  superconducting quantum error correction},\ }\href
  {https://doi.org/10.1038/s41467-021-21982-y} {\bibfield  {journal} {\bibinfo
  {journal} {Nature Communications}\ }\textbf {\bibinfo {volume} {12}},\
  \bibinfo {pages} {1761} (\bibinfo {year} {2021})}\BibitemShut {NoStop}%
\bibitem [{\citenamefont {Strauch}\ \emph {et~al.}(2003)\citenamefont
  {Strauch}, \citenamefont {Johnson}, \citenamefont {Dragt}, \citenamefont
  {Lobb}, \citenamefont {Anderson},\ and\ \citenamefont
  {Wellstood}}]{Strauch2003}%
  \BibitemOpen
  \bibfield  {author} {\bibinfo {author} {\bibfnamefont {F.~W.}\ \bibnamefont
  {Strauch}}, \bibinfo {author} {\bibfnamefont {P.~R.}\ \bibnamefont
  {Johnson}}, \bibinfo {author} {\bibfnamefont {A.~J.}\ \bibnamefont {Dragt}},
  \bibinfo {author} {\bibfnamefont {C.~J.}\ \bibnamefont {Lobb}}, \bibinfo
  {author} {\bibfnamefont {J.~R.}\ \bibnamefont {Anderson}},\ and\ \bibinfo
  {author} {\bibfnamefont {F.~C.}\ \bibnamefont {Wellstood}},\ }\bibfield
  {title} {\bibinfo {title} {Quantum logic gates for coupled superconducting
  phase qubits},\ }\href {https://doi.org/10.1103/PhysRevLett.91.167005}
  {\bibfield  {journal} {\bibinfo  {journal} {Phys. Rev. Lett.}\ }\textbf
  {\bibinfo {volume} {91}},\ \bibinfo {pages} {167005} (\bibinfo {year}
  {2003})}\BibitemShut {NoStop}%
\bibitem [{\citenamefont {DiCarlo}\ \emph {et~al.}(2009)\citenamefont
  {DiCarlo}, \citenamefont {Chow}, \citenamefont {Gambetta}, \citenamefont
  {Bishop}, \citenamefont {Johnson}, \citenamefont {Schuster}, \citenamefont
  {Majer}, \citenamefont {Blais}, \citenamefont {Frunzio}, \citenamefont
  {Girvin},\ and\ \citenamefont {Schoelkopf}}]{DiCarlo2009}%
  \BibitemOpen
  \bibfield  {author} {\bibinfo {author} {\bibfnamefont {L.}~\bibnamefont
  {DiCarlo}}, \bibinfo {author} {\bibfnamefont {J.~M.}\ \bibnamefont {Chow}},
  \bibinfo {author} {\bibfnamefont {J.~M.}\ \bibnamefont {Gambetta}}, \bibinfo
  {author} {\bibfnamefont {L.~S.}\ \bibnamefont {Bishop}}, \bibinfo {author}
  {\bibfnamefont {B.~R.}\ \bibnamefont {Johnson}}, \bibinfo {author}
  {\bibfnamefont {D.~I.}\ \bibnamefont {Schuster}}, \bibinfo {author}
  {\bibfnamefont {J.}~\bibnamefont {Majer}}, \bibinfo {author} {\bibfnamefont
  {A.}~\bibnamefont {Blais}}, \bibinfo {author} {\bibfnamefont
  {L.}~\bibnamefont {Frunzio}}, \bibinfo {author} {\bibfnamefont {S.~M.}\
  \bibnamefont {Girvin}},\ and\ \bibinfo {author} {\bibfnamefont {R.~J.}\
  \bibnamefont {Schoelkopf}},\ }\bibfield  {title} {\bibinfo {title}
  {Demonstration of two-qubit algorithms with a superconducting quantum
  processor},\ }\href {https://doi.org/10.1038/nature08121} {\bibfield
  {journal} {\bibinfo  {journal} {Nature}\ }\textbf {\bibinfo {volume} {460}},\
  \bibinfo {pages} {240} (\bibinfo {year} {2009})}\BibitemShut {NoStop}%
\bibitem [{\citenamefont {Rol}\ \emph {et~al.}(2019)\citenamefont {Rol},
  \citenamefont {Battistel}, \citenamefont {Malinowski}, \citenamefont
  {Bultink}, \citenamefont {Tarasinski}, \citenamefont {Vollmer}, \citenamefont
  {Haider}, \citenamefont {Muthusubramanian}, \citenamefont {Bruno},
  \citenamefont {Terhal},\ and\ \citenamefont {DiCarlo}}]{Rol2019}%
  \BibitemOpen
  \bibfield  {author} {\bibinfo {author} {\bibfnamefont {M.~A.}\ \bibnamefont
  {Rol}}, \bibinfo {author} {\bibfnamefont {F.}~\bibnamefont {Battistel}},
  \bibinfo {author} {\bibfnamefont {F.~K.}\ \bibnamefont {Malinowski}},
  \bibinfo {author} {\bibfnamefont {C.~C.}\ \bibnamefont {Bultink}}, \bibinfo
  {author} {\bibfnamefont {B.~M.}\ \bibnamefont {Tarasinski}}, \bibinfo
  {author} {\bibfnamefont {R.}~\bibnamefont {Vollmer}}, \bibinfo {author}
  {\bibfnamefont {N.}~\bibnamefont {Haider}}, \bibinfo {author} {\bibfnamefont
  {N.}~\bibnamefont {Muthusubramanian}}, \bibinfo {author} {\bibfnamefont
  {A.}~\bibnamefont {Bruno}}, \bibinfo {author} {\bibfnamefont {B.~M.}\
  \bibnamefont {Terhal}},\ and\ \bibinfo {author} {\bibfnamefont
  {L.}~\bibnamefont {DiCarlo}},\ }\bibfield  {title} {\bibinfo {title} {Fast,
  high-fidelity conditional-phase gate exploiting leakage interference in
  weakly anharmonic superconducting qubits},\ }\href
  {https://doi.org/10.1103/PhysRevLett.123.120502} {\bibfield  {journal}
  {\bibinfo  {journal} {Phys. Rev. Lett.}\ }\textbf {\bibinfo {volume} {123}},\
  \bibinfo {pages} {120502} (\bibinfo {year} {2019})}\BibitemShut {NoStop}%
\bibitem [{\citenamefont {Negirneac}\ \emph {et~al.}(2021)\citenamefont
  {Negirneac}, \citenamefont {Ali}, \citenamefont {Muthusubramanian},
  \citenamefont {Battistel}, \citenamefont {Sagastizabal}, \citenamefont
  {Moreira}, \citenamefont {Marques}, \citenamefont {Vlothuizen}, \citenamefont
  {Beekman}, \citenamefont {Zachariadis}, \citenamefont {Haider}, \citenamefont
  {Bruno},\ and\ \citenamefont {DiCarlo}}]{Negirneac2021}%
  \BibitemOpen
  \bibfield  {author} {\bibinfo {author} {\bibfnamefont {V.}~\bibnamefont
  {Negirneac}}, \bibinfo {author} {\bibfnamefont {H.}~\bibnamefont {Ali}},
  \bibinfo {author} {\bibfnamefont {N.}~\bibnamefont {Muthusubramanian}},
  \bibinfo {author} {\bibfnamefont {F.}~\bibnamefont {Battistel}}, \bibinfo
  {author} {\bibfnamefont {R.}~\bibnamefont {Sagastizabal}}, \bibinfo {author}
  {\bibfnamefont {M.~S.}\ \bibnamefont {Moreira}}, \bibinfo {author}
  {\bibfnamefont {J.~F.}\ \bibnamefont {Marques}}, \bibinfo {author}
  {\bibfnamefont {W.~J.}\ \bibnamefont {Vlothuizen}}, \bibinfo {author}
  {\bibfnamefont {M.}~\bibnamefont {Beekman}}, \bibinfo {author} {\bibfnamefont
  {C.}~\bibnamefont {Zachariadis}}, \bibinfo {author} {\bibfnamefont
  {N.}~\bibnamefont {Haider}}, \bibinfo {author} {\bibfnamefont
  {A.}~\bibnamefont {Bruno}},\ and\ \bibinfo {author} {\bibfnamefont
  {L.}~\bibnamefont {DiCarlo}},\ }\bibfield  {title} {\bibinfo {title}
  {High-fidelity controlled-${Z}$ gate with maximal intermediate leakage
  operating at the speed limit in a superconducting quantum processor},\ }\href
  {https://doi.org/10.1103/PhysRevLett.126.220502} {\bibfield  {journal}
  {\bibinfo  {journal} {Phys. Rev. Lett.}\ }\textbf {\bibinfo {volume} {126}},\
  \bibinfo {pages} {220502} (\bibinfo {year} {2021})}\BibitemShut {NoStop}%
\bibitem [{\citenamefont {Royer}\ \emph {et~al.}(2017)\citenamefont {Royer},
  \citenamefont {Grimsmo}, \citenamefont {Didier},\ and\ \citenamefont
  {Blais}}]{Royer2017}%
  \BibitemOpen
  \bibfield  {author} {\bibinfo {author} {\bibfnamefont {B.}~\bibnamefont
  {Royer}}, \bibinfo {author} {\bibfnamefont {A.~L.}\ \bibnamefont {Grimsmo}},
  \bibinfo {author} {\bibfnamefont {N.}~\bibnamefont {Didier}},\ and\ \bibinfo
  {author} {\bibfnamefont {A.}~\bibnamefont {Blais}},\ }\bibfield  {title}
  {\bibinfo {title} {Fast and high-fidelity entangling gate through
  parametrically modulated longitudinal coupling},\ }\href
  {https://doi.org/10.22331/q-2017-05-11-11} {\bibfield  {journal} {\bibinfo
  {journal} {Quantum}\ }\textbf {\bibinfo {volume} {1}},\ \bibinfo {pages} {11}
  (\bibinfo {year} {2017})}\BibitemShut {NoStop}%
\bibitem [{\citenamefont {Caldwell}\ \emph {et~al.}(2018)\citenamefont
  {Caldwell}, \citenamefont {Didier}, \citenamefont {Ryan}, \citenamefont
  {Sete}, \citenamefont {Hudson}, \citenamefont {Karalekas}, \citenamefont
  {Manenti}, \citenamefont {da~Silva}, \citenamefont {Sinclair}, \citenamefont
  {Acala}, \citenamefont {Alidoust}, \citenamefont {Angeles}, \citenamefont
  {Bestwick}, \citenamefont {Block}, \citenamefont {Bloom}, \citenamefont
  {Bradley}, \citenamefont {Bui}, \citenamefont {Capelluto}, \citenamefont
  {Chilcott}, \citenamefont {Cordova}, \citenamefont {Crossman}, \citenamefont
  {Curtis}, \citenamefont {Deshpande}, \citenamefont {Bouayadi}, \citenamefont
  {Girshovich}, \citenamefont {Hong}, \citenamefont {Kuang}, \citenamefont
  {Lenihan}, \citenamefont {Manning}, \citenamefont {Marchenkov}, \citenamefont
  {Marshall}, \citenamefont {Maydra}, \citenamefont {Mohan}, \citenamefont
  {O'Brien}, \citenamefont {Osborn}, \citenamefont {Otterbach}, \citenamefont
  {Papageorge}, \citenamefont {Paquette}, \citenamefont {Pelstring},
  \citenamefont {Polloreno}, \citenamefont {Prawiroatmodjo}, \citenamefont
  {Rawat}, \citenamefont {Reagor}, \citenamefont {Renzas}, \citenamefont
  {Rubin}, \citenamefont {Russell}, \citenamefont {Rust}, \citenamefont
  {Scarabelli}, \citenamefont {Scheer}, \citenamefont {Selvanayagam},
  \citenamefont {Smith}, \citenamefont {Staley}, \citenamefont {Suska},
  \citenamefont {Tezak}, \citenamefont {Thompson}, \citenamefont {To},
  \citenamefont {Vahidpour}, \citenamefont {Vodrahalli}, \citenamefont
  {Whyland}, \citenamefont {Yadav}, \citenamefont {Zeng},\ and\ \citenamefont
  {Rigetti}}]{Caldwell2018}%
  \BibitemOpen
  \bibfield  {author} {\bibinfo {author} {\bibfnamefont {S.~A.}\ \bibnamefont
  {Caldwell}}, \bibinfo {author} {\bibfnamefont {N.}~\bibnamefont {Didier}},
  \bibinfo {author} {\bibfnamefont {C.~A.}\ \bibnamefont {Ryan}}, \bibinfo
  {author} {\bibfnamefont {E.~A.}\ \bibnamefont {Sete}}, \bibinfo {author}
  {\bibfnamefont {A.}~\bibnamefont {Hudson}}, \bibinfo {author} {\bibfnamefont
  {P.}~\bibnamefont {Karalekas}}, \bibinfo {author} {\bibfnamefont
  {R.}~\bibnamefont {Manenti}}, \bibinfo {author} {\bibfnamefont {M.~P.}\
  \bibnamefont {da~Silva}}, \bibinfo {author} {\bibfnamefont {R.}~\bibnamefont
  {Sinclair}}, \bibinfo {author} {\bibfnamefont {E.}~\bibnamefont {Acala}},
  \bibinfo {author} {\bibfnamefont {N.}~\bibnamefont {Alidoust}}, \bibinfo
  {author} {\bibfnamefont {J.}~\bibnamefont {Angeles}}, \bibinfo {author}
  {\bibfnamefont {A.}~\bibnamefont {Bestwick}}, \bibinfo {author}
  {\bibfnamefont {M.}~\bibnamefont {Block}}, \bibinfo {author} {\bibfnamefont
  {B.}~\bibnamefont {Bloom}}, \bibinfo {author} {\bibfnamefont
  {A.}~\bibnamefont {Bradley}}, \bibinfo {author} {\bibfnamefont
  {C.}~\bibnamefont {Bui}}, \bibinfo {author} {\bibfnamefont {L.}~\bibnamefont
  {Capelluto}}, \bibinfo {author} {\bibfnamefont {R.}~\bibnamefont {Chilcott}},
  \bibinfo {author} {\bibfnamefont {J.}~\bibnamefont {Cordova}}, \bibinfo
  {author} {\bibfnamefont {G.}~\bibnamefont {Crossman}}, \bibinfo {author}
  {\bibfnamefont {M.}~\bibnamefont {Curtis}}, \bibinfo {author} {\bibfnamefont
  {S.}~\bibnamefont {Deshpande}}, \bibinfo {author} {\bibfnamefont {T.~E.}\
  \bibnamefont {Bouayadi}}, \bibinfo {author} {\bibfnamefont {D.}~\bibnamefont
  {Girshovich}}, \bibinfo {author} {\bibfnamefont {S.}~\bibnamefont {Hong}},
  \bibinfo {author} {\bibfnamefont {K.}~\bibnamefont {Kuang}}, \bibinfo
  {author} {\bibfnamefont {M.}~\bibnamefont {Lenihan}}, \bibinfo {author}
  {\bibfnamefont {T.}~\bibnamefont {Manning}}, \bibinfo {author} {\bibfnamefont
  {A.}~\bibnamefont {Marchenkov}}, \bibinfo {author} {\bibfnamefont
  {J.}~\bibnamefont {Marshall}}, \bibinfo {author} {\bibfnamefont
  {R.}~\bibnamefont {Maydra}}, \bibinfo {author} {\bibfnamefont
  {Y.}~\bibnamefont {Mohan}}, \bibinfo {author} {\bibfnamefont
  {W.}~\bibnamefont {O'Brien}}, \bibinfo {author} {\bibfnamefont
  {C.}~\bibnamefont {Osborn}}, \bibinfo {author} {\bibfnamefont
  {J.}~\bibnamefont {Otterbach}}, \bibinfo {author} {\bibfnamefont
  {A.}~\bibnamefont {Papageorge}}, \bibinfo {author} {\bibfnamefont {J.-P.}\
  \bibnamefont {Paquette}}, \bibinfo {author} {\bibfnamefont {M.}~\bibnamefont
  {Pelstring}}, \bibinfo {author} {\bibfnamefont {A.}~\bibnamefont
  {Polloreno}}, \bibinfo {author} {\bibfnamefont {G.}~\bibnamefont
  {Prawiroatmodjo}}, \bibinfo {author} {\bibfnamefont {V.}~\bibnamefont
  {Rawat}}, \bibinfo {author} {\bibfnamefont {M.}~\bibnamefont {Reagor}},
  \bibinfo {author} {\bibfnamefont {R.}~\bibnamefont {Renzas}}, \bibinfo
  {author} {\bibfnamefont {N.}~\bibnamefont {Rubin}}, \bibinfo {author}
  {\bibfnamefont {D.}~\bibnamefont {Russell}}, \bibinfo {author} {\bibfnamefont
  {M.}~\bibnamefont {Rust}}, \bibinfo {author} {\bibfnamefont {D.}~\bibnamefont
  {Scarabelli}}, \bibinfo {author} {\bibfnamefont {M.}~\bibnamefont {Scheer}},
  \bibinfo {author} {\bibfnamefont {M.}~\bibnamefont {Selvanayagam}}, \bibinfo
  {author} {\bibfnamefont {R.}~\bibnamefont {Smith}}, \bibinfo {author}
  {\bibfnamefont {A.}~\bibnamefont {Staley}}, \bibinfo {author} {\bibfnamefont
  {M.}~\bibnamefont {Suska}}, \bibinfo {author} {\bibfnamefont
  {N.}~\bibnamefont {Tezak}}, \bibinfo {author} {\bibfnamefont {D.~C.}\
  \bibnamefont {Thompson}}, \bibinfo {author} {\bibfnamefont {T.-W.}\
  \bibnamefont {To}}, \bibinfo {author} {\bibfnamefont {M.}~\bibnamefont
  {Vahidpour}}, \bibinfo {author} {\bibfnamefont {N.}~\bibnamefont
  {Vodrahalli}}, \bibinfo {author} {\bibfnamefont {T.}~\bibnamefont {Whyland}},
  \bibinfo {author} {\bibfnamefont {K.}~\bibnamefont {Yadav}}, \bibinfo
  {author} {\bibfnamefont {W.}~\bibnamefont {Zeng}},\ and\ \bibinfo {author}
  {\bibfnamefont {C.}~\bibnamefont {Rigetti}},\ }\bibfield  {title} {\bibinfo
  {title} {Parametrically activated entangling gates using transmon qubits},\
  }\href {https://doi.org/10.1103/PhysRevApplied.10.034050} {\bibfield
  {journal} {\bibinfo  {journal} {Phys. Rev. Applied}\ }\textbf {\bibinfo
  {volume} {10}},\ \bibinfo {pages} {034050} (\bibinfo {year}
  {2018})}\BibitemShut {NoStop}%
\bibitem [{\citenamefont {Chen}\ \emph {et~al.}(2014)\citenamefont {Chen},
  \citenamefont {Neill}, \citenamefont {Roushan}, \citenamefont {Leung},
  \citenamefont {Fang}, \citenamefont {Barends}, \citenamefont {Kelly},
  \citenamefont {Campbell}, \citenamefont {Chen}, \citenamefont {Chiaro},
  \citenamefont {Dunsworth}, \citenamefont {Jeffrey}, \citenamefont {Megrant},
  \citenamefont {Mutus}, \citenamefont {O'Malley}, \citenamefont {Quintana},
  \citenamefont {Sank}, \citenamefont {Vainsencher}, \citenamefont {Wenner},
  \citenamefont {White}, \citenamefont {Geller}, \citenamefont {Cleland},\ and\
  \citenamefont {Martinis}}]{Chen2014m}%
  \BibitemOpen
  \bibfield  {author} {\bibinfo {author} {\bibfnamefont {Y.}~\bibnamefont
  {Chen}}, \bibinfo {author} {\bibfnamefont {C.}~\bibnamefont {Neill}},
  \bibinfo {author} {\bibfnamefont {P.}~\bibnamefont {Roushan}}, \bibinfo
  {author} {\bibfnamefont {N.}~\bibnamefont {Leung}}, \bibinfo {author}
  {\bibfnamefont {M.}~\bibnamefont {Fang}}, \bibinfo {author} {\bibfnamefont
  {R.}~\bibnamefont {Barends}}, \bibinfo {author} {\bibfnamefont
  {J.}~\bibnamefont {Kelly}}, \bibinfo {author} {\bibfnamefont
  {B.}~\bibnamefont {Campbell}}, \bibinfo {author} {\bibfnamefont
  {Z.}~\bibnamefont {Chen}}, \bibinfo {author} {\bibfnamefont {B.}~\bibnamefont
  {Chiaro}}, \bibinfo {author} {\bibfnamefont {A.}~\bibnamefont {Dunsworth}},
  \bibinfo {author} {\bibfnamefont {E.}~\bibnamefont {Jeffrey}}, \bibinfo
  {author} {\bibfnamefont {A.}~\bibnamefont {Megrant}}, \bibinfo {author}
  {\bibfnamefont {J.~Y.}\ \bibnamefont {Mutus}}, \bibinfo {author}
  {\bibfnamefont {P.~J.~J.}\ \bibnamefont {O'Malley}}, \bibinfo {author}
  {\bibfnamefont {C.~M.}\ \bibnamefont {Quintana}}, \bibinfo {author}
  {\bibfnamefont {D.}~\bibnamefont {Sank}}, \bibinfo {author} {\bibfnamefont
  {A.}~\bibnamefont {Vainsencher}}, \bibinfo {author} {\bibfnamefont
  {J.}~\bibnamefont {Wenner}}, \bibinfo {author} {\bibfnamefont {T.~C.}\
  \bibnamefont {White}}, \bibinfo {author} {\bibfnamefont {M.~R.}\ \bibnamefont
  {Geller}}, \bibinfo {author} {\bibfnamefont {A.~N.}\ \bibnamefont
  {Cleland}},\ and\ \bibinfo {author} {\bibfnamefont {J.~M.}\ \bibnamefont
  {Martinis}},\ }\bibfield  {title} {\bibinfo {title} {Qubit architecture with
  high coherence and fast tunable coupling},\ }\href
  {https://doi.org/10.1103/PhysRevLett.113.220502} {\bibfield  {journal}
  {\bibinfo  {journal} {Phys. Rev. Lett.}\ }\textbf {\bibinfo {volume} {113}},\
  \bibinfo {pages} {220502} (\bibinfo {year} {2014})}\BibitemShut {NoStop}%
\bibitem [{\citenamefont {Yan}\ \emph {et~al.}(2018)\citenamefont {Yan},
  \citenamefont {Krantz}, \citenamefont {Sung}, \citenamefont {Kjaergaard},
  \citenamefont {Campbell}, \citenamefont {Orlando}, \citenamefont
  {Gustavsson},\ and\ \citenamefont {Oliver}}]{Yan2018b}%
  \BibitemOpen
  \bibfield  {author} {\bibinfo {author} {\bibfnamefont {F.}~\bibnamefont
  {Yan}}, \bibinfo {author} {\bibfnamefont {P.}~\bibnamefont {Krantz}},
  \bibinfo {author} {\bibfnamefont {Y.}~\bibnamefont {Sung}}, \bibinfo {author}
  {\bibfnamefont {M.}~\bibnamefont {Kjaergaard}}, \bibinfo {author}
  {\bibfnamefont {D.~L.}\ \bibnamefont {Campbell}}, \bibinfo {author}
  {\bibfnamefont {T.~P.}\ \bibnamefont {Orlando}}, \bibinfo {author}
  {\bibfnamefont {S.}~\bibnamefont {Gustavsson}},\ and\ \bibinfo {author}
  {\bibfnamefont {W.~D.}\ \bibnamefont {Oliver}},\ }\bibfield  {title}
  {\bibinfo {title} {Tunable coupling scheme for implementing high-fidelity
  two-qubit gates},\ }\href {https://doi.org/10.1103/PhysRevApplied.10.054062}
  {\bibfield  {journal} {\bibinfo  {journal} {Phys. Rev. Applied}\ }\textbf
  {\bibinfo {volume} {10}},\ \bibinfo {pages} {054062} (\bibinfo {year}
  {2018})}\BibitemShut {NoStop}%
\bibitem [{\citenamefont {Sung}\ \emph {et~al.}(2021)\citenamefont {Sung},
  \citenamefont {Ding}, \citenamefont {Braum\"uller}, \citenamefont
  {Veps\"al\"ainen}, \citenamefont {Kannan}, \citenamefont {Kjaergaard},
  \citenamefont {Greene}, \citenamefont {Samach}, \citenamefont {McNally},
  \citenamefont {Kim}, \citenamefont {Melville}, \citenamefont {Niedzielski},
  \citenamefont {Schwartz}, \citenamefont {Yoder}, \citenamefont {Orlando},
  \citenamefont {Gustavsson},\ and\ \citenamefont {Oliver}}]{Sung2021a}%
  \BibitemOpen
  \bibfield  {author} {\bibinfo {author} {\bibfnamefont {Y.}~\bibnamefont
  {Sung}}, \bibinfo {author} {\bibfnamefont {L.}~\bibnamefont {Ding}}, \bibinfo
  {author} {\bibfnamefont {J.}~\bibnamefont {Braum\"uller}}, \bibinfo {author}
  {\bibfnamefont {A.}~\bibnamefont {Veps\"al\"ainen}}, \bibinfo {author}
  {\bibfnamefont {B.}~\bibnamefont {Kannan}}, \bibinfo {author} {\bibfnamefont
  {M.}~\bibnamefont {Kjaergaard}}, \bibinfo {author} {\bibfnamefont
  {A.}~\bibnamefont {Greene}}, \bibinfo {author} {\bibfnamefont {G.~O.}\
  \bibnamefont {Samach}}, \bibinfo {author} {\bibfnamefont {C.}~\bibnamefont
  {McNally}}, \bibinfo {author} {\bibfnamefont {D.}~\bibnamefont {Kim}},
  \bibinfo {author} {\bibfnamefont {A.}~\bibnamefont {Melville}}, \bibinfo
  {author} {\bibfnamefont {B.~M.}\ \bibnamefont {Niedzielski}}, \bibinfo
  {author} {\bibfnamefont {M.~E.}\ \bibnamefont {Schwartz}}, \bibinfo {author}
  {\bibfnamefont {J.~L.}\ \bibnamefont {Yoder}}, \bibinfo {author}
  {\bibfnamefont {T.~P.}\ \bibnamefont {Orlando}}, \bibinfo {author}
  {\bibfnamefont {S.}~\bibnamefont {Gustavsson}},\ and\ \bibinfo {author}
  {\bibfnamefont {W.~D.}\ \bibnamefont {Oliver}},\ }\bibfield  {title}
  {\bibinfo {title} {Realization of high-fidelity {CZ} and {$ZZ$}-free {iSWAP}
  gates with a tunable coupler},\ }\href
  {https://doi.org/10.1103/PhysRevX.11.021058} {\bibfield  {journal} {\bibinfo
  {journal} {Phys. Rev. X}\ }\textbf {\bibinfo {volume} {11}},\ \bibinfo
  {pages} {021058} (\bibinfo {year} {2021})}\BibitemShut {NoStop}%
\bibitem [{\citenamefont {Xu}\ \emph {et~al.}(2020)\citenamefont {Xu},
  \citenamefont {Chu}, \citenamefont {Yuan}, \citenamefont {Qiu}, \citenamefont
  {Zhou}, \citenamefont {Zhang}, \citenamefont {Tan}, \citenamefont {Yu},
  \citenamefont {Liu}, \citenamefont {Li}, \citenamefont {Yan},\ and\
  \citenamefont {Yu}}]{Xu2020c}%
  \BibitemOpen
  \bibfield  {author} {\bibinfo {author} {\bibfnamefont {Y.}~\bibnamefont
  {Xu}}, \bibinfo {author} {\bibfnamefont {J.}~\bibnamefont {Chu}}, \bibinfo
  {author} {\bibfnamefont {J.}~\bibnamefont {Yuan}}, \bibinfo {author}
  {\bibfnamefont {J.}~\bibnamefont {Qiu}}, \bibinfo {author} {\bibfnamefont
  {Y.}~\bibnamefont {Zhou}}, \bibinfo {author} {\bibfnamefont {L.}~\bibnamefont
  {Zhang}}, \bibinfo {author} {\bibfnamefont {X.}~\bibnamefont {Tan}}, \bibinfo
  {author} {\bibfnamefont {Y.}~\bibnamefont {Yu}}, \bibinfo {author}
  {\bibfnamefont {S.}~\bibnamefont {Liu}}, \bibinfo {author} {\bibfnamefont
  {J.}~\bibnamefont {Li}}, \bibinfo {author} {\bibfnamefont {F.}~\bibnamefont
  {Yan}},\ and\ \bibinfo {author} {\bibfnamefont {D.}~\bibnamefont {Yu}},\
  }\bibfield  {title} {\bibinfo {title} {High-fidelity, high-scalability
  two-qubit gate scheme for superconducting qubits},\ }\href
  {https://doi.org/10.1103/PhysRevLett.125.240503} {\bibfield  {journal}
  {\bibinfo  {journal} {Phys. Rev. Lett.}\ }\textbf {\bibinfo {volume} {125}},\
  \bibinfo {pages} {240503} (\bibinfo {year} {2020})}\BibitemShut {NoStop}%
\bibitem [{\citenamefont {Collodo}\ \emph {et~al.}(2020)\citenamefont
  {Collodo}, \citenamefont {Herrmann}, \citenamefont {Lacroix}, \citenamefont
  {Andersen}, \citenamefont {Remm}, \citenamefont {Lazar}, \citenamefont
  {Besse}, \citenamefont {Walter}, \citenamefont {Wallraff},\ and\
  \citenamefont {Eichler}}]{Collodo2020}%
  \BibitemOpen
  \bibfield  {author} {\bibinfo {author} {\bibfnamefont {M.~C.}\ \bibnamefont
  {Collodo}}, \bibinfo {author} {\bibfnamefont {J.}~\bibnamefont {Herrmann}},
  \bibinfo {author} {\bibfnamefont {N.}~\bibnamefont {Lacroix}}, \bibinfo
  {author} {\bibfnamefont {C.~K.}\ \bibnamefont {Andersen}}, \bibinfo {author}
  {\bibfnamefont {A.}~\bibnamefont {Remm}}, \bibinfo {author} {\bibfnamefont
  {S.}~\bibnamefont {Lazar}}, \bibinfo {author} {\bibfnamefont {J.-C.}\
  \bibnamefont {Besse}}, \bibinfo {author} {\bibfnamefont {T.}~\bibnamefont
  {Walter}}, \bibinfo {author} {\bibfnamefont {A.}~\bibnamefont {Wallraff}},\
  and\ \bibinfo {author} {\bibfnamefont {C.}~\bibnamefont {Eichler}},\
  }\bibfield  {title} {\bibinfo {title} {Implementation of conditional phase
  gates based on tunable {$ZZ$} interactions},\ }\href
  {https://doi.org/10.1103/PhysRevLett.125.240502} {\bibfield  {journal}
  {\bibinfo  {journal} {Phys. Rev. Lett.}\ }\textbf {\bibinfo {volume} {125}},\
  \bibinfo {pages} {240502} (\bibinfo {year} {2020})}\BibitemShut {NoStop}%
\bibitem [{\citenamefont {Foxen}\ \emph {et~al.}(2020)\citenamefont {Foxen},
  \citenamefont {Neill}, \citenamefont {Dunsworth}, \citenamefont {Roushan},
  \citenamefont {Chiaro}, \citenamefont {Megrant}, \citenamefont {Kelly},
  \citenamefont {Chen}, \citenamefont {Satzinger}, \citenamefont {Barends},
  \citenamefont {Arute}, \citenamefont {Arya}, \citenamefont {Babbush},
  \citenamefont {Bacon}, \citenamefont {Bardin}, \citenamefont {Boixo},
  \citenamefont {Buell}, \citenamefont {Burkett}, \citenamefont {Chen},
  \citenamefont {Collins}, \citenamefont {Farhi}, \citenamefont {Fowler},
  \citenamefont {Gidney}, \citenamefont {Giustina}, \citenamefont {Graff},
  \citenamefont {Harrigan}, \citenamefont {Huang}, \citenamefont {Isakov},
  \citenamefont {Jeffrey}, \citenamefont {Jiang}, \citenamefont {Kafri},
  \citenamefont {Kechedzhi}, \citenamefont {Klimov}, \citenamefont {Korotkov},
  \citenamefont {Kostritsa}, \citenamefont {Landhuis}, \citenamefont {Lucero},
  \citenamefont {McClean}, \citenamefont {McEwen}, \citenamefont {Mi},
  \citenamefont {Mohseni}, \citenamefont {Mutus}, \citenamefont {Naaman},
  \citenamefont {Neeley}, \citenamefont {Niu}, \citenamefont {Petukhov},
  \citenamefont {Quintana}, \citenamefont {Rubin}, \citenamefont {Sank},
  \citenamefont {Smelyanskiy}, \citenamefont {Vainsencher}, \citenamefont
  {White}, \citenamefont {Yao}, \citenamefont {Yeh}, \citenamefont {Zalcman},
  \citenamefont {Neven},\ and\ \citenamefont {Martinis}}]{Foxen2020}%
  \BibitemOpen
  \bibfield  {author} {\bibinfo {author} {\bibfnamefont {B.}~\bibnamefont
  {Foxen}}, \bibinfo {author} {\bibfnamefont {C.}~\bibnamefont {Neill}},
  \bibinfo {author} {\bibfnamefont {A.}~\bibnamefont {Dunsworth}}, \bibinfo
  {author} {\bibfnamefont {P.}~\bibnamefont {Roushan}}, \bibinfo {author}
  {\bibfnamefont {B.}~\bibnamefont {Chiaro}}, \bibinfo {author} {\bibfnamefont
  {A.}~\bibnamefont {Megrant}}, \bibinfo {author} {\bibfnamefont
  {J.}~\bibnamefont {Kelly}}, \bibinfo {author} {\bibfnamefont
  {Z.}~\bibnamefont {Chen}}, \bibinfo {author} {\bibfnamefont {K.}~\bibnamefont
  {Satzinger}}, \bibinfo {author} {\bibfnamefont {R.}~\bibnamefont {Barends}},
  \bibinfo {author} {\bibfnamefont {F.}~\bibnamefont {Arute}}, \bibinfo
  {author} {\bibfnamefont {K.}~\bibnamefont {Arya}}, \bibinfo {author}
  {\bibfnamefont {R.}~\bibnamefont {Babbush}}, \bibinfo {author} {\bibfnamefont
  {D.}~\bibnamefont {Bacon}}, \bibinfo {author} {\bibfnamefont {J.~C.}\
  \bibnamefont {Bardin}}, \bibinfo {author} {\bibfnamefont {S.}~\bibnamefont
  {Boixo}}, \bibinfo {author} {\bibfnamefont {D.}~\bibnamefont {Buell}},
  \bibinfo {author} {\bibfnamefont {B.}~\bibnamefont {Burkett}}, \bibinfo
  {author} {\bibfnamefont {Y.}~\bibnamefont {Chen}}, \bibinfo {author}
  {\bibfnamefont {R.}~\bibnamefont {Collins}}, \bibinfo {author} {\bibfnamefont
  {E.}~\bibnamefont {Farhi}}, \bibinfo {author} {\bibfnamefont
  {A.}~\bibnamefont {Fowler}}, \bibinfo {author} {\bibfnamefont
  {C.}~\bibnamefont {Gidney}}, \bibinfo {author} {\bibfnamefont
  {M.}~\bibnamefont {Giustina}}, \bibinfo {author} {\bibfnamefont
  {R.}~\bibnamefont {Graff}}, \bibinfo {author} {\bibfnamefont
  {M.}~\bibnamefont {Harrigan}}, \bibinfo {author} {\bibfnamefont
  {T.}~\bibnamefont {Huang}}, \bibinfo {author} {\bibfnamefont {S.~V.}\
  \bibnamefont {Isakov}}, \bibinfo {author} {\bibfnamefont {E.}~\bibnamefont
  {Jeffrey}}, \bibinfo {author} {\bibfnamefont {Z.}~\bibnamefont {Jiang}},
  \bibinfo {author} {\bibfnamefont {D.}~\bibnamefont {Kafri}}, \bibinfo
  {author} {\bibfnamefont {K.}~\bibnamefont {Kechedzhi}}, \bibinfo {author}
  {\bibfnamefont {P.}~\bibnamefont {Klimov}}, \bibinfo {author} {\bibfnamefont
  {A.}~\bibnamefont {Korotkov}}, \bibinfo {author} {\bibfnamefont
  {F.}~\bibnamefont {Kostritsa}}, \bibinfo {author} {\bibfnamefont
  {D.}~\bibnamefont {Landhuis}}, \bibinfo {author} {\bibfnamefont
  {E.}~\bibnamefont {Lucero}}, \bibinfo {author} {\bibfnamefont
  {J.}~\bibnamefont {McClean}}, \bibinfo {author} {\bibfnamefont
  {M.}~\bibnamefont {McEwen}}, \bibinfo {author} {\bibfnamefont
  {X.}~\bibnamefont {Mi}}, \bibinfo {author} {\bibfnamefont {M.}~\bibnamefont
  {Mohseni}}, \bibinfo {author} {\bibfnamefont {J.~Y.}\ \bibnamefont {Mutus}},
  \bibinfo {author} {\bibfnamefont {O.}~\bibnamefont {Naaman}}, \bibinfo
  {author} {\bibfnamefont {M.}~\bibnamefont {Neeley}}, \bibinfo {author}
  {\bibfnamefont {M.}~\bibnamefont {Niu}}, \bibinfo {author} {\bibfnamefont
  {A.}~\bibnamefont {Petukhov}}, \bibinfo {author} {\bibfnamefont
  {C.}~\bibnamefont {Quintana}}, \bibinfo {author} {\bibfnamefont
  {N.}~\bibnamefont {Rubin}}, \bibinfo {author} {\bibfnamefont
  {D.}~\bibnamefont {Sank}}, \bibinfo {author} {\bibfnamefont {V.}~\bibnamefont
  {Smelyanskiy}}, \bibinfo {author} {\bibfnamefont {A.}~\bibnamefont
  {Vainsencher}}, \bibinfo {author} {\bibfnamefont {T.~C.}\ \bibnamefont
  {White}}, \bibinfo {author} {\bibfnamefont {Z.}~\bibnamefont {Yao}}, \bibinfo
  {author} {\bibfnamefont {P.}~\bibnamefont {Yeh}}, \bibinfo {author}
  {\bibfnamefont {A.}~\bibnamefont {Zalcman}}, \bibinfo {author} {\bibfnamefont
  {H.}~\bibnamefont {Neven}},\ and\ \bibinfo {author} {\bibfnamefont {J.~M.}\
  \bibnamefont {Martinis}},\ }\bibfield  {title} {\bibinfo {title}
  {Demonstrating a continuous set of two-qubit gates for near-term quantum
  algorithms},\ }\href {https://doi.org/10.1103/physrevlett.125.120504}
  {\bibfield  {journal} {\bibinfo  {journal} {Phys. Rev. Lett.}\ }\textbf
  {\bibinfo {volume} {125}},\ \bibinfo {pages} {120504} (\bibinfo {year}
  {2020})}\BibitemShut {NoStop}%
\bibitem [{\citenamefont {Marques}\ \emph {et~al.}(2022)\citenamefont
  {Marques}, \citenamefont {Varbanov}, \citenamefont {Moreira}, \citenamefont
  {Ali}, \citenamefont {Muthusubramanian}, \citenamefont {Zachariadis},
  \citenamefont {Battistel}, \citenamefont {Beekman}, \citenamefont {Haider},
  \citenamefont {Vlothuizen}, \citenamefont {Bruno}, \citenamefont {Terhal},\
  and\ \citenamefont {DiCarlo}}]{Marques2021}%
  \BibitemOpen
  \bibfield  {author} {\bibinfo {author} {\bibfnamefont {J.~F.}\ \bibnamefont
  {Marques}}, \bibinfo {author} {\bibfnamefont {B.~M.}\ \bibnamefont
  {Varbanov}}, \bibinfo {author} {\bibfnamefont {M.~S.}\ \bibnamefont
  {Moreira}}, \bibinfo {author} {\bibfnamefont {H.}~\bibnamefont {Ali}},
  \bibinfo {author} {\bibfnamefont {N.}~\bibnamefont {Muthusubramanian}},
  \bibinfo {author} {\bibfnamefont {C.}~\bibnamefont {Zachariadis}}, \bibinfo
  {author} {\bibfnamefont {F.}~\bibnamefont {Battistel}}, \bibinfo {author}
  {\bibfnamefont {M.}~\bibnamefont {Beekman}}, \bibinfo {author} {\bibfnamefont
  {N.}~\bibnamefont {Haider}}, \bibinfo {author} {\bibfnamefont
  {W.}~\bibnamefont {Vlothuizen}}, \bibinfo {author} {\bibfnamefont
  {A.}~\bibnamefont {Bruno}}, \bibinfo {author} {\bibfnamefont {B.~M.}\
  \bibnamefont {Terhal}},\ and\ \bibinfo {author} {\bibfnamefont
  {L.}~\bibnamefont {DiCarlo}},\ }\bibfield  {title} {\bibinfo {title}
  {Logical-qubit operations in an error-detecting surface code},\ }\href
  {https://doi.org/10.1038/s41567-021-01423-9} {\bibfield  {journal} {\bibinfo
  {journal} {Nature Physics}\ }\textbf {\bibinfo {volume} {18}},\ \bibinfo
  {pages} {80} (\bibinfo {year} {2022})}\BibitemShut {NoStop}%
\bibitem [{\citenamefont {{Google Quantum AI}}(2023)}]{Acharya2023}%
  \BibitemOpen
  \bibfield  {author} {\bibinfo {author} {\bibnamefont {{Google Quantum AI}}},\
  }\bibfield  {title} {\bibinfo {title} {Suppressing quantum errors by scaling
  a surface code logical qubit},\ }\href
  {https://doi.org/10.1038/s41586-022-05434-1} {\bibfield  {journal} {\bibinfo
  {journal} {Nature}\ }\textbf {\bibinfo {volume} {614}},\ \bibinfo {pages}
  {676} (\bibinfo {year} {2023})}\BibitemShut {NoStop}%
\bibitem [{\citenamefont {Sundaresan}\ \emph {et~al.}(2023)\citenamefont
  {Sundaresan}, \citenamefont {Yoder}, \citenamefont {Kim}, \citenamefont {Li},
  \citenamefont {Chen}, \citenamefont {Harper}, \citenamefont {Thorbeck},
  \citenamefont {Cross}, \citenamefont {C{\'{o}}rcoles},\ and\ \citenamefont
  {Takita}}]{Sundaresan2023}%
  \BibitemOpen
  \bibfield  {author} {\bibinfo {author} {\bibfnamefont {N.}~\bibnamefont
  {Sundaresan}}, \bibinfo {author} {\bibfnamefont {T.~J.}\ \bibnamefont
  {Yoder}}, \bibinfo {author} {\bibfnamefont {Y.}~\bibnamefont {Kim}}, \bibinfo
  {author} {\bibfnamefont {M.}~\bibnamefont {Li}}, \bibinfo {author}
  {\bibfnamefont {E.~H.}\ \bibnamefont {Chen}}, \bibinfo {author}
  {\bibfnamefont {G.}~\bibnamefont {Harper}}, \bibinfo {author} {\bibfnamefont
  {T.}~\bibnamefont {Thorbeck}}, \bibinfo {author} {\bibfnamefont {A.~W.}\
  \bibnamefont {Cross}}, \bibinfo {author} {\bibfnamefont {A.~D.}\ \bibnamefont
  {C{\'{o}}rcoles}},\ and\ \bibinfo {author} {\bibfnamefont {M.}~\bibnamefont
  {Takita}},\ }\bibfield  {title} {\bibinfo {title} {Demonstrating multi-round
  subsystem quantum error correction using matching and maximum likelihood
  decoders},\ }\href {https://doi.org/10.1038/s41467-023-38247-5} {\bibfield
  {journal} {\bibinfo  {journal} {Nature Communications}\ }\textbf {\bibinfo
  {volume} {14}},\ \bibinfo {pages} {2852} (\bibinfo {year}
  {2023})}\BibitemShut {NoStop}%
\bibitem [{\citenamefont {Acharya}\ \emph {et~al.}(2024)\citenamefont
  {Acharya}, \citenamefont {Aghababaie-Beni}, \citenamefont {Aleiner},
  \citenamefont {Andersen}, \citenamefont {Ansmann}, \citenamefont {Arute},
  \citenamefont {Arya}, \citenamefont {Asfaw}, \citenamefont {Astrakhantsev},
  \citenamefont {Atalaya}, \citenamefont {Babbush}, \citenamefont {Bacon},
  \citenamefont {Ballard}, \citenamefont {Bardin}, \citenamefont {Bausch},
  \citenamefont {Bengtsson}, \citenamefont {Bilmes}, \citenamefont {Blackwell},
  \citenamefont {Boixo}, \citenamefont {Bortoli}, \citenamefont {Bourassa},
  \citenamefont {Bovaird}, \citenamefont {Brill}, \citenamefont {Broughton},
  \citenamefont {Browne}, \citenamefont {Buchea}, \citenamefont {Buckley},
  \citenamefont {Buell}, \citenamefont {Burger}, \citenamefont {Burkett},
  \citenamefont {Bushnell}, \citenamefont {Cabrera}, \citenamefont {Campero},
  \citenamefont {Chang}, \citenamefont {Chen}, \citenamefont {Chen},
  \citenamefont {Chiaro}, \citenamefont {Chik}, \citenamefont {Chou},
  \citenamefont {Claes}, \citenamefont {Cleland}, \citenamefont {Cogan},
  \citenamefont {Collins}, \citenamefont {Conner}, \citenamefont {Courtney},
  \citenamefont {Crook}, \citenamefont {Curtin}, \citenamefont {Das},
  \citenamefont {Davies}, \citenamefont {De~Lorenzo}, \citenamefont {Debroy},
  \citenamefont {Demura}, \citenamefont {Devoret}, \citenamefont {Di~Paolo},
  \citenamefont {Donohoe}, \citenamefont {Drozdov}, \citenamefont {Dunsworth},
  \citenamefont {Earle}, \citenamefont {Edlich}, \citenamefont {Eickbusch},
  \citenamefont {Elbag}, \citenamefont {Elzouka}, \citenamefont {Erickson},
  \citenamefont {Faoro}, \citenamefont {Farhi}, \citenamefont {Ferreira},
  \citenamefont {Burgos}, \citenamefont {Forati}, \citenamefont {Fowler},
  \citenamefont {Foxen}, \citenamefont {Ganjam}, \citenamefont {Garcia},
  \citenamefont {Gasca}, \citenamefont {Genois}, \citenamefont {Giang},
  \citenamefont {Gidney}, \citenamefont {Gilboa}, \citenamefont {Gosula},
  \citenamefont {Dau}, \citenamefont {Graumann}, \citenamefont {Greene},
  \citenamefont {Gross}, \citenamefont {Habegger}, \citenamefont {Hall},
  \citenamefont {Hamilton}, \citenamefont {Hansen}, \citenamefont {Harrigan},
  \citenamefont {Harrington}, \citenamefont {Heras}, \citenamefont {Heslin},
  \citenamefont {Heu}, \citenamefont {Higgott}, \citenamefont {Hill},
  \citenamefont {Hilton}, \citenamefont {Holland}, \citenamefont {Hong},
  \citenamefont {Huang}, \citenamefont {Huff}, \citenamefont {Huggins},
  \citenamefont {Ioffe}, \citenamefont {Isakov}, \citenamefont {Iveland},
  \citenamefont {Jeffrey}, \citenamefont {Jiang}, \citenamefont {Jones},
  \citenamefont {Jordan}, \citenamefont {Joshi}, \citenamefont {Juhas},
  \citenamefont {Kafri}, \citenamefont {Kang}, \citenamefont {Karamlou},
  \citenamefont {Kechedzhi}, \citenamefont {Kelly}, \citenamefont {Khaire},
  \citenamefont {Khattar}, \citenamefont {Khezri}, \citenamefont {Kim},
  \citenamefont {Klimov}, \citenamefont {Klots}, \citenamefont {Kobrin},
  \citenamefont {Kohli}, \citenamefont {Korotkov}, \citenamefont {Kostritsa},
  \citenamefont {Kothari}, \citenamefont {Kozlovskii}, \citenamefont
  {Kreikebaum}, \citenamefont {Kurilovich}, \citenamefont {Lacroix},
  \citenamefont {Landhuis}, \citenamefont {Lange-Dei}, \citenamefont {Langley},
  \citenamefont {Laptev}, \citenamefont {Lau}, \citenamefont {Guevel},
  \citenamefont {Ledford}, \citenamefont {Lee}, \citenamefont {Lensky},
  \citenamefont {Leon}, \citenamefont {Lester}, \citenamefont {Li},
  \citenamefont {Li}, \citenamefont {Lill}, \citenamefont {Liu}, \citenamefont
  {Livingston}, \citenamefont {Locharla}, \citenamefont {Lucero}, \citenamefont
  {Lundahl}, \citenamefont {Lunt}, \citenamefont {Madhuk}, \citenamefont
  {Malone}, \citenamefont {Maloney}, \citenamefont {Mandr\'a}, \citenamefont
  {Martin}, \citenamefont {Martin}, \citenamefont {Martin}, \citenamefont
  {Maxfield}, \citenamefont {McClean}, \citenamefont {McEwen}, \citenamefont
  {Meeks}, \citenamefont {Megrant}, \citenamefont {Mi}, \citenamefont {Miao},
  \citenamefont {Mieszala}, \citenamefont {Molavi}, \citenamefont {Molina},
  \citenamefont {Montazeri}, \citenamefont {Morvan}, \citenamefont {Movassagh},
  \citenamefont {Mruczkiewicz}, \citenamefont {Naaman}, \citenamefont {Neeley},
  \citenamefont {Neill}, \citenamefont {Nersisyan}, \citenamefont {Neven},
  \citenamefont {Newman}, \citenamefont {Ng}, \citenamefont {Nguyen},
  \citenamefont {Nguyen}, \citenamefont {Ni}, \citenamefont {O'Brien},
  \citenamefont {Oliver}, \citenamefont {Opremcak}, \citenamefont {Ottosson},
  \citenamefont {Petukhov}, \citenamefont {Pizzuto}, \citenamefont {Platt},
  \citenamefont {Potter}, \citenamefont {Pritchard}, \citenamefont {Pryadko},
  \citenamefont {Quintana}, \citenamefont {Ramachandran}, \citenamefont
  {Reagor}, \citenamefont {Rhodes}, \citenamefont {Roberts}, \citenamefont
  {Rosenberg}, \citenamefont {Rosenfeld}, \citenamefont {Roushan},
  \citenamefont {Rubin}, \citenamefont {Saei}, \citenamefont {Sank},
  \citenamefont {Sankaragomathi}, \citenamefont {Satzinger}, \citenamefont
  {Schurkus}, \citenamefont {Schuster}, \citenamefont {Senior}, \citenamefont
  {Shearn}, \citenamefont {Shorter}, \citenamefont {Shutty}, \citenamefont
  {Shvarts}, \citenamefont {Singh}, \citenamefont {Sivak}, \citenamefont
  {Skruzny}, \citenamefont {Small}, \citenamefont {Smelyanskiy}, \citenamefont
  {Smith}, \citenamefont {Somma}, \citenamefont {Springer}, \citenamefont
  {Sterling}, \citenamefont {Strain}, \citenamefont {Suchard}, \citenamefont
  {Szasz}, \citenamefont {Sztein}, \citenamefont {Thor}, \citenamefont
  {Torres}, \citenamefont {Torunbalci}, \citenamefont {Vaishnav}, \citenamefont
  {Vargas}, \citenamefont {Vdovichev}, \citenamefont {Vidal}, \citenamefont
  {Villalonga}, \citenamefont {Heidweiller}, \citenamefont {Waltman},
  \citenamefont {Wang}, \citenamefont {Ware}, \citenamefont {Weber},
  \citenamefont {White}, \citenamefont {Wong}, \citenamefont {Woo},
  \citenamefont {Xing}, \citenamefont {Yao}, \citenamefont {Yeh}, \citenamefont
  {Ying}, \citenamefont {Yoo}, \citenamefont {Yosri}, \citenamefont {Young},
  \citenamefont {Zalcman}, \citenamefont {Zhang}, \citenamefont {Zhu},\ and\
  \citenamefont {Zobrist}}]{Acharya2024a}%
  \BibitemOpen
  \bibfield  {author} {\bibinfo {author} {\bibfnamefont {R.}~\bibnamefont
  {Acharya}}, \bibinfo {author} {\bibfnamefont {L.}~\bibnamefont
  {Aghababaie-Beni}}, \bibinfo {author} {\bibfnamefont {I.}~\bibnamefont
  {Aleiner}}, \bibinfo {author} {\bibfnamefont {T.~I.}\ \bibnamefont
  {Andersen}}, \bibinfo {author} {\bibfnamefont {M.}~\bibnamefont {Ansmann}},
  \bibinfo {author} {\bibfnamefont {F.}~\bibnamefont {Arute}}, \bibinfo
  {author} {\bibfnamefont {K.}~\bibnamefont {Arya}}, \bibinfo {author}
  {\bibfnamefont {A.}~\bibnamefont {Asfaw}}, \bibinfo {author} {\bibfnamefont
  {N.}~\bibnamefont {Astrakhantsev}}, \bibinfo {author} {\bibfnamefont
  {J.}~\bibnamefont {Atalaya}}, \bibinfo {author} {\bibfnamefont
  {R.}~\bibnamefont {Babbush}}, \bibinfo {author} {\bibfnamefont
  {D.}~\bibnamefont {Bacon}}, \bibinfo {author} {\bibfnamefont
  {B.}~\bibnamefont {Ballard}}, \bibinfo {author} {\bibfnamefont {J.~C.}\
  \bibnamefont {Bardin}}, \bibinfo {author} {\bibfnamefont {J.}~\bibnamefont
  {Bausch}}, \bibinfo {author} {\bibfnamefont {A.}~\bibnamefont {Bengtsson}},
  \bibinfo {author} {\bibfnamefont {A.}~\bibnamefont {Bilmes}}, \bibinfo
  {author} {\bibfnamefont {S.}~\bibnamefont {Blackwell}}, \bibinfo {author}
  {\bibfnamefont {S.}~\bibnamefont {Boixo}}, \bibinfo {author} {\bibfnamefont
  {G.}~\bibnamefont {Bortoli}}, \bibinfo {author} {\bibfnamefont
  {A.}~\bibnamefont {Bourassa}}, \bibinfo {author} {\bibfnamefont
  {J.}~\bibnamefont {Bovaird}}, \bibinfo {author} {\bibfnamefont
  {L.}~\bibnamefont {Brill}}, \bibinfo {author} {\bibfnamefont
  {M.}~\bibnamefont {Broughton}}, \bibinfo {author} {\bibfnamefont {D.~A.}\
  \bibnamefont {Browne}}, \bibinfo {author} {\bibfnamefont {B.}~\bibnamefont
  {Buchea}}, \bibinfo {author} {\bibfnamefont {B.~B.}\ \bibnamefont {Buckley}},
  \bibinfo {author} {\bibfnamefont {D.~A.}\ \bibnamefont {Buell}}, \bibinfo
  {author} {\bibfnamefont {T.}~\bibnamefont {Burger}}, \bibinfo {author}
  {\bibfnamefont {B.}~\bibnamefont {Burkett}}, \bibinfo {author} {\bibfnamefont
  {N.}~\bibnamefont {Bushnell}}, \bibinfo {author} {\bibfnamefont
  {A.}~\bibnamefont {Cabrera}}, \bibinfo {author} {\bibfnamefont
  {J.}~\bibnamefont {Campero}}, \bibinfo {author} {\bibfnamefont
  {H.}~\bibnamefont {Chang}}, \bibinfo {author} {\bibfnamefont
  {Y.}~\bibnamefont {Chen}}, \bibinfo {author} {\bibfnamefont {Z.}~\bibnamefont
  {Chen}}, \bibinfo {author} {\bibfnamefont {B.}~\bibnamefont {Chiaro}},
  \bibinfo {author} {\bibfnamefont {D.}~\bibnamefont {Chik}}, \bibinfo {author}
  {\bibfnamefont {C.}~\bibnamefont {Chou}}, \bibinfo {author} {\bibfnamefont
  {J.}~\bibnamefont {Claes}}, \bibinfo {author} {\bibfnamefont {A.~Y.}\
  \bibnamefont {Cleland}}, \bibinfo {author} {\bibfnamefont {J.}~\bibnamefont
  {Cogan}}, \bibinfo {author} {\bibfnamefont {R.}~\bibnamefont {Collins}},
  \bibinfo {author} {\bibfnamefont {P.}~\bibnamefont {Conner}}, \bibinfo
  {author} {\bibfnamefont {W.}~\bibnamefont {Courtney}}, \bibinfo {author}
  {\bibfnamefont {A.~L.}\ \bibnamefont {Crook}}, \bibinfo {author}
  {\bibfnamefont {B.}~\bibnamefont {Curtin}}, \bibinfo {author} {\bibfnamefont
  {S.}~\bibnamefont {Das}}, \bibinfo {author} {\bibfnamefont {A.}~\bibnamefont
  {Davies}}, \bibinfo {author} {\bibfnamefont {L.}~\bibnamefont {De~Lorenzo}},
  \bibinfo {author} {\bibfnamefont {D.~M.}\ \bibnamefont {Debroy}}, \bibinfo
  {author} {\bibfnamefont {S.}~\bibnamefont {Demura}}, \bibinfo {author}
  {\bibfnamefont {M.}~\bibnamefont {Devoret}}, \bibinfo {author} {\bibfnamefont
  {A.}~\bibnamefont {Di~Paolo}}, \bibinfo {author} {\bibfnamefont
  {P.}~\bibnamefont {Donohoe}}, \bibinfo {author} {\bibfnamefont
  {I.}~\bibnamefont {Drozdov}}, \bibinfo {author} {\bibfnamefont
  {A.}~\bibnamefont {Dunsworth}}, \bibinfo {author} {\bibfnamefont
  {C.}~\bibnamefont {Earle}}, \bibinfo {author} {\bibfnamefont
  {T.}~\bibnamefont {Edlich}}, \bibinfo {author} {\bibfnamefont
  {A.}~\bibnamefont {Eickbusch}}, \bibinfo {author} {\bibfnamefont {A.~M.}\
  \bibnamefont {Elbag}}, \bibinfo {author} {\bibfnamefont {M.}~\bibnamefont
  {Elzouka}}, \bibinfo {author} {\bibfnamefont {C.}~\bibnamefont {Erickson}},
  \bibinfo {author} {\bibfnamefont {L.}~\bibnamefont {Faoro}}, \bibinfo
  {author} {\bibfnamefont {E.}~\bibnamefont {Farhi}}, \bibinfo {author}
  {\bibfnamefont {V.~S.}\ \bibnamefont {Ferreira}}, \bibinfo {author}
  {\bibfnamefont {L.~F.}\ \bibnamefont {Burgos}}, \bibinfo {author}
  {\bibfnamefont {E.}~\bibnamefont {Forati}}, \bibinfo {author} {\bibfnamefont
  {A.~G.}\ \bibnamefont {Fowler}}, \bibinfo {author} {\bibfnamefont
  {B.}~\bibnamefont {Foxen}}, \bibinfo {author} {\bibfnamefont
  {S.}~\bibnamefont {Ganjam}}, \bibinfo {author} {\bibfnamefont
  {G.}~\bibnamefont {Garcia}}, \bibinfo {author} {\bibfnamefont
  {R.}~\bibnamefont {Gasca}}, \bibinfo {author} {\bibfnamefont
  {E.}~\bibnamefont {Genois}}, \bibinfo {author} {\bibfnamefont
  {W.}~\bibnamefont {Giang}}, \bibinfo {author} {\bibfnamefont
  {C.}~\bibnamefont {Gidney}}, \bibinfo {author} {\bibfnamefont
  {D.}~\bibnamefont {Gilboa}}, \bibinfo {author} {\bibfnamefont
  {R.}~\bibnamefont {Gosula}}, \bibinfo {author} {\bibfnamefont {A.~G.}\
  \bibnamefont {Dau}}, \bibinfo {author} {\bibfnamefont {D.}~\bibnamefont
  {Graumann}}, \bibinfo {author} {\bibfnamefont {A.}~\bibnamefont {Greene}},
  \bibinfo {author} {\bibfnamefont {J.~A.}\ \bibnamefont {Gross}}, \bibinfo
  {author} {\bibfnamefont {S.}~\bibnamefont {Habegger}}, \bibinfo {author}
  {\bibfnamefont {J.}~\bibnamefont {Hall}}, \bibinfo {author} {\bibfnamefont
  {M.~C.}\ \bibnamefont {Hamilton}}, \bibinfo {author} {\bibfnamefont
  {M.}~\bibnamefont {Hansen}}, \bibinfo {author} {\bibfnamefont {M.~P.}\
  \bibnamefont {Harrigan}}, \bibinfo {author} {\bibfnamefont {S.~D.}\
  \bibnamefont {Harrington}}, \bibinfo {author} {\bibfnamefont {F.~J.~H.}\
  \bibnamefont {Heras}}, \bibinfo {author} {\bibfnamefont {S.}~\bibnamefont
  {Heslin}}, \bibinfo {author} {\bibfnamefont {P.}~\bibnamefont {Heu}},
  \bibinfo {author} {\bibfnamefont {O.}~\bibnamefont {Higgott}}, \bibinfo
  {author} {\bibfnamefont {G.}~\bibnamefont {Hill}}, \bibinfo {author}
  {\bibfnamefont {J.}~\bibnamefont {Hilton}}, \bibinfo {author} {\bibfnamefont
  {G.}~\bibnamefont {Holland}}, \bibinfo {author} {\bibfnamefont
  {S.}~\bibnamefont {Hong}}, \bibinfo {author} {\bibfnamefont {H.}~\bibnamefont
  {Huang}}, \bibinfo {author} {\bibfnamefont {A.}~\bibnamefont {Huff}},
  \bibinfo {author} {\bibfnamefont {W.~J.}\ \bibnamefont {Huggins}}, \bibinfo
  {author} {\bibfnamefont {L.~B.}\ \bibnamefont {Ioffe}}, \bibinfo {author}
  {\bibfnamefont {S.~V.}\ \bibnamefont {Isakov}}, \bibinfo {author}
  {\bibfnamefont {J.}~\bibnamefont {Iveland}}, \bibinfo {author} {\bibfnamefont
  {E.}~\bibnamefont {Jeffrey}}, \bibinfo {author} {\bibfnamefont
  {Z.}~\bibnamefont {Jiang}}, \bibinfo {author} {\bibfnamefont
  {C.}~\bibnamefont {Jones}}, \bibinfo {author} {\bibfnamefont
  {S.}~\bibnamefont {Jordan}}, \bibinfo {author} {\bibfnamefont
  {C.}~\bibnamefont {Joshi}}, \bibinfo {author} {\bibfnamefont
  {P.}~\bibnamefont {Juhas}}, \bibinfo {author} {\bibfnamefont
  {D.}~\bibnamefont {Kafri}}, \bibinfo {author} {\bibfnamefont
  {H.}~\bibnamefont {Kang}}, \bibinfo {author} {\bibfnamefont {A.~H.}\
  \bibnamefont {Karamlou}}, \bibinfo {author} {\bibfnamefont {K.}~\bibnamefont
  {Kechedzhi}}, \bibinfo {author} {\bibfnamefont {J.}~\bibnamefont {Kelly}},
  \bibinfo {author} {\bibfnamefont {T.}~\bibnamefont {Khaire}}, \bibinfo
  {author} {\bibfnamefont {T.}~\bibnamefont {Khattar}}, \bibinfo {author}
  {\bibfnamefont {M.}~\bibnamefont {Khezri}}, \bibinfo {author} {\bibfnamefont
  {S.}~\bibnamefont {Kim}}, \bibinfo {author} {\bibfnamefont {P.~V.}\
  \bibnamefont {Klimov}}, \bibinfo {author} {\bibfnamefont {A.~R.}\
  \bibnamefont {Klots}}, \bibinfo {author} {\bibfnamefont {B.}~\bibnamefont
  {Kobrin}}, \bibinfo {author} {\bibfnamefont {P.}~\bibnamefont {Kohli}},
  \bibinfo {author} {\bibfnamefont {A.~N.}\ \bibnamefont {Korotkov}}, \bibinfo
  {author} {\bibfnamefont {F.}~\bibnamefont {Kostritsa}}, \bibinfo {author}
  {\bibfnamefont {R.}~\bibnamefont {Kothari}}, \bibinfo {author} {\bibfnamefont
  {B.}~\bibnamefont {Kozlovskii}}, \bibinfo {author} {\bibfnamefont {J.~M.}\
  \bibnamefont {Kreikebaum}}, \bibinfo {author} {\bibfnamefont {V.~D.}\
  \bibnamefont {Kurilovich}}, \bibinfo {author} {\bibfnamefont
  {N.}~\bibnamefont {Lacroix}}, \bibinfo {author} {\bibfnamefont
  {D.}~\bibnamefont {Landhuis}}, \bibinfo {author} {\bibfnamefont
  {T.}~\bibnamefont {Lange-Dei}}, \bibinfo {author} {\bibfnamefont {B.~W.}\
  \bibnamefont {Langley}}, \bibinfo {author} {\bibfnamefont {P.}~\bibnamefont
  {Laptev}}, \bibinfo {author} {\bibfnamefont {K.}~\bibnamefont {Lau}},
  \bibinfo {author} {\bibfnamefont {L.~L.}\ \bibnamefont {Guevel}}, \bibinfo
  {author} {\bibfnamefont {J.}~\bibnamefont {Ledford}}, \bibinfo {author}
  {\bibfnamefont {K.}~\bibnamefont {Lee}}, \bibinfo {author} {\bibfnamefont
  {Y.~D.}\ \bibnamefont {Lensky}}, \bibinfo {author} {\bibfnamefont
  {S.}~\bibnamefont {Leon}}, \bibinfo {author} {\bibfnamefont {B.~J.}\
  \bibnamefont {Lester}}, \bibinfo {author} {\bibfnamefont {W.~Y.}\
  \bibnamefont {Li}}, \bibinfo {author} {\bibfnamefont {Y.}~\bibnamefont {Li}},
  \bibinfo {author} {\bibfnamefont {A.~T.}\ \bibnamefont {Lill}}, \bibinfo
  {author} {\bibfnamefont {W.}~\bibnamefont {Liu}}, \bibinfo {author}
  {\bibfnamefont {W.~P.}\ \bibnamefont {Livingston}}, \bibinfo {author}
  {\bibfnamefont {A.}~\bibnamefont {Locharla}}, \bibinfo {author}
  {\bibfnamefont {E.}~\bibnamefont {Lucero}}, \bibinfo {author} {\bibfnamefont
  {D.}~\bibnamefont {Lundahl}}, \bibinfo {author} {\bibfnamefont
  {A.}~\bibnamefont {Lunt}}, \bibinfo {author} {\bibfnamefont {S.}~\bibnamefont
  {Madhuk}}, \bibinfo {author} {\bibfnamefont {F.~D.}\ \bibnamefont {Malone}},
  \bibinfo {author} {\bibfnamefont {A.}~\bibnamefont {Maloney}}, \bibinfo
  {author} {\bibfnamefont {S.}~\bibnamefont {Mandr\'a}}, \bibinfo {author}
  {\bibfnamefont {L.~S.}\ \bibnamefont {Martin}}, \bibinfo {author}
  {\bibfnamefont {S.}~\bibnamefont {Martin}}, \bibinfo {author} {\bibfnamefont
  {O.}~\bibnamefont {Martin}}, \bibinfo {author} {\bibfnamefont
  {C.}~\bibnamefont {Maxfield}}, \bibinfo {author} {\bibfnamefont {J.~R.}\
  \bibnamefont {McClean}}, \bibinfo {author} {\bibfnamefont {M.}~\bibnamefont
  {McEwen}}, \bibinfo {author} {\bibfnamefont {S.}~\bibnamefont {Meeks}},
  \bibinfo {author} {\bibfnamefont {A.}~\bibnamefont {Megrant}}, \bibinfo
  {author} {\bibfnamefont {X.}~\bibnamefont {Mi}}, \bibinfo {author}
  {\bibfnamefont {K.~C.}\ \bibnamefont {Miao}}, \bibinfo {author}
  {\bibfnamefont {A.}~\bibnamefont {Mieszala}}, \bibinfo {author}
  {\bibfnamefont {R.}~\bibnamefont {Molavi}}, \bibinfo {author} {\bibfnamefont
  {S.}~\bibnamefont {Molina}}, \bibinfo {author} {\bibfnamefont
  {S.}~\bibnamefont {Montazeri}}, \bibinfo {author} {\bibfnamefont
  {A.}~\bibnamefont {Morvan}}, \bibinfo {author} {\bibfnamefont
  {R.}~\bibnamefont {Movassagh}}, \bibinfo {author} {\bibfnamefont
  {W.}~\bibnamefont {Mruczkiewicz}}, \bibinfo {author} {\bibfnamefont
  {O.}~\bibnamefont {Naaman}}, \bibinfo {author} {\bibfnamefont
  {M.}~\bibnamefont {Neeley}}, \bibinfo {author} {\bibfnamefont
  {C.}~\bibnamefont {Neill}}, \bibinfo {author} {\bibfnamefont
  {A.}~\bibnamefont {Nersisyan}}, \bibinfo {author} {\bibfnamefont
  {H.}~\bibnamefont {Neven}}, \bibinfo {author} {\bibfnamefont
  {M.}~\bibnamefont {Newman}}, \bibinfo {author} {\bibfnamefont {J.~H.}\
  \bibnamefont {Ng}}, \bibinfo {author} {\bibfnamefont {A.}~\bibnamefont
  {Nguyen}}, \bibinfo {author} {\bibfnamefont {M.}~\bibnamefont {Nguyen}},
  \bibinfo {author} {\bibfnamefont {C.}~\bibnamefont {Ni}}, \bibinfo {author}
  {\bibfnamefont {T.~E.}\ \bibnamefont {O'Brien}}, \bibinfo {author}
  {\bibfnamefont {W.~D.}\ \bibnamefont {Oliver}}, \bibinfo {author}
  {\bibfnamefont {A.}~\bibnamefont {Opremcak}}, \bibinfo {author}
  {\bibfnamefont {K.}~\bibnamefont {Ottosson}}, \bibinfo {author}
  {\bibfnamefont {A.}~\bibnamefont {Petukhov}}, \bibinfo {author}
  {\bibfnamefont {A.}~\bibnamefont {Pizzuto}}, \bibinfo {author} {\bibfnamefont
  {J.}~\bibnamefont {Platt}}, \bibinfo {author} {\bibfnamefont
  {R.}~\bibnamefont {Potter}}, \bibinfo {author} {\bibfnamefont
  {O.}~\bibnamefont {Pritchard}}, \bibinfo {author} {\bibfnamefont {L.~P.}\
  \bibnamefont {Pryadko}}, \bibinfo {author} {\bibfnamefont {C.}~\bibnamefont
  {Quintana}}, \bibinfo {author} {\bibfnamefont {G.}~\bibnamefont
  {Ramachandran}}, \bibinfo {author} {\bibfnamefont {M.~J.}\ \bibnamefont
  {Reagor}}, \bibinfo {author} {\bibfnamefont {D.~M.}\ \bibnamefont {Rhodes}},
  \bibinfo {author} {\bibfnamefont {G.}~\bibnamefont {Roberts}}, \bibinfo
  {author} {\bibfnamefont {E.}~\bibnamefont {Rosenberg}}, \bibinfo {author}
  {\bibfnamefont {E.}~\bibnamefont {Rosenfeld}}, \bibinfo {author}
  {\bibfnamefont {P.}~\bibnamefont {Roushan}}, \bibinfo {author} {\bibfnamefont
  {N.~C.}\ \bibnamefont {Rubin}}, \bibinfo {author} {\bibfnamefont
  {N.}~\bibnamefont {Saei}}, \bibinfo {author} {\bibfnamefont {D.}~\bibnamefont
  {Sank}}, \bibinfo {author} {\bibfnamefont {K.}~\bibnamefont
  {Sankaragomathi}}, \bibinfo {author} {\bibfnamefont {K.~J.}\ \bibnamefont
  {Satzinger}}, \bibinfo {author} {\bibfnamefont {H.~F.}\ \bibnamefont
  {Schurkus}}, \bibinfo {author} {\bibfnamefont {C.}~\bibnamefont {Schuster}},
  \bibinfo {author} {\bibfnamefont {A.~W.}\ \bibnamefont {Senior}}, \bibinfo
  {author} {\bibfnamefont {M.~J.}\ \bibnamefont {Shearn}}, \bibinfo {author}
  {\bibfnamefont {A.}~\bibnamefont {Shorter}}, \bibinfo {author} {\bibfnamefont
  {N.}~\bibnamefont {Shutty}}, \bibinfo {author} {\bibfnamefont
  {V.}~\bibnamefont {Shvarts}}, \bibinfo {author} {\bibfnamefont
  {S.}~\bibnamefont {Singh}}, \bibinfo {author} {\bibfnamefont
  {V.}~\bibnamefont {Sivak}}, \bibinfo {author} {\bibfnamefont
  {J.}~\bibnamefont {Skruzny}}, \bibinfo {author} {\bibfnamefont
  {S.}~\bibnamefont {Small}}, \bibinfo {author} {\bibfnamefont
  {V.}~\bibnamefont {Smelyanskiy}}, \bibinfo {author} {\bibfnamefont {W.~C.}\
  \bibnamefont {Smith}}, \bibinfo {author} {\bibfnamefont {R.~D.}\ \bibnamefont
  {Somma}}, \bibinfo {author} {\bibfnamefont {S.}~\bibnamefont {Springer}},
  \bibinfo {author} {\bibfnamefont {G.}~\bibnamefont {Sterling}}, \bibinfo
  {author} {\bibfnamefont {D.}~\bibnamefont {Strain}}, \bibinfo {author}
  {\bibfnamefont {J.}~\bibnamefont {Suchard}}, \bibinfo {author} {\bibfnamefont
  {A.}~\bibnamefont {Szasz}}, \bibinfo {author} {\bibfnamefont
  {A.}~\bibnamefont {Sztein}}, \bibinfo {author} {\bibfnamefont
  {D.}~\bibnamefont {Thor}}, \bibinfo {author} {\bibfnamefont {A.}~\bibnamefont
  {Torres}}, \bibinfo {author} {\bibfnamefont {M.~M.}\ \bibnamefont
  {Torunbalci}}, \bibinfo {author} {\bibfnamefont {A.}~\bibnamefont
  {Vaishnav}}, \bibinfo {author} {\bibfnamefont {J.}~\bibnamefont {Vargas}},
  \bibinfo {author} {\bibfnamefont {S.}~\bibnamefont {Vdovichev}}, \bibinfo
  {author} {\bibfnamefont {G.}~\bibnamefont {Vidal}}, \bibinfo {author}
  {\bibfnamefont {B.}~\bibnamefont {Villalonga}}, \bibinfo {author}
  {\bibfnamefont {C.~V.}\ \bibnamefont {Heidweiller}}, \bibinfo {author}
  {\bibfnamefont {S.}~\bibnamefont {Waltman}}, \bibinfo {author} {\bibfnamefont
  {S.~X.}\ \bibnamefont {Wang}}, \bibinfo {author} {\bibfnamefont
  {B.}~\bibnamefont {Ware}}, \bibinfo {author} {\bibfnamefont {K.}~\bibnamefont
  {Weber}}, \bibinfo {author} {\bibfnamefont {T.}~\bibnamefont {White}},
  \bibinfo {author} {\bibfnamefont {K.}~\bibnamefont {Wong}}, \bibinfo {author}
  {\bibfnamefont {B.~W.~K.}\ \bibnamefont {Woo}}, \bibinfo {author}
  {\bibfnamefont {C.}~\bibnamefont {Xing}}, \bibinfo {author} {\bibfnamefont
  {Z.~J.}\ \bibnamefont {Yao}}, \bibinfo {author} {\bibfnamefont
  {P.}~\bibnamefont {Yeh}}, \bibinfo {author} {\bibfnamefont {B.}~\bibnamefont
  {Ying}}, \bibinfo {author} {\bibfnamefont {J.}~\bibnamefont {Yoo}}, \bibinfo
  {author} {\bibfnamefont {N.}~\bibnamefont {Yosri}}, \bibinfo {author}
  {\bibfnamefont {G.}~\bibnamefont {Young}}, \bibinfo {author} {\bibfnamefont
  {A.}~\bibnamefont {Zalcman}}, \bibinfo {author} {\bibfnamefont
  {Y.}~\bibnamefont {Zhang}}, \bibinfo {author} {\bibfnamefont
  {N.}~\bibnamefont {Zhu}},\ and\ \bibinfo {author} {\bibfnamefont
  {N.}~\bibnamefont {Zobrist}},\ }\bibfield  {title} {\bibinfo {title} {Quantum
  error correction below the surface code threshold},\ }\href
  {https://arxiv.org/abs/2408.13687} {\bibfield  {journal} {\bibinfo  {journal}
  {arXiv:2408.13687}\ } (\bibinfo {year} {2024})}\BibitemShut {NoStop}%
\bibitem [{\citenamefont {Rol}\ \emph {et~al.}(2020)\citenamefont {Rol},
  \citenamefont {Ciorciaro}, \citenamefont {Malinowski}, \citenamefont
  {Tarasinski}, \citenamefont {Sagastizabal}, \citenamefont {Bultink},
  \citenamefont {Salathe}, \citenamefont {Haandbaek}, \citenamefont {Sedivy},\
  and\ \citenamefont {DiCarlo}}]{Rol2020}%
  \BibitemOpen
  \bibfield  {author} {\bibinfo {author} {\bibfnamefont {M.~A.}\ \bibnamefont
  {Rol}}, \bibinfo {author} {\bibfnamefont {L.}~\bibnamefont {Ciorciaro}},
  \bibinfo {author} {\bibfnamefont {F.~K.}\ \bibnamefont {Malinowski}},
  \bibinfo {author} {\bibfnamefont {B.~M.}\ \bibnamefont {Tarasinski}},
  \bibinfo {author} {\bibfnamefont {R.~E.}\ \bibnamefont {Sagastizabal}},
  \bibinfo {author} {\bibfnamefont {C.~C.}\ \bibnamefont {Bultink}}, \bibinfo
  {author} {\bibfnamefont {Y.}~\bibnamefont {Salathe}}, \bibinfo {author}
  {\bibfnamefont {N.}~\bibnamefont {Haandbaek}}, \bibinfo {author}
  {\bibfnamefont {J.}~\bibnamefont {Sedivy}},\ and\ \bibinfo {author}
  {\bibfnamefont {L.}~\bibnamefont {DiCarlo}},\ }\bibfield  {title} {\bibinfo
  {title} {Time-domain characterization and correction of on-chip distortion of
  control pulses in a quantum processor},\ }\href
  {https://doi.org/10.1063/1.5133894} {\bibfield  {journal} {\bibinfo
  {journal} {Appl. Phys. Lett.}\ }\textbf {\bibinfo {volume} {116}},\ \bibinfo
  {pages} {054001} (\bibinfo {year} {2020})}\BibitemShut {NoStop}%
\bibitem [{\citenamefont {Foxen}\ \emph {et~al.}(2018)\citenamefont {Foxen},
  \citenamefont {Mutus}, \citenamefont {Lucero}, \citenamefont {Jeffrey},
  \citenamefont {Sank}, \citenamefont {Barends}, \citenamefont {Arya},
  \citenamefont {Burkett}, \citenamefont {Chen}, \citenamefont {Chen},
  \citenamefont {Chiaro}, \citenamefont {Dunsworth}, \citenamefont {Fowler},
  \citenamefont {Gidney}, \citenamefont {Giustina}, \citenamefont {Graff},
  \citenamefont {Huang}, \citenamefont {Kelly}, \citenamefont {Klimov},
  \citenamefont {Megrant}, \citenamefont {Naaman}, \citenamefont {Neeley},
  \citenamefont {Neill}, \citenamefont {Quintana}, \citenamefont {Roushan},
  \citenamefont {Vainsencher}, \citenamefont {Wenner}, \citenamefont {White},\
  and\ \citenamefont {Martinis}}]{Foxen2018a}%
  \BibitemOpen
  \bibfield  {author} {\bibinfo {author} {\bibfnamefont {B.}~\bibnamefont
  {Foxen}}, \bibinfo {author} {\bibfnamefont {J.~Y.}\ \bibnamefont {Mutus}},
  \bibinfo {author} {\bibfnamefont {E.}~\bibnamefont {Lucero}}, \bibinfo
  {author} {\bibfnamefont {E.}~\bibnamefont {Jeffrey}}, \bibinfo {author}
  {\bibfnamefont {D.}~\bibnamefont {Sank}}, \bibinfo {author} {\bibfnamefont
  {R.}~\bibnamefont {Barends}}, \bibinfo {author} {\bibfnamefont
  {K.}~\bibnamefont {Arya}}, \bibinfo {author} {\bibfnamefont {B.}~\bibnamefont
  {Burkett}}, \bibinfo {author} {\bibfnamefont {Y.}~\bibnamefont {Chen}},
  \bibinfo {author} {\bibfnamefont {Z.}~\bibnamefont {Chen}}, \bibinfo {author}
  {\bibfnamefont {B.}~\bibnamefont {Chiaro}}, \bibinfo {author} {\bibfnamefont
  {A.}~\bibnamefont {Dunsworth}}, \bibinfo {author} {\bibfnamefont
  {A.}~\bibnamefont {Fowler}}, \bibinfo {author} {\bibfnamefont
  {C.}~\bibnamefont {Gidney}}, \bibinfo {author} {\bibfnamefont
  {M.}~\bibnamefont {Giustina}}, \bibinfo {author} {\bibfnamefont
  {R.}~\bibnamefont {Graff}}, \bibinfo {author} {\bibfnamefont
  {T.}~\bibnamefont {Huang}}, \bibinfo {author} {\bibfnamefont
  {J.}~\bibnamefont {Kelly}}, \bibinfo {author} {\bibfnamefont
  {P.}~\bibnamefont {Klimov}}, \bibinfo {author} {\bibfnamefont
  {A.}~\bibnamefont {Megrant}}, \bibinfo {author} {\bibfnamefont
  {O.}~\bibnamefont {Naaman}}, \bibinfo {author} {\bibfnamefont
  {M.}~\bibnamefont {Neeley}}, \bibinfo {author} {\bibfnamefont
  {C.}~\bibnamefont {Neill}}, \bibinfo {author} {\bibfnamefont
  {C.}~\bibnamefont {Quintana}}, \bibinfo {author} {\bibfnamefont
  {P.}~\bibnamefont {Roushan}}, \bibinfo {author} {\bibfnamefont
  {A.}~\bibnamefont {Vainsencher}}, \bibinfo {author} {\bibfnamefont
  {J.}~\bibnamefont {Wenner}}, \bibinfo {author} {\bibfnamefont {T.~C.}\
  \bibnamefont {White}},\ and\ \bibinfo {author} {\bibfnamefont {J.~M.}\
  \bibnamefont {Martinis}},\ }\bibfield  {title} {\bibinfo {title} {High speed
  flux sampling for tunable superconducting qubits with an embedded cryogenic
  transducer},\ }\href {https://doi.org/10.1088/1361-6668/aaf048} {\bibfield
  {journal} {\bibinfo  {journal} {Superconductor Science and Technology}\
  }\textbf {\bibinfo {volume} {32}},\ \bibinfo {pages} {015012} (\bibinfo
  {year} {2018})}\BibitemShut {NoStop}%
\bibitem [{\citenamefont {Johnson}(2011)}]{Johnson2011PhD}%
  \BibitemOpen
  \bibfield  {author} {\bibinfo {author} {\bibfnamefont {B.}~\bibnamefont
  {Johnson}},\ }\emph {\bibinfo {title} {Controlling Photons in Superconducting
  Electrical Circuits}},\ \href@noop {} {Ph.D. thesis},\ \bibinfo  {school}
  {Yale} (\bibinfo {year} {2011})\BibitemShut {NoStop}%
\bibitem [{\citenamefont {Barends}\ \emph {et~al.}(2014)\citenamefont
  {Barends}, \citenamefont {Kelly}, \citenamefont {Megrant}, \citenamefont
  {Veitia}, \citenamefont {Sank}, \citenamefont {Jeffrey}, \citenamefont
  {White}, \citenamefont {Mutus}, \citenamefont {Fowler}, \citenamefont
  {Campbell}, \citenamefont {Chen}, \citenamefont {Chen}, \citenamefont
  {Chiaro}, \citenamefont {Dunsworth}, \citenamefont {Neill}, \citenamefont
  {{O\'Malley}}, \citenamefont {Roushan}, \citenamefont {Vainsencher},
  \citenamefont {Wenner}, \citenamefont {Korotkov}, \citenamefont {Cleland},\
  and\ \citenamefont {Martinis}}]{Barends2014}%
  \BibitemOpen
  \bibfield  {author} {\bibinfo {author} {\bibfnamefont {R.}~\bibnamefont
  {Barends}}, \bibinfo {author} {\bibfnamefont {J.}~\bibnamefont {Kelly}},
  \bibinfo {author} {\bibfnamefont {A.}~\bibnamefont {Megrant}}, \bibinfo
  {author} {\bibfnamefont {A.}~\bibnamefont {Veitia}}, \bibinfo {author}
  {\bibfnamefont {D.}~\bibnamefont {Sank}}, \bibinfo {author} {\bibfnamefont
  {E.}~\bibnamefont {Jeffrey}}, \bibinfo {author} {\bibfnamefont {T.~C.}\
  \bibnamefont {White}}, \bibinfo {author} {\bibfnamefont {J.}~\bibnamefont
  {Mutus}}, \bibinfo {author} {\bibfnamefont {A.~G.}\ \bibnamefont {Fowler}},
  \bibinfo {author} {\bibfnamefont {B.}~\bibnamefont {Campbell}}, \bibinfo
  {author} {\bibfnamefont {Y.}~\bibnamefont {Chen}}, \bibinfo {author}
  {\bibfnamefont {Z.}~\bibnamefont {Chen}}, \bibinfo {author} {\bibfnamefont
  {B.}~\bibnamefont {Chiaro}}, \bibinfo {author} {\bibfnamefont
  {A.}~\bibnamefont {Dunsworth}}, \bibinfo {author} {\bibfnamefont
  {C.}~\bibnamefont {Neill}}, \bibinfo {author} {\bibfnamefont
  {P.}~\bibnamefont {{O\'Malley}}}, \bibinfo {author} {\bibfnamefont
  {P.}~\bibnamefont {Roushan}}, \bibinfo {author} {\bibfnamefont
  {A.}~\bibnamefont {Vainsencher}}, \bibinfo {author} {\bibfnamefont
  {J.}~\bibnamefont {Wenner}}, \bibinfo {author} {\bibfnamefont {A.~N.}\
  \bibnamefont {Korotkov}}, \bibinfo {author} {\bibfnamefont {A.~N.}\
  \bibnamefont {Cleland}},\ and\ \bibinfo {author} {\bibfnamefont {J.~M.}\
  \bibnamefont {Martinis}},\ }\bibfield  {title} {\bibinfo {title}
  {Superconducting quantum circuits at the surface code threshold for fault
  tolerance},\ }\href {https://doi.org/10.1038/nature13171} {\bibfield
  {journal} {\bibinfo  {journal} {Nature}\ }\textbf {\bibinfo {volume} {508}},\
  \bibinfo {pages} {500} (\bibinfo {year} {2014})}\BibitemShut {NoStop}%
\bibitem [{\citenamefont {Kelly}(2015)}]{Kelly2015a}%
  \BibitemOpen
  \bibfield  {author} {\bibinfo {author} {\bibfnamefont {J.~S.}\ \bibnamefont
  {Kelly}},\ }\emph {\bibinfo {title} {Fault-tolerant superconducting
  qubits}},\ \href
  {https://web.physics.ucsb.edu/~martinisgroup/theses/Kelly2015.pdf} {Ph.D.
  thesis},\ \bibinfo  {school} {University of California Santa Barbara}
  (\bibinfo {year} {2015})\BibitemShut {NoStop}%
\bibitem [{\citenamefont {Hofheinz}\ \emph {et~al.}(2009)\citenamefont
  {Hofheinz}, \citenamefont {Wang}, \citenamefont {Ansmann}, \citenamefont
  {Bialczak}, \citenamefont {Lucero}, \citenamefont {Neeley}, \citenamefont
  {O'Connell}, \citenamefont {Sank}, \citenamefont {Wenner}, \citenamefont
  {Martinis},\ and\ \citenamefont {Cleland}}]{Hofheinz2009}%
  \BibitemOpen
  \bibfield  {author} {\bibinfo {author} {\bibfnamefont {M.}~\bibnamefont
  {Hofheinz}}, \bibinfo {author} {\bibfnamefont {H.}~\bibnamefont {Wang}},
  \bibinfo {author} {\bibfnamefont {M.}~\bibnamefont {Ansmann}}, \bibinfo
  {author} {\bibfnamefont {R.~C.}\ \bibnamefont {Bialczak}}, \bibinfo {author}
  {\bibfnamefont {E.}~\bibnamefont {Lucero}}, \bibinfo {author} {\bibfnamefont
  {M.}~\bibnamefont {Neeley}}, \bibinfo {author} {\bibfnamefont {A.~D.}\
  \bibnamefont {O'Connell}}, \bibinfo {author} {\bibfnamefont {D.}~\bibnamefont
  {Sank}}, \bibinfo {author} {\bibfnamefont {J.}~\bibnamefont {Wenner}},
  \bibinfo {author} {\bibfnamefont {J.~M.}\ \bibnamefont {Martinis}},\ and\
  \bibinfo {author} {\bibfnamefont {A.~N.}\ \bibnamefont {Cleland}},\
  }\bibfield  {title} {\bibinfo {title} {Synthesizing arbitrary quantum states
  in a superconducting resonator},\ }\href
  {https://doi.org/10.1038/nature08005} {\bibfield  {journal} {\bibinfo
  {journal} {Nature}\ }\textbf {\bibinfo {volume} {459}},\ \bibinfo {pages}
  {546} (\bibinfo {year} {2009})}\BibitemShut {NoStop}%
\bibitem [{\citenamefont {Chen}\ \emph
  {et~al.}(2022{\natexlab{a}})\citenamefont {Chen}, \citenamefont {Shi},
  \citenamefont {Xiang}, \citenamefont {Wang}, \citenamefont {Li},
  \citenamefont {Sun}, \citenamefont {He}, \citenamefont {Song}, \citenamefont
  {Zhao}, \citenamefont {Zheng}, \citenamefont {Xu},\ and\ \citenamefont
  {Fan}}]{Chen2022b}%
  \BibitemOpen
  \bibfield  {author} {\bibinfo {author} {\bibfnamefont {C.-T.}\ \bibnamefont
  {Chen}}, \bibinfo {author} {\bibfnamefont {Y.-H.}\ \bibnamefont {Shi}},
  \bibinfo {author} {\bibfnamefont {Z.}~\bibnamefont {Xiang}}, \bibinfo
  {author} {\bibfnamefont {Z.-A.}\ \bibnamefont {Wang}}, \bibinfo {author}
  {\bibfnamefont {T.-M.}\ \bibnamefont {Li}}, \bibinfo {author} {\bibfnamefont
  {H.-Y.}\ \bibnamefont {Sun}}, \bibinfo {author} {\bibfnamefont {T.-S.}\
  \bibnamefont {He}}, \bibinfo {author} {\bibfnamefont {X.}~\bibnamefont
  {Song}}, \bibinfo {author} {\bibfnamefont {S.}~\bibnamefont {Zhao}}, \bibinfo
  {author} {\bibfnamefont {D.}~\bibnamefont {Zheng}}, \bibinfo {author}
  {\bibfnamefont {K.}~\bibnamefont {Xu}},\ and\ \bibinfo {author}
  {\bibfnamefont {H.}~\bibnamefont {Fan}},\ }\bibfield  {title} {\bibinfo
  {title} {{ScQ} cloud quantum computation for generating
  {G}reenberger-{H}orne-{Z}eilinger states of up to 10 qubits},\ }\href
  {https://doi.org/10.1007/s11433-022-1972-1} {\bibfield  {journal} {\bibinfo
  {journal} {Science China Physics, Mechanics {\&} Astronomy}\ }\textbf
  {\bibinfo {volume} {65}},\ \bibinfo {pages} {110362} (\bibinfo {year}
  {2022}{\natexlab{a}})}\BibitemShut {NoStop}%
\bibitem [{\citenamefont {Chen}(2018)}]{Chen2018j}%
  \BibitemOpen
  \bibfield  {author} {\bibinfo {author} {\bibfnamefont {Z.}~\bibnamefont
  {Chen}},\ }\emph {\bibinfo {title} {Metrology of Quantum Control and
  Measurement in Superconducting Qubits}},\ \href
  {https://web.physics.ucsb.edu/~martinisgroup/theses/Chen2018.pdf} {Ph.D.
  thesis},\ \bibinfo  {school} {University of California Santa Barbara}
  (\bibinfo {year} {2018})\BibitemShut {NoStop}%
\bibitem [{\citenamefont {Neill}\ \emph {et~al.}(2018)\citenamefont {Neill},
  \citenamefont {Roushan}, \citenamefont {Kechedzhi}, \citenamefont {Boixo},
  \citenamefont {Isakov}, \citenamefont {Smelyanskiy}, \citenamefont {Megrant},
  \citenamefont {Chiaro}, \citenamefont {Dunsworth}, \citenamefont {Arya},
  \citenamefont {Barends}, \citenamefont {Burkett}, \citenamefont {Chen},
  \citenamefont {Chen}, \citenamefont {Fowler}, \citenamefont {Foxen},
  \citenamefont {Giustina}, \citenamefont {Graff}, \citenamefont {Jeffrey},
  \citenamefont {Huang}, \citenamefont {Kelly}, \citenamefont {Klimov},
  \citenamefont {Lucero}, \citenamefont {Mutus}, \citenamefont {Neeley},
  \citenamefont {Quintana}, \citenamefont {Sank}, \citenamefont {Vainsencher},
  \citenamefont {Wenner}, \citenamefont {White}, \citenamefont {Neven},\ and\
  \citenamefont {Martinis}}]{Neill2018}%
  \BibitemOpen
  \bibfield  {author} {\bibinfo {author} {\bibfnamefont {C.}~\bibnamefont
  {Neill}}, \bibinfo {author} {\bibfnamefont {P.}~\bibnamefont {Roushan}},
  \bibinfo {author} {\bibfnamefont {K.}~\bibnamefont {Kechedzhi}}, \bibinfo
  {author} {\bibfnamefont {S.}~\bibnamefont {Boixo}}, \bibinfo {author}
  {\bibfnamefont {S.~V.}\ \bibnamefont {Isakov}}, \bibinfo {author}
  {\bibfnamefont {V.}~\bibnamefont {Smelyanskiy}}, \bibinfo {author}
  {\bibfnamefont {A.}~\bibnamefont {Megrant}}, \bibinfo {author} {\bibfnamefont
  {B.}~\bibnamefont {Chiaro}}, \bibinfo {author} {\bibfnamefont
  {A.}~\bibnamefont {Dunsworth}}, \bibinfo {author} {\bibfnamefont
  {K.}~\bibnamefont {Arya}}, \bibinfo {author} {\bibfnamefont {R.}~\bibnamefont
  {Barends}}, \bibinfo {author} {\bibfnamefont {B.}~\bibnamefont {Burkett}},
  \bibinfo {author} {\bibfnamefont {Y.}~\bibnamefont {Chen}}, \bibinfo {author}
  {\bibfnamefont {Z.}~\bibnamefont {Chen}}, \bibinfo {author} {\bibfnamefont
  {A.}~\bibnamefont {Fowler}}, \bibinfo {author} {\bibfnamefont
  {B.}~\bibnamefont {Foxen}}, \bibinfo {author} {\bibfnamefont
  {M.}~\bibnamefont {Giustina}}, \bibinfo {author} {\bibfnamefont
  {R.}~\bibnamefont {Graff}}, \bibinfo {author} {\bibfnamefont
  {E.}~\bibnamefont {Jeffrey}}, \bibinfo {author} {\bibfnamefont
  {T.}~\bibnamefont {Huang}}, \bibinfo {author} {\bibfnamefont
  {J.}~\bibnamefont {Kelly}}, \bibinfo {author} {\bibfnamefont
  {P.}~\bibnamefont {Klimov}}, \bibinfo {author} {\bibfnamefont
  {E.}~\bibnamefont {Lucero}}, \bibinfo {author} {\bibfnamefont
  {J.}~\bibnamefont {Mutus}}, \bibinfo {author} {\bibfnamefont
  {M.}~\bibnamefont {Neeley}}, \bibinfo {author} {\bibfnamefont
  {C.}~\bibnamefont {Quintana}}, \bibinfo {author} {\bibfnamefont
  {D.}~\bibnamefont {Sank}}, \bibinfo {author} {\bibfnamefont {A.}~\bibnamefont
  {Vainsencher}}, \bibinfo {author} {\bibfnamefont {J.}~\bibnamefont {Wenner}},
  \bibinfo {author} {\bibfnamefont {T.~C.}\ \bibnamefont {White}}, \bibinfo
  {author} {\bibfnamefont {H.}~\bibnamefont {Neven}},\ and\ \bibinfo {author}
  {\bibfnamefont {J.~M.}\ \bibnamefont {Martinis}},\ }\bibfield  {title}
  {\bibinfo {title} {A blueprint for demonstrating quantum supremacy with
  superconducting qubits},\ }\href {https://doi.org/10.1126/science.aao4309}
  {\bibfield  {journal} {\bibinfo  {journal} {Science}\ }\textbf {\bibinfo
  {volume} {360}},\ \bibinfo {pages} {195} (\bibinfo {year}
  {2018})}\BibitemShut {NoStop}%
\bibitem [{\citenamefont {Braum{\"u}ller}\ \emph {et~al.}(2022)\citenamefont
  {Braum{\"u}ller}, \citenamefont {Karamlou}, \citenamefont {Yanay},
  \citenamefont {Kannan}, \citenamefont {Kim}, \citenamefont {Kjaergaard},
  \citenamefont {Melville}, \citenamefont {Niedzielski}, \citenamefont {Sung},
  \citenamefont {Veps{\"a}l{\"a}inen}, \citenamefont {Winik}, \citenamefont
  {Yoder}, \citenamefont {Orlando}, \citenamefont {Gustavsson}, \citenamefont
  {Tahan},\ and\ \citenamefont {Oliver}}]{Braumuller2022}%
  \BibitemOpen
  \bibfield  {author} {\bibinfo {author} {\bibfnamefont {J.}~\bibnamefont
  {Braum{\"u}ller}}, \bibinfo {author} {\bibfnamefont {A.~H.}\ \bibnamefont
  {Karamlou}}, \bibinfo {author} {\bibfnamefont {Y.}~\bibnamefont {Yanay}},
  \bibinfo {author} {\bibfnamefont {B.}~\bibnamefont {Kannan}}, \bibinfo
  {author} {\bibfnamefont {D.}~\bibnamefont {Kim}}, \bibinfo {author}
  {\bibfnamefont {M.}~\bibnamefont {Kjaergaard}}, \bibinfo {author}
  {\bibfnamefont {A.}~\bibnamefont {Melville}}, \bibinfo {author}
  {\bibfnamefont {B.~M.}\ \bibnamefont {Niedzielski}}, \bibinfo {author}
  {\bibfnamefont {Y.}~\bibnamefont {Sung}}, \bibinfo {author} {\bibfnamefont
  {A.}~\bibnamefont {Veps{\"a}l{\"a}inen}}, \bibinfo {author} {\bibfnamefont
  {R.}~\bibnamefont {Winik}}, \bibinfo {author} {\bibfnamefont {J.~L.}\
  \bibnamefont {Yoder}}, \bibinfo {author} {\bibfnamefont {T.~P.}\ \bibnamefont
  {Orlando}}, \bibinfo {author} {\bibfnamefont {S.}~\bibnamefont {Gustavsson}},
  \bibinfo {author} {\bibfnamefont {C.}~\bibnamefont {Tahan}},\ and\ \bibinfo
  {author} {\bibfnamefont {W.~D.}\ \bibnamefont {Oliver}},\ }\bibfield  {title}
  {\bibinfo {title} {Probing quantum information propagation with
  out-of-time-ordered correlators},\ }\href
  {https://doi.org/10.1038/s41567-021-01430-w} {\bibfield  {journal} {\bibinfo
  {journal} {Nature Physics}\ }\textbf {\bibinfo {volume} {18}},\ \bibinfo
  {pages} {172} (\bibinfo {year} {2022})}\BibitemShut {NoStop}%
\bibitem [{\citenamefont {Jerger}\ \emph {et~al.}(2019)\citenamefont {Jerger},
  \citenamefont {Kulikov}, \citenamefont {Vasselin},\ and\ \citenamefont
  {Fedorov}}]{Jerger2019}%
  \BibitemOpen
  \bibfield  {author} {\bibinfo {author} {\bibfnamefont {M.}~\bibnamefont
  {Jerger}}, \bibinfo {author} {\bibfnamefont {A.}~\bibnamefont {Kulikov}},
  \bibinfo {author} {\bibfnamefont {Z.}~\bibnamefont {Vasselin}},\ and\
  \bibinfo {author} {\bibfnamefont {A.}~\bibnamefont {Fedorov}},\ }\bibfield
  {title} {\bibinfo {title} {In situ characterization of qubit control lines: A
  qubit as a vector network analyzer},\ }\href
  {https://doi.org/10.1103/PhysRevLett.123.150501} {\bibfield  {journal}
  {\bibinfo  {journal} {Phys. Rev. Lett.}\ }\textbf {\bibinfo {volume} {123}},\
  \bibinfo {pages} {150501} (\bibinfo {year} {2019})}\BibitemShut {NoStop}%
\bibitem [{\citenamefont {Li}\ \emph {et~al.}(2025)\citenamefont {Li},
  \citenamefont {Zhang}, \citenamefont {Chen}, \citenamefont {Huang},
  \citenamefont {Liu}, \citenamefont {Xiao}, \citenamefont {Deng},
  \citenamefont {Liang}, \citenamefont {Chen}, \citenamefont {Liu},
  \citenamefont {Li}, \citenamefont {Bao}, \citenamefont {Zhao}, \citenamefont
  {Xu}, \citenamefont {Li}, \citenamefont {He}, \citenamefont {Liu},
  \citenamefont {Yu}, \citenamefont {Zhou}, \citenamefont {Liu}, \citenamefont
  {Song}, \citenamefont {Zheng}, \citenamefont {Xiang}, \citenamefont {Shi},
  \citenamefont {Xu},\ and\ \citenamefont {Fan}}]{Li2025c}%
  \BibitemOpen
  \bibfield  {author} {\bibinfo {author} {\bibfnamefont {T.-M.}\ \bibnamefont
  {Li}}, \bibinfo {author} {\bibfnamefont {J.-C.}\ \bibnamefont {Zhang}},
  \bibinfo {author} {\bibfnamefont {B.-J.}\ \bibnamefont {Chen}}, \bibinfo
  {author} {\bibfnamefont {K.}~\bibnamefont {Huang}}, \bibinfo {author}
  {\bibfnamefont {H.-T.}\ \bibnamefont {Liu}}, \bibinfo {author} {\bibfnamefont
  {Y.-X.}\ \bibnamefont {Xiao}}, \bibinfo {author} {\bibfnamefont {C.-L.}\
  \bibnamefont {Deng}}, \bibinfo {author} {\bibfnamefont {G.-H.}\ \bibnamefont
  {Liang}}, \bibinfo {author} {\bibfnamefont {C.-T.}\ \bibnamefont {Chen}},
  \bibinfo {author} {\bibfnamefont {Y.}~\bibnamefont {Liu}}, \bibinfo {author}
  {\bibfnamefont {H.}~\bibnamefont {Li}}, \bibinfo {author} {\bibfnamefont
  {Z.-T.}\ \bibnamefont {Bao}}, \bibinfo {author} {\bibfnamefont
  {K.}~\bibnamefont {Zhao}}, \bibinfo {author} {\bibfnamefont {Y.}~\bibnamefont
  {Xu}}, \bibinfo {author} {\bibfnamefont {L.}~\bibnamefont {Li}}, \bibinfo
  {author} {\bibfnamefont {Y.}~\bibnamefont {He}}, \bibinfo {author}
  {\bibfnamefont {Z.-H.}\ \bibnamefont {Liu}}, \bibinfo {author} {\bibfnamefont
  {Y.-H.}\ \bibnamefont {Yu}}, \bibinfo {author} {\bibfnamefont {S.-Y.}\
  \bibnamefont {Zhou}}, \bibinfo {author} {\bibfnamefont {Y.-J.}\ \bibnamefont
  {Liu}}, \bibinfo {author} {\bibfnamefont {X.}~\bibnamefont {Song}}, \bibinfo
  {author} {\bibfnamefont {D.}~\bibnamefont {Zheng}}, \bibinfo {author}
  {\bibfnamefont {Z.}~\bibnamefont {Xiang}}, \bibinfo {author} {\bibfnamefont
  {Y.-H.}\ \bibnamefont {Shi}}, \bibinfo {author} {\bibfnamefont
  {K.}~\bibnamefont {Xu}},\ and\ \bibinfo {author} {\bibfnamefont
  {H.}~\bibnamefont {Fan}},\ }\bibfield  {title} {\bibinfo {title}
  {High-precision pulse calibration of tunable couplers for high-fidelity
  two-qubit gates in superconducting quantum processors},\ }\href
  {https://doi.org/10.1103/PhysRevApplied.23.024059} {\bibfield  {journal}
  {\bibinfo  {journal} {Phys. Rev. Appl.}\ }\textbf {\bibinfo {volume} {23}},\
  \bibinfo {pages} {024059} (\bibinfo {year} {2025})}\BibitemShut {NoStop}%
\bibitem [{\citenamefont {Kelly}\ \emph {et~al.}(2014)\citenamefont {Kelly},
  \citenamefont {Barends}, \citenamefont {Campbell}, \citenamefont {Chen},
  \citenamefont {Chen}, \citenamefont {Chiaro}, \citenamefont {Dunsworth},
  \citenamefont {Fowler}, \citenamefont {Hoi}, \citenamefont {Jeffrey},
  \citenamefont {Megrant}, \citenamefont {Mutus}, \citenamefont {Neill},
  \citenamefont {O'Malley}, \citenamefont {Quintana}, \citenamefont {Roushan},
  \citenamefont {Sank}, \citenamefont {Vainsencher}, \citenamefont {Wenner},
  \citenamefont {White}, \citenamefont {Cleland},\ and\ \citenamefont
  {Martinis}}]{Kelly2014}%
  \BibitemOpen
  \bibfield  {author} {\bibinfo {author} {\bibfnamefont {J.}~\bibnamefont
  {Kelly}}, \bibinfo {author} {\bibfnamefont {R.}~\bibnamefont {Barends}},
  \bibinfo {author} {\bibfnamefont {B.}~\bibnamefont {Campbell}}, \bibinfo
  {author} {\bibfnamefont {Y.}~\bibnamefont {Chen}}, \bibinfo {author}
  {\bibfnamefont {Z.}~\bibnamefont {Chen}}, \bibinfo {author} {\bibfnamefont
  {B.}~\bibnamefont {Chiaro}}, \bibinfo {author} {\bibfnamefont
  {A.}~\bibnamefont {Dunsworth}}, \bibinfo {author} {\bibfnamefont {A.~G.}\
  \bibnamefont {Fowler}}, \bibinfo {author} {\bibfnamefont {I.-C.}\
  \bibnamefont {Hoi}}, \bibinfo {author} {\bibfnamefont {E.}~\bibnamefont
  {Jeffrey}}, \bibinfo {author} {\bibfnamefont {A.}~\bibnamefont {Megrant}},
  \bibinfo {author} {\bibfnamefont {J.}~\bibnamefont {Mutus}}, \bibinfo
  {author} {\bibfnamefont {C.}~\bibnamefont {Neill}}, \bibinfo {author}
  {\bibfnamefont {P.~J.~J.}\ \bibnamefont {O'Malley}}, \bibinfo {author}
  {\bibfnamefont {C.}~\bibnamefont {Quintana}}, \bibinfo {author}
  {\bibfnamefont {P.}~\bibnamefont {Roushan}}, \bibinfo {author} {\bibfnamefont
  {D.}~\bibnamefont {Sank}}, \bibinfo {author} {\bibfnamefont {A.}~\bibnamefont
  {Vainsencher}}, \bibinfo {author} {\bibfnamefont {J.}~\bibnamefont {Wenner}},
  \bibinfo {author} {\bibfnamefont {T.~C.}\ \bibnamefont {White}}, \bibinfo
  {author} {\bibfnamefont {A.~N.}\ \bibnamefont {Cleland}},\ and\ \bibinfo
  {author} {\bibfnamefont {J.~M.}\ \bibnamefont {Martinis}},\ }\bibfield
  {title} {\bibinfo {title} {Optimal quantum control using randomized
  benchmarking},\ }\href {https://doi.org/10.1103/PhysRevLett.112.240504}
  {\bibfield  {journal} {\bibinfo  {journal} {Phys. Rev. Lett.}\ }\textbf
  {\bibinfo {volume} {112}},\ \bibinfo {pages} {240504} (\bibinfo {year}
  {2014})}\BibitemShut {NoStop}%
\bibitem [{\citenamefont {Chen}\ \emph
  {et~al.}(2022{\natexlab{b}})\citenamefont {Chen}, \citenamefont
  {Kronowetter}, \citenamefont {Fesquet}, \citenamefont {Honasoge},
  \citenamefont {Nojiri}, \citenamefont {Renger}, \citenamefont {Fedorov},
  \citenamefont {Marx}, \citenamefont {Deppe},\ and\ \citenamefont
  {Gross}}]{Chen2022}%
  \BibitemOpen
  \bibfield  {author} {\bibinfo {author} {\bibfnamefont {Q.-M.}\ \bibnamefont
  {Chen}}, \bibinfo {author} {\bibfnamefont {F.}~\bibnamefont {Kronowetter}},
  \bibinfo {author} {\bibfnamefont {F.}~\bibnamefont {Fesquet}}, \bibinfo
  {author} {\bibfnamefont {K.~E.}\ \bibnamefont {Honasoge}}, \bibinfo {author}
  {\bibfnamefont {Y.}~\bibnamefont {Nojiri}}, \bibinfo {author} {\bibfnamefont
  {M.}~\bibnamefont {Renger}}, \bibinfo {author} {\bibfnamefont {K.~G.}\
  \bibnamefont {Fedorov}}, \bibinfo {author} {\bibfnamefont {A.}~\bibnamefont
  {Marx}}, \bibinfo {author} {\bibfnamefont {F.}~\bibnamefont {Deppe}},\ and\
  \bibinfo {author} {\bibfnamefont {R.}~\bibnamefont {Gross}},\ }\bibfield
  {title} {\bibinfo {title} {Tuning and amplifying the interactions in
  superconducting quantum circuits with subradiant qubits},\ }\href
  {https://doi.org/10.1103/PhysRevA.105.012405} {\bibfield  {journal} {\bibinfo
   {journal} {Phys. Rev. A}\ }\textbf {\bibinfo {volume} {105}},\ \bibinfo
  {pages} {012405} (\bibinfo {year} {2022}{\natexlab{b}})}\BibitemShut
  {NoStop}%
\bibitem [{\citenamefont {Ithier}\ \emph {et~al.}(2005)\citenamefont {Ithier},
  \citenamefont {Collin}, \citenamefont {Joyez}, \citenamefont {Meeson},
  \citenamefont {Vion}, \citenamefont {Esteve}, \citenamefont {Chiarello},
  \citenamefont {Shnirman}, \citenamefont {Makhlin}, \citenamefont {Schriefl},\
  and\ \citenamefont {Sch{\"o}n}}]{Ithier2005}%
  \BibitemOpen
  \bibfield  {author} {\bibinfo {author} {\bibfnamefont {G.}~\bibnamefont
  {Ithier}}, \bibinfo {author} {\bibfnamefont {E.}~\bibnamefont {Collin}},
  \bibinfo {author} {\bibfnamefont {P.}~\bibnamefont {Joyez}}, \bibinfo
  {author} {\bibfnamefont {P.~J.}\ \bibnamefont {Meeson}}, \bibinfo {author}
  {\bibfnamefont {D.}~\bibnamefont {Vion}}, \bibinfo {author} {\bibfnamefont
  {D.}~\bibnamefont {Esteve}}, \bibinfo {author} {\bibfnamefont
  {F.}~\bibnamefont {Chiarello}}, \bibinfo {author} {\bibfnamefont
  {A.}~\bibnamefont {Shnirman}}, \bibinfo {author} {\bibfnamefont
  {Y.}~\bibnamefont {Makhlin}}, \bibinfo {author} {\bibfnamefont
  {J.}~\bibnamefont {Schriefl}},\ and\ \bibinfo {author} {\bibfnamefont
  {G.}~\bibnamefont {Sch{\"o}n}},\ }\bibfield  {title} {\bibinfo {title}
  {Decoherence in a superconducting quantum bit circuit},\ }\href
  {https://doi.org/10.1103/PhysRevB.72.134519} {\bibfield  {journal} {\bibinfo
  {journal} {Phys. Rev. B}\ }\textbf {\bibinfo {volume} {72}},\ \bibinfo
  {pages} {134519} (\bibinfo {year} {2005})}\BibitemShut {NoStop}%
\bibitem [{\citenamefont {Rice}(2009)}]{Rice2009}%
  \BibitemOpen
  \bibfield  {author} {\bibinfo {author} {\bibfnamefont {M.}~\bibnamefont
  {Rice}},\ }\href@noop {} {\emph {\bibinfo {title} {Digital Communications: A
  Discrete-Time Approach}}}\ (\bibinfo  {publisher} {Pearson Prentice Hall},\
  \bibinfo {address} {Upper Saddle River, NJ, USA},\ \bibinfo {year}
  {2009})\BibitemShut {NoStop}%
\bibitem [{\citenamefont {Vitanov}\ \emph {et~al.}(2001)\citenamefont
  {Vitanov}, \citenamefont {Shore}, \citenamefont {Yatsenko}, \citenamefont
  {B\"ohmer}, \citenamefont {Halfmann}, \citenamefont {Rickes},\ and\
  \citenamefont {Bergmann}}]{Vitanov2001}%
  \BibitemOpen
  \bibfield  {author} {\bibinfo {author} {\bibfnamefont {N.}~\bibnamefont
  {Vitanov}}, \bibinfo {author} {\bibfnamefont {B.}~\bibnamefont {Shore}},
  \bibinfo {author} {\bibfnamefont {L.}~\bibnamefont {Yatsenko}}, \bibinfo
  {author} {\bibfnamefont {K.}~\bibnamefont {B\"ohmer}}, \bibinfo {author}
  {\bibfnamefont {T.}~\bibnamefont {Halfmann}}, \bibinfo {author}
  {\bibfnamefont {T.}~\bibnamefont {Rickes}},\ and\ \bibinfo {author}
  {\bibfnamefont {K.}~\bibnamefont {Bergmann}},\ }\bibfield  {title} {\bibinfo
  {title} {Power broadening revisited: theory and experiment},\ }\href
  {https://doi.org/10.1016/S0030-4018(01)01495-X} {\bibfield  {journal}
  {\bibinfo  {journal} {Optics Communications}\ }\textbf {\bibinfo {volume}
  {199}},\ \bibinfo {pages} {117} (\bibinfo {year} {2001})}\BibitemShut
  {NoStop}%
\bibitem [{\citenamefont {Chua}\ \emph {et~al.}(1987)\citenamefont {Chua},
  \citenamefont {Desoer},\ and\ \citenamefont {Kuh}}]{Chua1987}%
  \BibitemOpen
  \bibfield  {author} {\bibinfo {author} {\bibfnamefont {L.~O.}\ \bibnamefont
  {Chua}}, \bibinfo {author} {\bibfnamefont {C.~A.}\ \bibnamefont {Desoer}},\
  and\ \bibinfo {author} {\bibfnamefont {E.~S.}\ \bibnamefont {Kuh}},\
  }\href@noop {} {\emph {\bibinfo {title} {Linear and Nonlinear Circuits}}}\
  (\bibinfo  {publisher} {McGraw-Hill},\ \bibinfo {year} {1987})\BibitemShut
  {NoStop}%
\bibitem [{\citenamefont {Oppenheim}\ \emph {et~al.}(1999)\citenamefont
  {Oppenheim}, \citenamefont {Schafer},\ and\ \citenamefont
  {Buck}}]{Oppenheim1999}%
  \BibitemOpen
  \bibfield  {author} {\bibinfo {author} {\bibfnamefont {A.~V.}\ \bibnamefont
  {Oppenheim}}, \bibinfo {author} {\bibfnamefont {R.~W.}\ \bibnamefont
  {Schafer}},\ and\ \bibinfo {author} {\bibfnamefont {J.~R.}\ \bibnamefont
  {Buck}},\ }\href@noop {} {\emph {\bibinfo {title} {Discrete-Time Signal
  Processing}}},\ \bibinfo {edition} {2nd}\ ed.\ (\bibinfo  {publisher}
  {Prentice Hall},\ \bibinfo {address} {Upper Saddle River, NJ},\ \bibinfo
  {year} {1999})\BibitemShut {NoStop}%
\bibitem [{\citenamefont {Wing}(1991)}]{Milton1991}%
  \BibitemOpen
  \bibfield  {author} {\bibinfo {author} {\bibfnamefont {G.~M.}\ \bibnamefont
  {Wing}},\ }\href {https://doi.org/10.1137/1.9781611971675} {\emph {\bibinfo
  {title} {A Primer on Integral Equations of the First Kind}}}\ (\bibinfo
  {publisher} {Society for Industrial and Applied Mathematics},\ \bibinfo
  {year} {1991})\BibitemShut {NoStop}%
\bibitem [{\citenamefont {Borchers}\ and\ \citenamefont
  {Aster}(2013)}]{Borchers13}%
  \BibitemOpen
  \bibfield  {author} {\bibinfo {author} {\bibfnamefont {B.}~\bibnamefont
  {Borchers}}\ and\ \bibinfo {author} {\bibfnamefont {R.}~\bibnamefont
  {Aster}},\ }\href
  {http://www.ceri.memphis.edu/people/mwithers/CERI7106/aster/GEOP505/Docs/deconv.pdf}
  {\bibinfo {title} {Time series/data processing and analysis: Notes on
  deconvolution}} (\bibinfo {year} {2013}),\ \bibinfo {note} {lecture notes,
  University of Memphis}\BibitemShut {NoStop}%
\bibitem [{\citenamefont {Ferreira}\ \emph {et~al.}(2024)\citenamefont
  {Ferreira}, \citenamefont {Kim}, \citenamefont {Butler}, \citenamefont
  {Pichler},\ and\ \citenamefont {Painter}}]{Ferreira2024}%
  \BibitemOpen
  \bibfield  {author} {\bibinfo {author} {\bibfnamefont {V.~S.}\ \bibnamefont
  {Ferreira}}, \bibinfo {author} {\bibfnamefont {G.}~\bibnamefont {Kim}},
  \bibinfo {author} {\bibfnamefont {A.}~\bibnamefont {Butler}}, \bibinfo
  {author} {\bibfnamefont {H.}~\bibnamefont {Pichler}},\ and\ \bibinfo {author}
  {\bibfnamefont {O.}~\bibnamefont {Painter}},\ }\bibfield  {title} {\bibinfo
  {title} {Deterministic generation of multidimensional photonic cluster states
  with a single quantum emitter},\ }\href
  {https://doi.org/10.1038/s41567-024-02408-0} {\bibfield  {journal} {\bibinfo
  {journal} {Nature Physics}\ }\textbf {\bibinfo {volume} {20}},\ \bibinfo
  {pages} {865} (\bibinfo {year} {2024})}\BibitemShut {NoStop}%
\bibitem [{\citenamefont {Magesan}\ \emph {et~al.}(2012)\citenamefont
  {Magesan}, \citenamefont {Gambetta}, \citenamefont {Johnson}, \citenamefont
  {Ryan}, \citenamefont {Chow}, \citenamefont {Merkel}, \citenamefont
  {da~Silva}, \citenamefont {Keefe}, \citenamefont {Rothwell}, \citenamefont
  {Ohki}, \citenamefont {Ketchen},\ and\ \citenamefont
  {Steffen}}]{Magesan2012}%
  \BibitemOpen
  \bibfield  {author} {\bibinfo {author} {\bibfnamefont {E.}~\bibnamefont
  {Magesan}}, \bibinfo {author} {\bibfnamefont {J.~M.}\ \bibnamefont
  {Gambetta}}, \bibinfo {author} {\bibfnamefont {B.~R.}\ \bibnamefont
  {Johnson}}, \bibinfo {author} {\bibfnamefont {C.~A.}\ \bibnamefont {Ryan}},
  \bibinfo {author} {\bibfnamefont {J.~M.}\ \bibnamefont {Chow}}, \bibinfo
  {author} {\bibfnamefont {S.~T.}\ \bibnamefont {Merkel}}, \bibinfo {author}
  {\bibfnamefont {M.~P.}\ \bibnamefont {da~Silva}}, \bibinfo {author}
  {\bibfnamefont {G.~A.}\ \bibnamefont {Keefe}}, \bibinfo {author}
  {\bibfnamefont {M.~B.}\ \bibnamefont {Rothwell}}, \bibinfo {author}
  {\bibfnamefont {T.~A.}\ \bibnamefont {Ohki}}, \bibinfo {author}
  {\bibfnamefont {M.~B.}\ \bibnamefont {Ketchen}},\ and\ \bibinfo {author}
  {\bibfnamefont {M.}~\bibnamefont {Steffen}},\ }\bibfield  {title} {\bibinfo
  {title} {Efficient measurement of quantum gate error by interleaved
  randomized benchmarking},\ }\href
  {https://doi.org/10.1103/PhysRevLett.109.080505} {\bibfield  {journal}
  {\bibinfo  {journal} {Phys. Rev. Lett.}\ }\textbf {\bibinfo {volume} {109}},\
  \bibinfo {pages} {080505} (\bibinfo {year} {2012})}\BibitemShut {NoStop}%
\bibitem [{\citenamefont {C\'orcoles}\ \emph {et~al.}(2013)\citenamefont
  {C\'orcoles}, \citenamefont {Gambetta}, \citenamefont {Chow}, \citenamefont
  {Smolin}, \citenamefont {Ware}, \citenamefont {Strand}, \citenamefont
  {Plourde},\ and\ \citenamefont {Steffen}}]{Corcoles2013}%
  \BibitemOpen
  \bibfield  {author} {\bibinfo {author} {\bibfnamefont {A.~D.}\ \bibnamefont
  {C\'orcoles}}, \bibinfo {author} {\bibfnamefont {J.~M.}\ \bibnamefont
  {Gambetta}}, \bibinfo {author} {\bibfnamefont {J.~M.}\ \bibnamefont {Chow}},
  \bibinfo {author} {\bibfnamefont {J.~A.}\ \bibnamefont {Smolin}}, \bibinfo
  {author} {\bibfnamefont {M.}~\bibnamefont {Ware}}, \bibinfo {author}
  {\bibfnamefont {J.}~\bibnamefont {Strand}}, \bibinfo {author} {\bibfnamefont
  {B.~L.~T.}\ \bibnamefont {Plourde}},\ and\ \bibinfo {author} {\bibfnamefont
  {M.}~\bibnamefont {Steffen}},\ }\bibfield  {title} {\bibinfo {title} {Process
  verification of two-qubit quantum gates by randomized benchmarking},\ }\href
  {https://doi.org/10.1103/PhysRevA.87.030301} {\bibfield  {journal} {\bibinfo
  {journal} {Phys. Rev. A}\ }\textbf {\bibinfo {volume} {87}},\ \bibinfo
  {pages} {030301} (\bibinfo {year} {2013})}\BibitemShut {NoStop}%
\bibitem [{\citenamefont {Wood}\ and\ \citenamefont
  {Gambetta}(2018)}]{Wood2017}%
  \BibitemOpen
  \bibfield  {author} {\bibinfo {author} {\bibfnamefont {C.~J.}\ \bibnamefont
  {Wood}}\ and\ \bibinfo {author} {\bibfnamefont {J.~M.}\ \bibnamefont
  {Gambetta}},\ }\bibfield  {title} {\bibinfo {title} {Quantification and
  characterization of leakage errors},\ }\href
  {https://doi.org/10.1103/PhysRevA.97.032306} {\bibfield  {journal} {\bibinfo
  {journal} {Phys. Rev. A}\ }\textbf {\bibinfo {volume} {97}},\ \bibinfo
  {pages} {032306} (\bibinfo {year} {2018})}\BibitemShut {NoStop}%
\bibitem [{\citenamefont {Karamlou}\ \emph {et~al.}(2022)\citenamefont
  {Karamlou}, \citenamefont {Braum\"uller}, \citenamefont {Yanay},
  \citenamefont {Di~Paolo}, \citenamefont {Harrington}, \citenamefont {Kannan},
  \citenamefont {Kim}, \citenamefont {Kjaergaard}, \citenamefont {Melville},
  \citenamefont {Muschinske}, \citenamefont {Niedzielski}, \citenamefont
  {Veps\"al\"ainen}, \citenamefont {Winik}, \citenamefont {Yoder},
  \citenamefont {Schwartz}, \citenamefont {Tahan}, \citenamefont {Orlando},
  \citenamefont {Gustavsson},\ and\ \citenamefont {Oliver}}]{Karamlou2022}%
  \BibitemOpen
  \bibfield  {author} {\bibinfo {author} {\bibfnamefont {A.~H.}\ \bibnamefont
  {Karamlou}}, \bibinfo {author} {\bibfnamefont {J.}~\bibnamefont
  {Braum\"uller}}, \bibinfo {author} {\bibfnamefont {Y.}~\bibnamefont {Yanay}},
  \bibinfo {author} {\bibfnamefont {A.}~\bibnamefont {Di~Paolo}}, \bibinfo
  {author} {\bibfnamefont {P.~M.}\ \bibnamefont {Harrington}}, \bibinfo
  {author} {\bibfnamefont {B.}~\bibnamefont {Kannan}}, \bibinfo {author}
  {\bibfnamefont {D.}~\bibnamefont {Kim}}, \bibinfo {author} {\bibfnamefont
  {M.}~\bibnamefont {Kjaergaard}}, \bibinfo {author} {\bibfnamefont
  {A.}~\bibnamefont {Melville}}, \bibinfo {author} {\bibfnamefont
  {S.}~\bibnamefont {Muschinske}}, \bibinfo {author} {\bibfnamefont {B.~M.}\
  \bibnamefont {Niedzielski}}, \bibinfo {author} {\bibfnamefont
  {A.}~\bibnamefont {Veps\"al\"ainen}}, \bibinfo {author} {\bibfnamefont
  {R.}~\bibnamefont {Winik}}, \bibinfo {author} {\bibfnamefont {J.~L.}\
  \bibnamefont {Yoder}}, \bibinfo {author} {\bibfnamefont {M.}~\bibnamefont
  {Schwartz}}, \bibinfo {author} {\bibfnamefont {C.}~\bibnamefont {Tahan}},
  \bibinfo {author} {\bibfnamefont {T.~P.}\ \bibnamefont {Orlando}}, \bibinfo
  {author} {\bibfnamefont {S.}~\bibnamefont {Gustavsson}},\ and\ \bibinfo
  {author} {\bibfnamefont {W.~D.}\ \bibnamefont {Oliver}},\ }\bibfield  {title}
  {\bibinfo {title} {Quantum transport and localization in 1d and 2d
  tight-binding lattices},\ }\href {https://doi.org/10.1038/s41534-022-00528-0}
  {\bibfield  {journal} {\bibinfo  {journal} {npj Quantum Information}\
  }\textbf {\bibinfo {volume} {8}},\ \bibinfo {pages} {35} (\bibinfo {year}
  {2022})}\BibitemShut {NoStop}%
\bibitem [{\citenamefont {M\"{u}ller}\ \emph {et~al.}(2019)\citenamefont
  {M\"{u}ller}, \citenamefont {Cole},\ and\ \citenamefont
  {Lisenfeld}}]{Mueller2019}%
  \BibitemOpen
  \bibfield  {author} {\bibinfo {author} {\bibfnamefont {C.}~\bibnamefont
  {M\"{u}ller}}, \bibinfo {author} {\bibfnamefont {J.~H.}\ \bibnamefont
  {Cole}},\ and\ \bibinfo {author} {\bibfnamefont {J.}~\bibnamefont
  {Lisenfeld}},\ }\bibfield  {title} {\bibinfo {title} {Towards understanding
  two-level-systems in amorphous solids: insights from quantum circuits},\
  }\href {https://doi.org/10.1088/1361-6633/ab3a7e} {\bibfield  {journal}
  {\bibinfo  {journal} {Reports on Progress in Physics}\ }\textbf {\bibinfo
  {volume} {82}},\ \bibinfo {pages} {124501} (\bibinfo {year}
  {2019})}\BibitemShut {NoStop}%
\bibitem [{\citenamefont {Earnest}\ \emph {et~al.}(2018)\citenamefont
  {Earnest}, \citenamefont {Chakram}, \citenamefont {Lu}, \citenamefont
  {Irons}, \citenamefont {Naik}, \citenamefont {Leung}, \citenamefont {Ocola},
  \citenamefont {Czaplewski}, \citenamefont {Baker}, \citenamefont {Lawrence},
  \citenamefont {Koch},\ and\ \citenamefont {Schuster}}]{Earnest2018b}%
  \BibitemOpen
  \bibfield  {author} {\bibinfo {author} {\bibfnamefont {N.}~\bibnamefont
  {Earnest}}, \bibinfo {author} {\bibfnamefont {S.}~\bibnamefont {Chakram}},
  \bibinfo {author} {\bibfnamefont {Y.}~\bibnamefont {Lu}}, \bibinfo {author}
  {\bibfnamefont {N.}~\bibnamefont {Irons}}, \bibinfo {author} {\bibfnamefont
  {R.~K.}\ \bibnamefont {Naik}}, \bibinfo {author} {\bibfnamefont
  {N.}~\bibnamefont {Leung}}, \bibinfo {author} {\bibfnamefont
  {L.}~\bibnamefont {Ocola}}, \bibinfo {author} {\bibfnamefont {D.~A.}\
  \bibnamefont {Czaplewski}}, \bibinfo {author} {\bibfnamefont
  {B.}~\bibnamefont {Baker}}, \bibinfo {author} {\bibfnamefont
  {J.}~\bibnamefont {Lawrence}}, \bibinfo {author} {\bibfnamefont
  {J.}~\bibnamefont {Koch}},\ and\ \bibinfo {author} {\bibfnamefont {D.~I.}\
  \bibnamefont {Schuster}},\ }\bibfield  {title} {\bibinfo {title} {Realization
  of a $\mathrm{\ensuremath{\Lambda}}$ system with metastable states of a
  capacitively shunted fluxonium},\ }\href
  {https://doi.org/10.1103/PhysRevLett.120.150504} {\bibfield  {journal}
  {\bibinfo  {journal} {Phys. Rev. Lett.}\ }\textbf {\bibinfo {volume} {120}},\
  \bibinfo {pages} {150504} (\bibinfo {year} {2018})}\BibitemShut {NoStop}%
\bibitem [{\citenamefont {Grover}\ \emph {et~al.}(2020)\citenamefont {Grover},
  \citenamefont {Basham}, \citenamefont {Marakov}, \citenamefont {Disseler},
  \citenamefont {Hinkey}, \citenamefont {Khalil}, \citenamefont {Stegen},
  \citenamefont {Chamberlin}, \citenamefont {DeGottardi}, \citenamefont
  {Clarke}, \citenamefont {Medford}, \citenamefont {Strand}, \citenamefont
  {Stoutimore}, \citenamefont {Novikov}, \citenamefont {Ferguson},
  \citenamefont {Lidar}, \citenamefont {Zick},\ and\ \citenamefont
  {Przybysz}}]{Grover2020}%
  \BibitemOpen
  \bibfield  {author} {\bibinfo {author} {\bibfnamefont {J.~A.}\ \bibnamefont
  {Grover}}, \bibinfo {author} {\bibfnamefont {J.~I.}\ \bibnamefont {Basham}},
  \bibinfo {author} {\bibfnamefont {A.}~\bibnamefont {Marakov}}, \bibinfo
  {author} {\bibfnamefont {S.~M.}\ \bibnamefont {Disseler}}, \bibinfo {author}
  {\bibfnamefont {R.~T.}\ \bibnamefont {Hinkey}}, \bibinfo {author}
  {\bibfnamefont {M.}~\bibnamefont {Khalil}}, \bibinfo {author} {\bibfnamefont
  {Z.~A.}\ \bibnamefont {Stegen}}, \bibinfo {author} {\bibfnamefont
  {T.}~\bibnamefont {Chamberlin}}, \bibinfo {author} {\bibfnamefont
  {W.}~\bibnamefont {DeGottardi}}, \bibinfo {author} {\bibfnamefont {D.~J.}\
  \bibnamefont {Clarke}}, \bibinfo {author} {\bibfnamefont {J.~R.}\
  \bibnamefont {Medford}}, \bibinfo {author} {\bibfnamefont {J.~D.}\
  \bibnamefont {Strand}}, \bibinfo {author} {\bibfnamefont {M.~J.~A.}\
  \bibnamefont {Stoutimore}}, \bibinfo {author} {\bibfnamefont
  {S.}~\bibnamefont {Novikov}}, \bibinfo {author} {\bibfnamefont {D.~G.}\
  \bibnamefont {Ferguson}}, \bibinfo {author} {\bibfnamefont {D.}~\bibnamefont
  {Lidar}}, \bibinfo {author} {\bibfnamefont {K.~M.}\ \bibnamefont {Zick}},\
  and\ \bibinfo {author} {\bibfnamefont {A.~J.}\ \bibnamefont {Przybysz}},\
  }\bibfield  {title} {\bibinfo {title} {Fast, lifetime-preserving readout for
  high-coherence quantum annealers},\ }\href
  {https://doi.org/10.1103/PRXQuantum.01.020314} {\bibfield  {journal}
  {\bibinfo  {journal} {PRX Quantum}\ }\textbf {\bibinfo {volume} {1}},\
  \bibinfo {pages} {020314} (\bibinfo {year} {2020})}\BibitemShut {NoStop}%
\bibitem [{\citenamefont {Blais}\ \emph {et~al.}(2021)\citenamefont {Blais},
  \citenamefont {Grimsmo}, \citenamefont {Girvin},\ and\ \citenamefont
  {Wallraff}}]{Blais2021}%
  \BibitemOpen
  \bibfield  {author} {\bibinfo {author} {\bibfnamefont {A.}~\bibnamefont
  {Blais}}, \bibinfo {author} {\bibfnamefont {A.~L.}\ \bibnamefont {Grimsmo}},
  \bibinfo {author} {\bibfnamefont {S.~M.}\ \bibnamefont {Girvin}},\ and\
  \bibinfo {author} {\bibfnamefont {A.}~\bibnamefont {Wallraff}},\ }\bibfield
  {title} {\bibinfo {title} {Circuit quantum electrodynamics},\ }\href
  {https://doi.org/10.1103/RevModPhys.93.025005} {\bibfield  {journal}
  {\bibinfo  {journal} {Rev. Mod. Phys.}\ }\textbf {\bibinfo {volume} {93}},\
  \bibinfo {pages} {025005} (\bibinfo {year} {2021})}\BibitemShut {NoStop}%
\bibitem [{\citenamefont {Krinner}\ \emph {et~al.}(2019)\citenamefont
  {Krinner}, \citenamefont {Storz}, \citenamefont {Kurpiers}, \citenamefont
  {Magnard}, \citenamefont {Heinsoo}, \citenamefont {Keller}, \citenamefont
  {L{\"u}tolf}, \citenamefont {Eichler},\ and\ \citenamefont
  {Wallraff}}]{Krinner2019}%
  \BibitemOpen
  \bibfield  {author} {\bibinfo {author} {\bibfnamefont {S.}~\bibnamefont
  {Krinner}}, \bibinfo {author} {\bibfnamefont {S.}~\bibnamefont {Storz}},
  \bibinfo {author} {\bibfnamefont {P.}~\bibnamefont {Kurpiers}}, \bibinfo
  {author} {\bibfnamefont {P.}~\bibnamefont {Magnard}}, \bibinfo {author}
  {\bibfnamefont {J.}~\bibnamefont {Heinsoo}}, \bibinfo {author} {\bibfnamefont
  {R.}~\bibnamefont {Keller}}, \bibinfo {author} {\bibfnamefont
  {J.}~\bibnamefont {L{\"u}tolf}}, \bibinfo {author} {\bibfnamefont
  {C.}~\bibnamefont {Eichler}},\ and\ \bibinfo {author} {\bibfnamefont
  {A.}~\bibnamefont {Wallraff}},\ }\bibfield  {title} {\bibinfo {title}
  {Engineering cryogenic setups for 100-qubit scale superconducting circuit
  systems},\ }\href {https://doi.org/10.1140/epjqt/s40507-019-0072-0}
  {\bibfield  {journal} {\bibinfo  {journal} {EPJ Quantum Technology}\ }\textbf
  {\bibinfo {volume} {6}},\ \bibinfo {pages} {2} (\bibinfo {year}
  {2019})}\BibitemShut {NoStop}%
\bibitem [{\citenamefont {Macklin}\ \emph {et~al.}(2015)\citenamefont
  {Macklin}, \citenamefont {O'Brien}, \citenamefont {Hover}, \citenamefont
  {Schwartz}, \citenamefont {Bolkhovsky}, \citenamefont {Zhang}, \citenamefont
  {Oliver},\ and\ \citenamefont {Siddiqi}}]{Macklin2015}%
  \BibitemOpen
  \bibfield  {author} {\bibinfo {author} {\bibfnamefont {C.}~\bibnamefont
  {Macklin}}, \bibinfo {author} {\bibfnamefont {K.}~\bibnamefont {O'Brien}},
  \bibinfo {author} {\bibfnamefont {D.}~\bibnamefont {Hover}}, \bibinfo
  {author} {\bibfnamefont {M.~E.}\ \bibnamefont {Schwartz}}, \bibinfo {author}
  {\bibfnamefont {V.}~\bibnamefont {Bolkhovsky}}, \bibinfo {author}
  {\bibfnamefont {X.}~\bibnamefont {Zhang}}, \bibinfo {author} {\bibfnamefont
  {W.~D.}\ \bibnamefont {Oliver}},\ and\ \bibinfo {author} {\bibfnamefont
  {I.}~\bibnamefont {Siddiqi}},\ }\bibfield  {title} {\bibinfo {title} {A
  near-quantum-limited {Josephson} traveling-wave parametric amplifier},\
  }\href {https://doi.org/10.1126/science.aaa8525} {\bibfield  {journal}
  {\bibinfo  {journal} {Science}\ }\textbf {\bibinfo {volume} {350}},\ \bibinfo
  {pages} {307} (\bibinfo {year} {2015})}\BibitemShut {NoStop}%
\bibitem [{\citenamefont {Baldwin}\ and\ \citenamefont
  {Dubbert}(2002)}]{Sandia2002}%
  \BibitemOpen
  \bibfield  {author} {\bibinfo {author} {\bibfnamefont {J.~G.}\ \bibnamefont
  {Baldwin}}\ and\ \bibinfo {author} {\bibfnamefont {D.~F.}\ \bibnamefont
  {Dubbert}},\ }\href {https://www.osti.gov/servlets/purl/800958/} {\emph
  {\bibinfo {title} {Quadrature Mixer LO Leakage Suppression Through Quadrature
  {DC} Bias}}},\ \bibinfo {type} {Tech. Rep.}\ (\bibinfo  {institution} {Sandia
  National Laboratories},\ \bibinfo {year} {2002})\BibitemShut {NoStop}%
\bibitem [{\citenamefont {Herrmann}\ \emph {et~al.}(2022)\citenamefont
  {Herrmann}, \citenamefont {Hellings}, \citenamefont {Lazar}, \citenamefont
  {Pf\"affli}, \citenamefont {Haupt}, \citenamefont {Thiele}, \citenamefont
  {Zanuz}, \citenamefont {Norris}, \citenamefont {Heer}, \citenamefont
  {Eichler},\ and\ \citenamefont {Wallraff}}]{Herrmann2022a}%
  \BibitemOpen
  \bibfield  {author} {\bibinfo {author} {\bibfnamefont {J.}~\bibnamefont
  {Herrmann}}, \bibinfo {author} {\bibfnamefont {C.}~\bibnamefont {Hellings}},
  \bibinfo {author} {\bibfnamefont {S.}~\bibnamefont {Lazar}}, \bibinfo
  {author} {\bibfnamefont {F.}~\bibnamefont {Pf\"affli}}, \bibinfo {author}
  {\bibfnamefont {F.}~\bibnamefont {Haupt}}, \bibinfo {author} {\bibfnamefont
  {T.}~\bibnamefont {Thiele}}, \bibinfo {author} {\bibfnamefont {D.~C.}\
  \bibnamefont {Zanuz}}, \bibinfo {author} {\bibfnamefont {G.~J.}\ \bibnamefont
  {Norris}}, \bibinfo {author} {\bibfnamefont {F.}~\bibnamefont {Heer}},
  \bibinfo {author} {\bibfnamefont {C.}~\bibnamefont {Eichler}},\ and\ \bibinfo
  {author} {\bibfnamefont {A.}~\bibnamefont {Wallraff}},\ }\bibfield  {title}
  {\bibinfo {title} {Frequency up-conversion schemes for controlling
  superconducting qubits},\ }\href {https://arxiv.org/abs/2210.02513}
  {\bibfield  {journal} {\bibinfo  {journal} {arXiv:2210.02513}\ } (\bibinfo
  {year} {2022})}\BibitemShut {NoStop}%
\bibitem [{\citenamefont {Jolin}\ \emph {et~al.}(2020)\citenamefont {Jolin},
  \citenamefont {Borgani}, \citenamefont {Thol{\'{e}}n}, \citenamefont
  {Forchheimer},\ and\ \citenamefont {Haviland}}]{Jolin2020}%
  \BibitemOpen
  \bibfield  {author} {\bibinfo {author} {\bibfnamefont {S.~W.}\ \bibnamefont
  {Jolin}}, \bibinfo {author} {\bibfnamefont {R.}~\bibnamefont {Borgani}},
  \bibinfo {author} {\bibfnamefont {M.~O.}\ \bibnamefont {Thol{\'{e}}n}},
  \bibinfo {author} {\bibfnamefont {D.}~\bibnamefont {Forchheimer}},\ and\
  \bibinfo {author} {\bibfnamefont {D.~B.}\ \bibnamefont {Haviland}},\
  }\bibfield  {title} {\bibinfo {title} {Calibration of mixer amplitude and
  phase imbalance in superconducting circuits},\ }\href
  {https://doi.org/10.1063/5.0025836} {\bibfield  {journal} {\bibinfo
  {journal} {Rev. Sci. Instrum.}\ }\textbf {\bibinfo {volume} {91}},\ \bibinfo
  {pages} {124707} (\bibinfo {year} {2020})}\BibitemShut {NoStop}%
\bibitem [{\citenamefont {Abramowitz}\ and\ \citenamefont
  {Stegun}(1972)}]{Abramowitz1972}%
  \BibitemOpen
  \bibinfo {editor} {\bibfnamefont {M.}~\bibnamefont {Abramowitz}}\ and\
  \bibinfo {editor} {\bibfnamefont {I.}~\bibnamefont {Stegun}},\ eds.,\
  \href@noop {} {\emph {\bibinfo {title} {Handbook of Mathematical
  Functions}}}\ (\bibinfo  {publisher} {United States Department of Commerce},\
  \bibinfo {year} {1972})\BibitemShut {NoStop}%
\bibitem [{\citenamefont {Collins}\ and\ \citenamefont
  {Krandick}(1992)}]{Collins1992}%
  \BibitemOpen
  \bibfield  {author} {\bibinfo {author} {\bibfnamefont {G.~E.}\ \bibnamefont
  {Collins}}\ and\ \bibinfo {author} {\bibfnamefont {W.}~\bibnamefont
  {Krandick}},\ }\bibfield  {title} {\bibinfo {title} {An efficient algorithm
  for infallible polynomial complex root isolation},\ }in\ \href
  {https://doi.org/10.1145/143242.143308} {\emph {\bibinfo {booktitle}
  {Proceedings of International Symposium on Symbolic and Algebraic
  Computation}}}\ (\bibinfo  {publisher} {ACM Press},\ \bibinfo {year} {1992})\
  pp.\ \bibinfo {pages} {189--194}\BibitemShut {NoStop}%
\bibitem [{\citenamefont {Oppenheim}\ \emph {et~al.}(1997)\citenamefont
  {Oppenheim}, \citenamefont {Willsky},\ and\ \citenamefont
  {Nawab}}]{Oppenheim1997}%
  \BibitemOpen
  \bibfield  {author} {\bibinfo {author} {\bibfnamefont {A.~S.}\ \bibnamefont
  {Oppenheim}}, \bibinfo {author} {\bibfnamefont {A.~S.}\ \bibnamefont
  {Willsky}},\ and\ \bibinfo {author} {\bibfnamefont {S.~H.}\ \bibnamefont
  {Nawab}},\ }\href@noop {} {\emph {\bibinfo {title} {Signals and Systems}}}\
  (\bibinfo  {publisher} {Prentice Hall},\ \bibinfo {year} {1997})\BibitemShut
  {NoStop}%
\bibitem [{\citenamefont {Gavin}(2024)}]{Gavin2024}%
  \BibitemOpen
  \bibfield  {author} {\bibinfo {author} {\bibfnamefont {H.~P.}\ \bibnamefont
  {Gavin}},\ }\href {https://people.duke.edu/~hpgavin/lm.pdf} {\emph {\bibinfo
  {title} {The {L}evenberg-{M}arquardt algorithm for nonlinear least squares
  curve-fitting problems}}},\ \bibinfo {type} {Tech. Rep.}\ (\bibinfo
  {institution} {Department of Civil and Environmental Engineering, Duke
  University},\ \bibinfo {address} {Durham, North Carolina},\ \bibinfo {year}
  {2024})\BibitemShut {NoStop}%
\bibitem [{\citenamefont {Salath\'e}\ \emph {et~al.}(2018)\citenamefont
  {Salath\'e}, \citenamefont {Kurpiers}, \citenamefont {Karg}, \citenamefont
  {Lang}, \citenamefont {Andersen}, \citenamefont {Akin}, \citenamefont
  {Krinner}, \citenamefont {Eichler},\ and\ \citenamefont
  {Wallraff}}]{Salathe2018}%
  \BibitemOpen
  \bibfield  {author} {\bibinfo {author} {\bibfnamefont {Y.}~\bibnamefont
  {Salath\'e}}, \bibinfo {author} {\bibfnamefont {P.}~\bibnamefont {Kurpiers}},
  \bibinfo {author} {\bibfnamefont {T.}~\bibnamefont {Karg}}, \bibinfo {author}
  {\bibfnamefont {C.}~\bibnamefont {Lang}}, \bibinfo {author} {\bibfnamefont
  {C.~K.}\ \bibnamefont {Andersen}}, \bibinfo {author} {\bibfnamefont
  {A.}~\bibnamefont {Akin}}, \bibinfo {author} {\bibfnamefont {S.}~\bibnamefont
  {Krinner}}, \bibinfo {author} {\bibfnamefont {C.}~\bibnamefont {Eichler}},\
  and\ \bibinfo {author} {\bibfnamefont {A.}~\bibnamefont {Wallraff}},\
  }\bibfield  {title} {\bibinfo {title} {Low-latency digital signal processing
  for feedback and feedforward in quantum computing and communication},\ }\href
  {https://doi.org/10.1103/PhysRevApplied.9.034011} {\bibfield  {journal}
  {\bibinfo  {journal} {Phys. Rev. Appl.}\ }\textbf {\bibinfo {volume} {9}},\
  \bibinfo {pages} {034011} (\bibinfo {year} {2018})}\BibitemShut {NoStop}%
\bibitem [{\citenamefont {Riad}(1986)}]{Riad1986}%
  \BibitemOpen
  \bibfield  {author} {\bibinfo {author} {\bibfnamefont {S.~M.}\ \bibnamefont
  {Riad}},\ }\bibfield  {title} {\bibinfo {title} {{The deconvolution problem:
  An overview}},\ }\href {https://doi.org/10.1109/PROC.1986.13407} {\bibfield
  {journal} {\bibinfo  {journal} {Proceedings of the IEEE}\ }\textbf {\bibinfo
  {volume} {74}},\ \bibinfo {pages} {82} (\bibinfo {year} {1986})}\BibitemShut
  {NoStop}%
\bibitem [{\citenamefont {Neumaier}(1998)}]{Neumaier1998}%
  \BibitemOpen
  \bibfield  {author} {\bibinfo {author} {\bibfnamefont {A.}~\bibnamefont
  {Neumaier}},\ }\bibfield  {title} {\bibinfo {title} {Solving ill-conditioned
  and singular linear systems: A tutorial on regularization},\ }\href
  {http://www.jstor.org/stable/2653234} {\bibfield  {journal} {\bibinfo
  {journal} {SIAM Review}\ }\textbf {\bibinfo {volume} {40}},\ \bibinfo {pages}
  {636} (\bibinfo {year} {1998})}\BibitemShut {NoStop}%
\bibitem [{\citenamefont {Tikhonov}(1963)}]{Tikhonov1963}%
  \BibitemOpen
  \bibfield  {author} {\bibinfo {author} {\bibfnamefont {A.~N.}\ \bibnamefont
  {Tikhonov}},\ }\bibfield  {title} {\bibinfo {title} {On the solution of
  ill-posed problems and the method of regularization},\ }in\ \href@noop {}
  {\emph {\bibinfo {booktitle} {Doklady Akademii Nauk}}},\ Vol.\ \bibinfo
  {volume} {151}\ (\bibinfo {organization} {Russian Academy of Sciences},\
  \bibinfo {year} {1963})\ pp.\ \bibinfo {pages} {501--504}\BibitemShut
  {NoStop}%
\bibitem [{\citenamefont {Adams}\ and\ \citenamefont
  {Fournier}(2003)}]{Adams2003}%
  \BibitemOpen
  \bibfield  {author} {\bibinfo {author} {\bibfnamefont {R.}~\bibnamefont
  {Adams}}\ and\ \bibinfo {author} {\bibfnamefont {J.}~\bibnamefont
  {Fournier}},\ }\href
  {https://www.elsevier.com/books/sobolev-spaces/adams/978-0-12-044143-3}
  {\emph {\bibinfo {title} {{Sobolev Spaces}}}}\ (\bibinfo  {publisher}
  {Academic Press},\ \bibinfo {address} {Cambridge, MA, USA},\ \bibinfo {year}
  {2003})\BibitemShut {NoStop}%
\end{thebibliography}%

\end{document}